\documentclass[aps,prx,twocolumn,showpacs,superscriptaddress,floatfix]{revtex4-1}
\usepackage{comment}
\usepackage{graphics,amssymb,amsmath,epsfig,color,textgreek}
\usepackage{graphicx}
\usepackage[dvipsnames]{xcolor}
\usepackage{dcolumn}
\usepackage{bm}
\usepackage[colorlinks=true,citecolor=cyan]{hyperref}
\hypersetup{colorlinks=true,citecolor=cyan,linkcolor=red,urlcolor=magenta}
\usepackage{braket}
\usepackage[normalem]{ulem}
\usepackage{cancel}
\usepackage{diagbox}
\usepackage{lipsum}
\usepackage{svg}

\usepackage{ORCIDinREVTeX}

\usepackage[dvipsnames]{xcolor}

\newcommand{\op}[1]{\mathbf{#1}}

\newcommand{\ham}{\mathcal{H}}

\begin{document}

\title{A linear response framework for simulating bosonic and fermionic correlation functions illustrated on quantum computers}

\author{Efekan K\"okc\"u}
\orcid{0000-0002-7323-7274}
\affiliation{Department of Physics, North Carolina State University, Raleigh, North Carolina 27695, USA}

\author{Heba A. Labib}
\orcid{0000-0002-6929-9114}
\affiliation{Department of Physics, North Carolina State University, Raleigh, North Carolina 27695, USA}

\author{J.~K.~Freericks}
\orcid{0000-0002-6232-9165}
\affiliation{Department of Physics, Georgetown University, 37th and O Sts. NW, Washington, DC 20057 USA}

\author{A.~F.~Kemper}
\orcid{0000-0002-5426-5181}
\email{akemper@ncsu.edu}
\affiliation{Department of Physics, North Carolina State University, Raleigh, North Carolina 27695, USA}

\date{\today{}}

\begin{abstract}
Response functions are a fundamental aspect of physics; they represent the link between experimental observations and the underlying quantum many-body state. However, this link is often under-appreciated, as the Lehmann formalism for obtaining response functions in linear response has no direct link to experiment. Within the context of quantum computing, 
and by using a linear response framework, 
we restore this link by making the experiment an inextricable part of the quantum simulation. This method can be frequency- and momentum-selective, avoids limitations on operators that can be directly measured, and is ancilla-free.
As prototypical examples of response functions, we demonstrate that both bosonic and fermionic Green's functions can be obtained, and apply these ideas to the study of a charge-density-wave material on {\emph{ibm\_auckland}}. The linear response method provides a robust framework for using quantum computers to study systems in physics and chemistry. It also provides new paradigms for computing response functions on classical computers.

\end{abstract}

\maketitle

\section*{Introduction}

Quantum computers are showing promise as quantum simulators of many-body physics, with the
hope of being able to further our
understanding of complex
interacting systems.
In order to realize this promise, a key task is to compute response functions for a prepared
many-body state. They represent the experimental measurements that are performed
on the physical realizations of such systems,
and 
computing them via simulation is a critical step in connecting to
experiments and building an understanding of the physics they contain. 
Examples of experiments that measure response functions 
are neutron scattering, optical spectroscopy, and angle-resolved photoemission spectroscopy (ARPES), which measure
the spin-spin correlation, current-current correlation, and single-particle Green's function, respectively\cite{Mahan,stefanucci_nonequilibrium_2013}. The first
two are bosonic correlation functions, while the
latter is a fermionic correlation function.
Both of these contain valuable information ---
both have direct links to experiments, and in
addition the electronic Green's function
is a key ingredient in
hybrid-classical algorithms such as dynamical
mean field theory\cite{georges1996dynamical,Zgid2011,
rungger2019dynamical,keen2020quantum, steckmann2021simulating,jamet2022quantum}.

There are several techniques for
computing correlation functions on quantum computers. The primary tool is based on Hadamard
test circuits\cite{chiesa2019quantum,Roggero19,francis2020quantum,Kosugi20,Kosugi20linear,endo2020calculation,libbi2022effective}; alternatives
include variational approaches\cite{chen2021variational,gyawali2021insights,lee2022variational,jensen2022nearterm,huang2022variational},
spectral decomposition\cite{Ciavarella20,Roggero20,keen2021quantum}, and linear systems
of equation solvers\cite{Tong21}. Each
of these has their own advantages and 
disadvantages, based on the particular quantum
algorithms and hardware at hand. For example, one of the challenges with the Hadamard test is the need to maintain coherence between the ancilla and the system for the potentially long length of time in the measured correlation function in the presence of decoherence and noise.

\begin{figure}[t]
\includegraphics[width=0.48\textwidth]{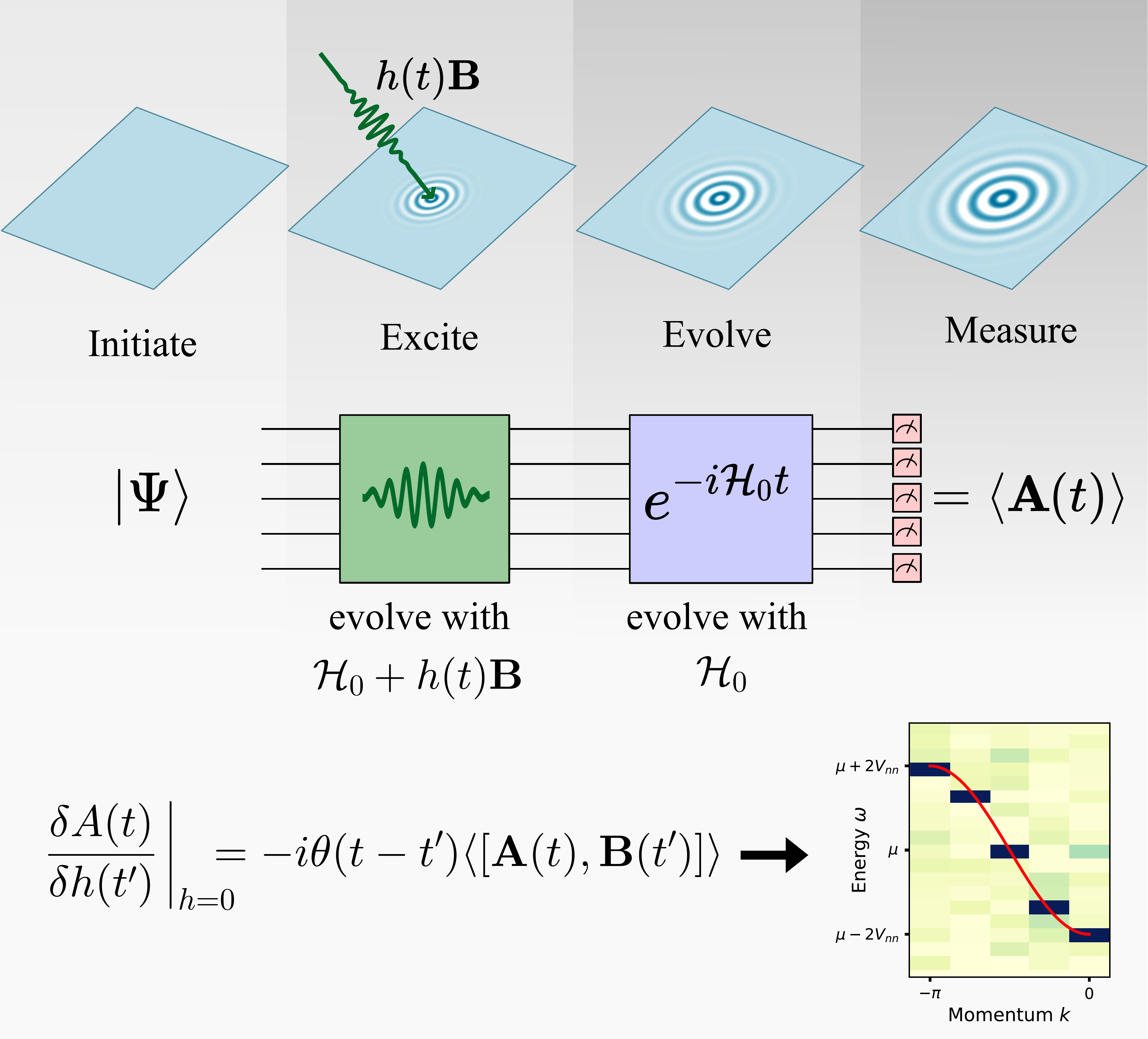}
\caption{\textbf{Linear response method.} 
We establish an equivalence between the experimental
measurement of a response function and an
ancilla-free quantum
simulation under a time dependent Hamiltonian
that includes the perturbative excitation
$h(t) \op{B}$.
Following excitation, the system is evolved
under $\ham_0$, and $\op{A}$ is measured.
The functional derivative of $A(t) = \braket{\op{A}(t)}$ with respect to $h(t')$ yields the retarded response function shown in the
figure. The data shown is taken from Fig.~\ref{fig:ssh_model}.
}
\label{fig:summary}
\end{figure}

In this work, we outline a method for calculating correlation functions
based on a \emph{linear response} framework that is
in direct correspondence to experiments, as
schematically illustrated in 
Fig.~\ref{fig:summary}.
The quantum state is driven with an
applied field with specific temporal and spatial structure, and the response
of the system to that field is measured as a 
function of
space and time. 
The proportionality between
the field and the response then yield the desired correlation function(s).

The linear response framework has several advantages.
First, a judicious choice of the applied field enables
frequency- and momentum-selectivity in obtaining the desired correlation function(s); in particular, momentum-selectivity can significantly reduce circuit
noise.
Second, for systems that conserve total momentum, 
correlation functions in momentum space can be obtained with a single
quantum circuit. And finally, the operators in the correlation function do not have to be unitary,
and can even be non-Hermitian through block encoding\cite{gilyen2019quantum}.

We demonstrate the power of the linear response
framework by applying it to the study of a
model charge density wave system --- the
Su-Schrieffer-Heeger model\cite{ssh1}.
We use the two fermionic methods with a 
momentum-selective field to obtain
the electronic spectrum as would be measured
by ARPES on IBM
quantum hardware, and on a noisy simulator to
compare the linear-response method to
the Hadamard-test method.
We next use the bosonic method and frequency
selectivity
to obtain the density-density response
function of the same model system,
as would be measured by
momentum-resolved
electron energy loss spectroscopy (M-EELS).
These developments make significant inroads
to being able to use near-term quantum computers
in real-world applications.

This work also has impact on classical computing via a quantum inspired algorithm. The approach described below allows for one to compute response functions by simply running time evolution on a classical computer. This provides a different paradigm for computing response functions in exact diagonalization (and potentially other approaches, including matrix-product states) by mapping the problem onto time evolution with a time-dependent Hamiltonian. While our work here does not focus on this application, it should be clear that the approach developed here can be directly applied much more broadly.

\section*{Results}
\label{sec:formalism}
Correlation functions are composed of expectation values of the form
$\langle \op{A}(r,t) \op{B}(r',t') \pm \op{B}(r',t') \op{A}(r,t) \rangle$ --- the 
amplitude of the operator $\op{A}$ at spacetime
point $(r,t)$ given that $\op{B}$ acted on the system at spacetime point $(r',t')$. The amplitudes are substracted in the case of bosonic correlation functions, whereas they are added in the case of fermionic correlation functions. Both can be calculated via the \emph{linear response} method that we present here. We will first describe the formalism for bosonic correlation functions and describe how to apply momentum and frequency selectivity. Then, we shall describe two different ways to apply the linear-response formalism to calculate fermionic correlation functions for Hamiltonians that conserve particle count parity (maintain even or odd numbers of electrons).

\subsection*{Bosonic (commutator) correlation functions}\label{sec:bosonic}
The methodology employs the standard results from 
linear response in many-body
physics (see e.g. Refs.~\onlinecite{stefanucci_nonequilibrium_2013,bruus,Mahan}), as we develop below.
We are interested in the 
expectation value of
the operator $\op{A}(t)$ 
measured in a prepared many-body state $\ket{\psi_0}$ and time-evolved in the Hamiltonian plus the
applied (Hermitian) field;
$h(t) \op{B}$ 
that is, $\ham(t) = \ham_0 + h(t) \op{B}$. 
Then, $A(t)$ is given by
\begin{subequations}
\begin{align}
    A(t) 
    &= \braket{ \psi_0 | U(t)^\dagger \op{A} U(t) | \psi_0} \\
    U(t) &= \mathcal{T}_t e^{-i\int^t \left[ \ham_0+ \op{B}h(\bar{t})\right]  d\bar t },
    \label{eq:time_evo_op}
\end{align}
\end{subequations}
where $U(t)$, in Eq.~\ref{eq:time_evo_op}, is the time ordered
exponential for time evolution
with respect to the time-dependent Hamiltonian plus field.  
Expanding $A(t)$ with respect to $h(t)$, we find
\begin{align}
    A(t) = \int dt' \chi^R(t,t') h(t') + \mathcal{O}(h^2).
\end{align}
Here, $\chi^R(t,t')$ is defined to be the functional derivative of $A(t)$ with respect to $h(t')$, which is given by
\begin{align}\label{eq:functional_derivative}
    \frac{\delta A(t)}{\delta h(t')} \bigg|_{h=0} = -i 
    \theta(t-t')
    \braket{\psi_0 | \left[ \op{A}(t), \op{B}(t') \right] | \psi_0}.
\end{align}
In this result, we used the fact that $\op{A}(t) := e^{it\ham_0}\op{A}e^{it\ham_0}$ in the limit of vanishing field. 
The $\theta$-function arises because in Eq.~\ref{eq:time_evo_op} the integration region on the time ordered exponents is limited to $\bar{t}$ values that are smaller than $t$. Since $\ham_0$ is time independent, the response function $\chi^R(t,t')$ only depends on the time difference $t-t'$. Fourier transforming from time to frequency, and using the convolution theorem, yields
\begin{align}\label{eq:freq_func_deriv}
    A(\omega) = \chi^R(\omega) h(\omega) + \mathcal{O}(h^2).
\end{align}
Thus, if the amplitude of the signal $h(t)$ is chosen to be small enough, the higher-order terms can be neglected and the response function can be calculated as a simple ratio. 

\textit{\textbf{Frequency selectivity:}} One might be interested in the response function centered in a specific frequency interval and want to improve the signal-to-noise ratio of the calculation. This is achieved by choosing the frequency support of $h(t)$ to be most concentrated within the desired frequency interval. 

\textit{\textbf{Momentum selectivity:}} By choosing $\op{A}$ and $\op{B}$ as operators with definite momentum, we can directly calculate the response function in the momentum basis. For example, for creation of a fermion with momentum $k$, we pick $\op{B} = \sum_r e^{ikr}\tilde X_r + \text{H.C.} = \sum_r \cos (kr) \tilde X_r$,
where $\tilde X_r$ is $X_rZ_{r-1}\ldots Z_0$.
This can be directly implemented 
when $|h(t)|\Delta t \ll 1$ with the expansion of the evolution for a single time step
\begin{align}\label{eq:momentum_selective_trotter}
    e^{-ih(t)\op{B} \Delta t} = \prod_r e^{-2ih(t)\Delta t \cos(kr) \tilde X_r} + \mathcal{O}(h^2).
\end{align}
We can use a similar form for $\op{A}$ as we use for $\op{B}$, but since
it is directly measured (rather than appearing in the
time evolution), this can be achieved instead with multiple circuits.
However, if $\ham_0$ is translation invariant, only one circuit is sufficient to calculate the response function $\chi^R$ in momentum space because it satisfies 
\begin{align}
    \chi^R_{k,k'}(t-t') = \delta_{k,k'} \chi^R_{k,k}(t-t'),
\end{align}
that is, it is diagonal in momentum.

Both momentum and frequency selectivity allow us to immediately focus the signal we obtain from the quantum computer into desired ranges of momentum or frequency.
This frequency selectivity is not possible in the Hadamard test (as well as other approaches)\cite{gustafson_large_2021, uhrich}.
Moreover, 
implementing a momentum selective operator can only be achieved
via costly circuit modifications such as embedding techniques.
To avoid this, other approaches
require each real space pair $(r_1,r_2)$ to be measured separately
with independent circuits; these are then Fourier transformed
to obtain a momentum response function. On a fault-tolerant computer this might not have any difference, but on a noisy device, systematic errors can add from the different measurements, reducing the precision of the final result.
In the following sections, we show that momentum selectivity in our approach significantly reduces noise in the measured signal.

\subsection*{Fermionic (anti-commutator) correlation functions}

The most important fermionic correlation function is the retarded electronic Green's function given by
\begin{align}
G^R(r_i,t;r_j,t') = -i \theta(t-t') \braket{\psi_0 | \{ c_i(t), c_j^\dagger(t')\} | \psi_0},
\label{eq:electron_gf}
\end{align}
where $c_i$ and $c_j^\dagger$ are the fermionic annihilation and creation operators at $r=r_{i}$ and $r_j$, respectively.
Note that Eq.~\ref{eq:electron_gf} is the correlation
function with respect to a single many-body
state $\ket{\psi_0}$.
For the Green's function at $T=0$ in standard many-body theory $\ket{\psi_0}$ is the ground state.
At finite temperatures the expectation value
has to be additionally averaged over a thermal
distribution of states, which can be achieved
via classical averaging of eigenstates\cite{white2009minimally,verdon2019quantum,cohn2020minimal}
or by going over
to a density matrix representation\cite{poulin2009sampling,gilyen2019quantum,motta2020determining,metcalf2020engineered,polla2019quantum,zhang2020continuous,metcalf2021quantum}.
The formalism below is applicable
for any of these cases.

The functional derivative method does not directly carry over,
because it requires adding a Grassman number valued
field, which cannot be easily realized in a numerical simulation.
This has thus far limited
the potential of ancilla-free methods to bosonic correlation functions only\cite{gustafson_large_2021, uhrich}.
To overcome this, we introduce two complementary approaches.  The first uses an auxiliary
operator $\op{P}$ which anti-commutes with $\op{B}$, while the second uses simple post-selection.

\subsubsection*{Auxiliary Operator Method}

We consider the fermionic version of Eq.~\eqref{eq:functional_derivative}, and denote this by $G(t,t'):$

\begin{align}
    G(t,t') = -i
    \theta(t-t')
    \braket{\psi_0 | \left\{ \op{A}(t), \op{B}(t') \right\} | \psi_0}.
\end{align}
In order to produce an anticommutator, we introduce an additional operator $\op{P}$ which satisfies the following properties
\begin{enumerate}
    \item $\op{P} \ket{\psi_0} = s \ket{\psi_0}$ with $s \neq 0$.
    \item $\{\op{B}(t),\op{P} \} = 0$ for all times $t$.
    \item $[\ham_0, \op{P}] = 0$, or $\op{P}$ has no time dependence.
\end{enumerate}
With these properties, it is straightforward to show that
\begin{align}
    G(t,t') = \frac{i}{s}\theta(t-t') \braket{\psi_0 | \left[ \op{A}(t)\op{P}(t), \op{B}(t') \right]  | \psi_0}.
    \label{eq:fermion_gf}
\end{align}
This is of the form of ~Eq.~\eqref{eq:functional_derivative} with $\op{A}(t)$ replaced by $\op{A}(t) \op{P}(t)$; therefore, the bosonic linear response method can be directly used.

Even though the assumptions on $\op{P}$ appear to be restrictive, when $G(t,t')$ is
the retarded electronic Green's function, as in  Eq.~(\ref{eq:electron_gf}), the assumptions are
satisfied by the parity operator for Hamiltonians that preserve particle parity;
this covers a vast class of Hamiltonians of interest in quantum chemistry, condensed matter physics and quantum field theory. 
If the Hamiltonian of interest conserves the parity of the electron number,
then the parity operator $\op{P} = Z_1Z_2...Z_n$ satisfies second and third conditions, where we use the spin representation (obtained after Jordan-Wigner transformation) to represent the parity operator. 
 The fermionic operators, 
 $c_i$ and $c_i^\dagger$, in their spin representation, have a Jordan-Wigner string attached;
 that is, they are composed of  
 $i-1$ consecutive $Z$ operators followed by a $X \pm i Y$. 
 In this case both $c_i$ and $c^\dagger_i$ anticommute with
 the parity operator $\op{P} = Z_1Z_2...Z_n$, which satisfies the second condition.
 With this, $G(t,t')$ can be obtained by measuring Eq.~\eqref{eq:fermion_gf} 
upon replacing $\op{A}$ with $X_i \op{P}$ (and/or $Y_i \op{P}$) and $\op{B}$ with $X_j$ (and/or $Y_j$). 
 
 We can choose $h(t)$ and $\op{B}$ to have frequency and momentum selectivity in the same way as we did for bosonic correlation functions.
 Thus, we can directly calculate the fermionic Green's function in momentum space,
\begin{align}
G^R(k,t;k',t') = -i \theta(t-t') \braket{\psi_0 | \{ c_k(t), c_{k'}^\dagger(t')\} | \psi_0},
\end{align}
by selecting $\op{A}$ as a Fourier combination of $X_i \op{P}$ (and/or $Y_i \op{P}$) with momentum $k$, and $\op{B}$ as a Fourier combination of $X_j$ (and/or $Y_j$) with momentum $k'$, and forming the appropriate linear combination to select the desired $c/c^\dagger$
terms. Similarly, by choosing an appropriate frequency support for $h(t)$, we can calculate $G^R$ in a desired frequency range. 

\subsubsection*{Post-selection for single-particle Green's functions}

When the desired anti-commutator is the single-particle Green's function (Eq.~\ref{eq:electron_gf})
for a particle number conserving Hamiltonian,
i.e. $\ket{\psi_0}$ is an
$N$-particle wave function, a powerful alternate approach
exists. A complete derivation in shown in App.~\ref{app:postselection};
we outline the salient parts here. 
Let us specify our perturbing field as
\begin{align}
    \op{B} = \sum_m \alpha_m \tilde{X}_m = \sum_m \alpha_m \left( c_m + c_m^\dagger\right),
\end{align}
where 
$\tilde{X}_m = Z_1 ... Z_{m-1} X_m$.
Position or momentum selectivity can be imposed by the
choice of $\alpha_m$.
Starting from a wavefunction
with $N$ particles 
and evolving
with $\ham_0+h(t)\op{B}$, the system will be in a superposition of the $N-1$, $N$,
and $N+1$ particle sectors to linear order in $h(t)$.
For clarity, let us choose $h(t) = \eta \delta(t)$ where $\eta \ll 1$ and $\delta(t)$ is a Dirac delta pulse.
This choice is not necessary, 
we can choose $h(t)$ more generally to
achieve frequency selectivity. In order to measure the Green's function
we apply a rotation about $y$ (or $x$) to enable measurement of $c_1 \pm c_1^\dagger$ on the
first qubit, which generates $N-2$ and $N+2$ particle states as well. Denoting $\ket{\Phi^y_M}$ (or $\ket{\Phi^x_M}$) as the $M$ particle component of this final state, we observe that a simple post-selection that picks out
one of the fixed particle number sectors yields
\begin{widetext}
\begin{subequations}\label{eq:postselection}
\begin{align}
    \braket{\Phi_{N-1}^y|\Phi_{N-1}^y} +  &\braket{\Phi_{N+1}^y|\Phi_{N+1}^y}  = 
        \frac{1}{2} + \eta \sum_m \alpha_m \mathrm{Re }\left[ G^>_{1m}(t) - G^<_{1m}(t) \right]
        = \frac{1}{2} + \eta \sum_m \alpha_m  \mathrm{Re\,}G^R_{1m}(t) \\
    \braket{\Phi_{N}^y|c^\dagger_1 c_1 | \Phi_{N}^y} +  &\braket{\Phi_{N+1}^y|\Phi_{N+1}^y} =
         \frac{1}{2} + \eta \sum_m \alpha_m  \mathrm{Re }\left[ G^>_{1m}(t) + G^<_{1m}(t) \right] \\
    \braket{\Phi_{N-1}^x|\Phi_{N-1}^x} +  &\braket{\Phi_{N+1}^x|\Phi_{N+1}^x}  = 
        \frac{1}{2} +\eta \sum_m \alpha_m  \mathrm{Im }\left[ G^>_{1m}(t) - G^<_{1m}(t) \right]
        = \frac{1}{2} + \eta \sum_m \alpha_m  \mathrm{Im\,}G^R_{1m}(t) \\
    \braket{\Phi_{N}^x|c^\dagger_1 c_1 | \Phi_{N}^x} +  &\braket{\Phi_{N+1}^x|\Phi_{N+1}^x} = 
        \frac{1}{2} + \eta \sum_m \alpha_m  \mathrm{Im }\left[ G^>_{1m}(t) + G^<_{1m}(t) \right]
\end{align}
\end{subequations}
\end{widetext}
where the fermionic Green's functions are\cite{Mahan},
\begin{align}\label{eq:postprocessing}
\begin{split}
    G_{ij}^<(t) &= i \braket{\psi_0|c_j^\dagger(0)c_i(t)|\psi_0} \\
    G_{ij}^>(t) &= -i \braket{\psi_0|c_i(t)c_j^\dagger(0)|\psi_0} \\
    G_{ij}^R(t) &= -i \theta(t) \braket{\psi_0| \{c_i(t), c_j^\dagger(0)\}|\psi_0}.
\end{split}
\end{align}
The quantities in Eq.~\ref{eq:postselection} can be obtained simply by considering the probabilities of states with specific particle number.
While this is limited to particle-conserving Hamiltonians,
this is a relatively mild restriction as all fermionic Hamiltonians that do not
have superconducting terms satisfy this restriction. 

\begin{figure*}[htpb]
    \centering
    \includegraphics[width=1.0\textwidth]{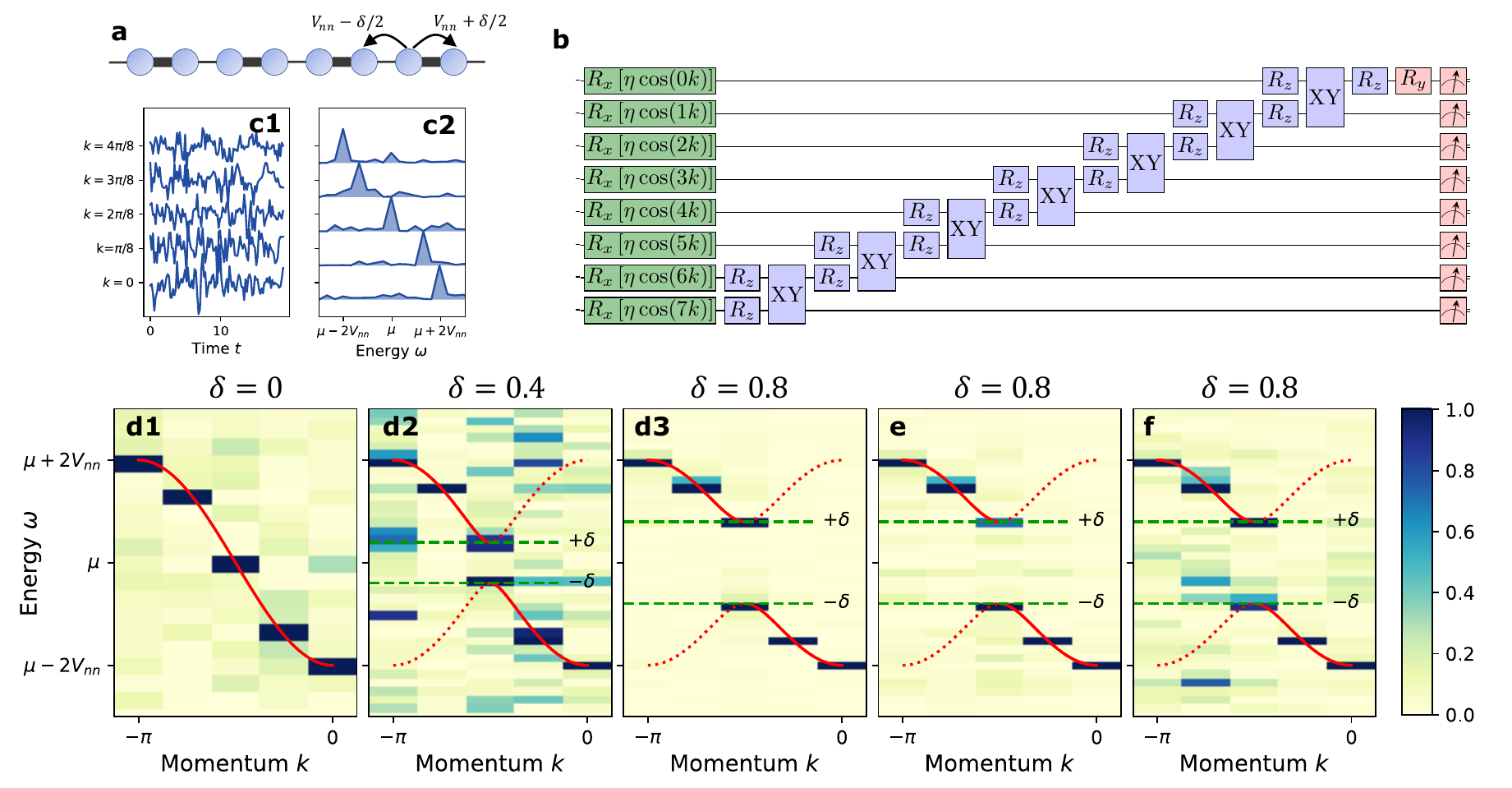}
    \caption{\textbf{Electronic Green's function for
    the Su-Schrieffer-Heeger (SSH) model.} 
    \textbf{a.} Lattice and hopping structure
    of the SSH model.
    \textbf{b.}  Compressed linear response method quantum circuit run on \emph{ibm\_auckland}. $XY$ indicates a rotation about $XX$ followed by $YY$\cite{kokcu2022algebraic,camps2022algebraic}.    \textbf{c1.} Fermionic
    correlation function $\mathcal{L}_k(t) = 2\:\mathrm{Re\:}G_k(t)$ for $\delta=0$ using the
    commutator method.
    Data for
    other values of $\delta$ are available
    in the SI.
    \textbf{c2.} Normalized power spectrum $|\mathcal{L}_k(\omega)|^2$
    \textbf{d.} Normalized false-color plots of $|\mathcal{L}_k(\omega)|^2$ for $\delta=\{0,0.4,0.8\}$.
    Green dashed
    lines indicate the expected bounds of
    the gap, and the red lines the analytically obtained spectrum.
    \textbf{e, f.} 
    Normalized false-color plot of post-selected $\braket{\Phi^y_0|\Phi^y_0}$ and $\braket{\Phi^y_1|\Phi^y_1}$, respectively (see text for definition).  
    The projected norms contain the same spectral information as $\mathcal{L}_k(\omega)$.
    }
    \label{fig:ssh_model}
\end{figure*}

\subsection*{Algorithmic and analysis details}
Here, we outline the details of the implementation
and the signal analysis. One noteworthy aspect is
the use of a damping function $g(t)$. In many-body
physics, an $\exp(-\gamma t)$ convergence factor 
is often use to regularize otherwise divergent
Fourier integrals, where $\gamma > 0$ \cite{bruus,Mahan,stefanucci_nonequilibrium_2013}. Moreover, in realistic materials, 
sharp peaks in the
spectrum are broadened due to the natural interactions
that occur. The damping function $g(t)$ is an
effective way to incorporate these effects.
Practically speaking, enforcing the signal to decay
has a benefit from the quantum circuit perspective:
namely, it limits the maximum simulation time necessary,
which in turn limits the circuit depth. 

Here, we consider it an adjustable function
that softens the Fourier spectra by ensuring 
the signal has compact support in the time domain.
Applying an exponential decay factor to
the signal is equivalent to a Lorentzian broadening
in the frequency domain, and sets the effective resolution
of this approach. Similarly, the
natural noise inherent in quantum hardware where the circuit
depth grows with increasing time may perform
a similar function.

The procedure to obtain
the correlation function given a state of interest $\ket{\psi_0}$ is as follows
\begin{enumerate}
    \item[1.] Evolve $\ket{\psi_0}$ with the perturbed Hamiltonian $\ham(t) = \ham_0 + h(t)\op{B}$
    during the time where $h(t)$ is finite.
    $h(t)$ should be a small field in order to ensure the simulation is in the linear response regime.  This can be tested by repeating the simulation with larger/smaller $h(t)$ and checking that the response scales similarly.
    \item [2.] Continue to evolve with the unperturbed
    Hamiltonian $\ham$. The maximum length of time needed is
    set by the desired minimum energy resolution.
    \item [3.] At each time of interest $t$, measure 
    $A(t)=\braket{\op{A}(t)}$.
    \item [4.] Apply a semi-phenomenological damping
    function such as $g(t)=\exp(-t/\tau)$ to obtain $\tilde{A}(t)=g(t)A(t)$. This sets an effective energy
    resolution $\tau^{-1}$ in the susceptibility.
    This approach was recently shown to additionally
    help by limiting the circuit depth required\cite{lee2022compact}.
    \item [5.] Fourier transform $\tilde{A}(t)$ to $A(\omega)$ and divide by $h(\omega$) to obtain $\chi(\omega)$, thus performing the (numerical) functional differentiation.
\end{enumerate}

\subsection*{Green's function of the SSH model.}
\label{sec:ssh_model}

We demonstrate the linear response approach by calculating 
the fermionic Green's function as would be measured by ARPES (angle-resolved
photoemission spectroscopy).
We study a minimal model for a charge density wave known as the
Su-Schrieffer-Heeger (SSH) model --- an N-site 1D free fermionic
chain with nearest-neighbor bond-dependent hoppings (see Fig.~\ref{fig:ssh_model}\textbf{a}) ---
in the limit where the lattice distortion
is static,
\begin{align}
\mathcal{H}_0 = -\sum_{\langle i,j \rangle} \left[V_{nn} + \left(-1\right)^i\delta/2 \right] c^\dagger_i c^{\phantom{\dagger}}_j
- \mu \sum_{i} c^\dagger_i c^{\phantom{\dagger}}_i.
\end{align}
For finite $\delta$ this model exhibits a charge density wave, with a gap
proportional to $\delta$.
Since this is a free fermionic system, the spectrum is easily obtained by starting from
the vacuum state, so we set
$\mu = 5$ to suppress the initial total electron number. 

We use a momentum-selective instantaneous (and thus broadband)
driving field coupled
to the particle creation and annihilation operators that act on all the sites $i$,
\begin{align}\label{eq:momentum_pulse}
    \op{B} = \sum_i 2\cos(kr_i) \left[c_i + c^\dagger_i\right],
\end{align}
with a pulse $h(t) = \eta \delta(t)$,
where we used $\eta dt = 0.04$.
{We measure $X_0 = c_0 + c_0^\dagger$ which is local in position, and includes all momentum modes.
Because $\ham_0$ for the SSH model conserves
momentum,} by measuring $X_0$ 
we obtain $\mathcal{L}_k(t) = 2\: \mathrm{Re\:}G_k(t)$ which has the full information of the single particle spectral function; in the frequency basis, this is (see Appendix \ref{app:circuit} 
for details)
\begin{align}
    \mathcal{L}_k(\omega) = G_k(\omega) + G_k(-\omega)^*.
\end{align}
{Even though our method is capable of measuring $G_k(\omega)$, isolating it from $G_k(-\omega)^*$ requires running the same circuit and measuring $Y_0$ as well.}
Since $\mu = 5$, for this model the single particle energies are manifestly positive, and the interference between $G_k(\omega)$ and $G_k(-\omega)$ is negligible. Thus,  $|\mathcal{L}_k(\omega)|^2$  
{tracks the quasi-particle peaks in $\mathrm{Im}\:G_k(\omega)$, and measuring $\mathcal{L}_k(\omega)$ is sufficient to obtain the single-particle spectrum.}

On the quantum computer, the driving field is implemented ain a single Trotter step; a set of
single-qubit $x$-rotations with an amplitude
$2h_0 \cos(kr_j)$ on the $j$-th qubit. The
subsequent evolution uses compressed free fermionic evolution~\cite{kokcu2022algebraic,camps2022algebraic}. 
To minimize the weight of the measured Pauli string (and thus reduce measurement noise)
we perform the measurement on the 1st qubit. To further
mitigate error, we use Pauli twirling
and dynamic decoupling\cite{Qiskit}. Additional
details of the quantum computation may be
found in the supplementary material.

We performed the calculation on \emph{ibm\_auckland} for an $N=8$-site chain, which has allowed
momentum values 
$k=\frac{2\pi}{N}j, j \in \left\{0 \ldots 7\right\}$.
Since 
the driving field $\op{B}$ is symmetric in $k$, both $k$ and
$-k$ are obtained at the same time. 
We used a compressed form of the quantum circuit
shown in panel \textbf{b} (check Appendix \ref{app:time_evolution_circuit} for details).
Fig.~\ref{fig:ssh_model} panel \textbf{c1} shows the raw data
for $\mathcal{L}_k(t)$ with $\delta=0$ at each unique $k$;
the data was
obtained from \emph{ibm\_auckland} via the parity operator method.
The power spectrum is shown in panels \textbf{c2}
and $\textbf{d1}$.
While the data from the quantum computer appears
quite noisy, in the frequency regime of 
interest
there is only a single peak present in the Fourier
transform, illustrating the remarkable strength of
a momentum-selective probe, which picks out the single energy at each momentum, together with Fourier filtering.
Upon increasing $\delta$ (panels \textbf{d2,d3}),
a gap opens up in the spectrum (time traces and
Fourier amplitudes are available in Appendix \ref{app:raw_data}).
The spectrum
for $\delta=0.4$ is noisier than the other two, which
we attribute to machine noise from those particular
measurements.
In panels \textbf{e,f}, we plot the norms of 0- and 1- particle components of the state right before the measurement, i.e. $\braket{\Phi^y_0|\Phi^y_0}$ and $\braket{\Phi^y_1|\Phi^y_1}$, where $\ket{\Phi^y_M}$ is defined above Eq.~\ref{eq:postselection}.
Both of these partial norms are equivalent to  $\mathcal{L}_k(t)$ (See Appendix \ref{sec:ssh_0_particle} for details). 
Both
methods faithfully reproduce the power spectrum,
with slightly higher levels of noise for post-selection on $N=1$.

\begin{figure}[htpb]
    \centering
    \includegraphics[clip=true,trim=0 0 0 0,
    width=.49\textwidth]{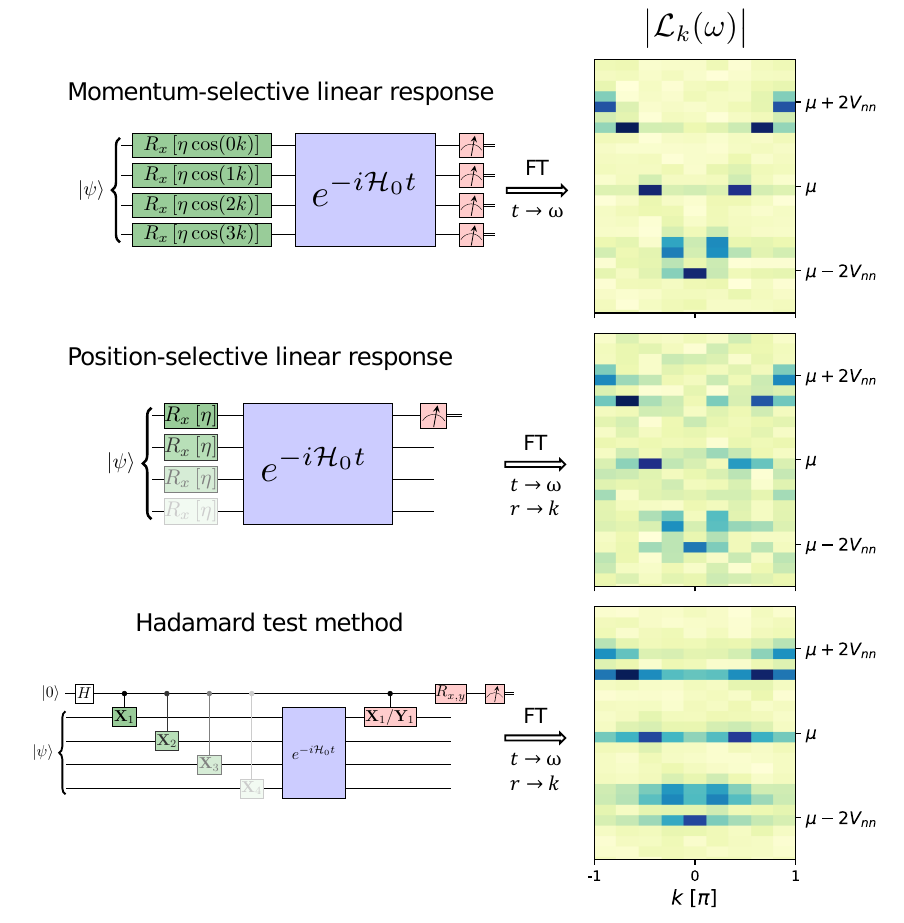}
    \caption{\textbf{Comparison of the momentum selective linear response, position selective linear response, and Hadamard test methods.}
    The circuit diagrams schematically represent
    the 3 approaches, which were run on a noisy
    simulators with one/two qubit noise of
    $1\%$ and $10\%$, respectively.
    While the momentum selective linear response method directly yields $\mathcal{L}_k(t)$, an additional spatial Fourier transformation
    is needed for the other two methods.
    }
    \label{fig:simulator}
\end{figure}

In order to further underscore the power of the
momentum-selective linear response approach,
we compare its effectiveness to a position-selective
linear response and Hadamard test methods in
Fig.~\ref{fig:simulator} on a noisy simulator
(see Appendix \ref{app:comparison} for details of the simulation and detailed
analysis).
Compared to the momentum-selective linear
response method, the position-selective one
is noisier, but without particular structure.
The Hadamard test, on the other hand, exhibits streaks
that arise from leakage of signal from one
momentum to the others.
There are two key reasons for the differences
seen in the figure. First, both position-selective
and Hadamard test methods involve excitations at each position ($X_i$ in the figure). These must be combined in the post-processing with a Fourier transform. But, because a Fourier
transform relies on constructive/destructive interference between signals, and we are performing this on noisy data, the interference is not perfect, which leads to leakage between momentum channels.
Second, the Hadamard test method introduces more
of the same problem because each $X_i$ is a separate circuit --- in addition to needing more circuits to be run and an additional ancilla. This further exacerbates the issue with the Fourier analysis.
The momentum-selectivity avoids these issues by
making a unique excitation and thus producing a response function
with a single large contribution.

\subsection*{Polarizability of the SSH model}
\label{sec:neutron_results}

\begin{figure}[b]
    \includegraphics[width=0.49\textwidth]{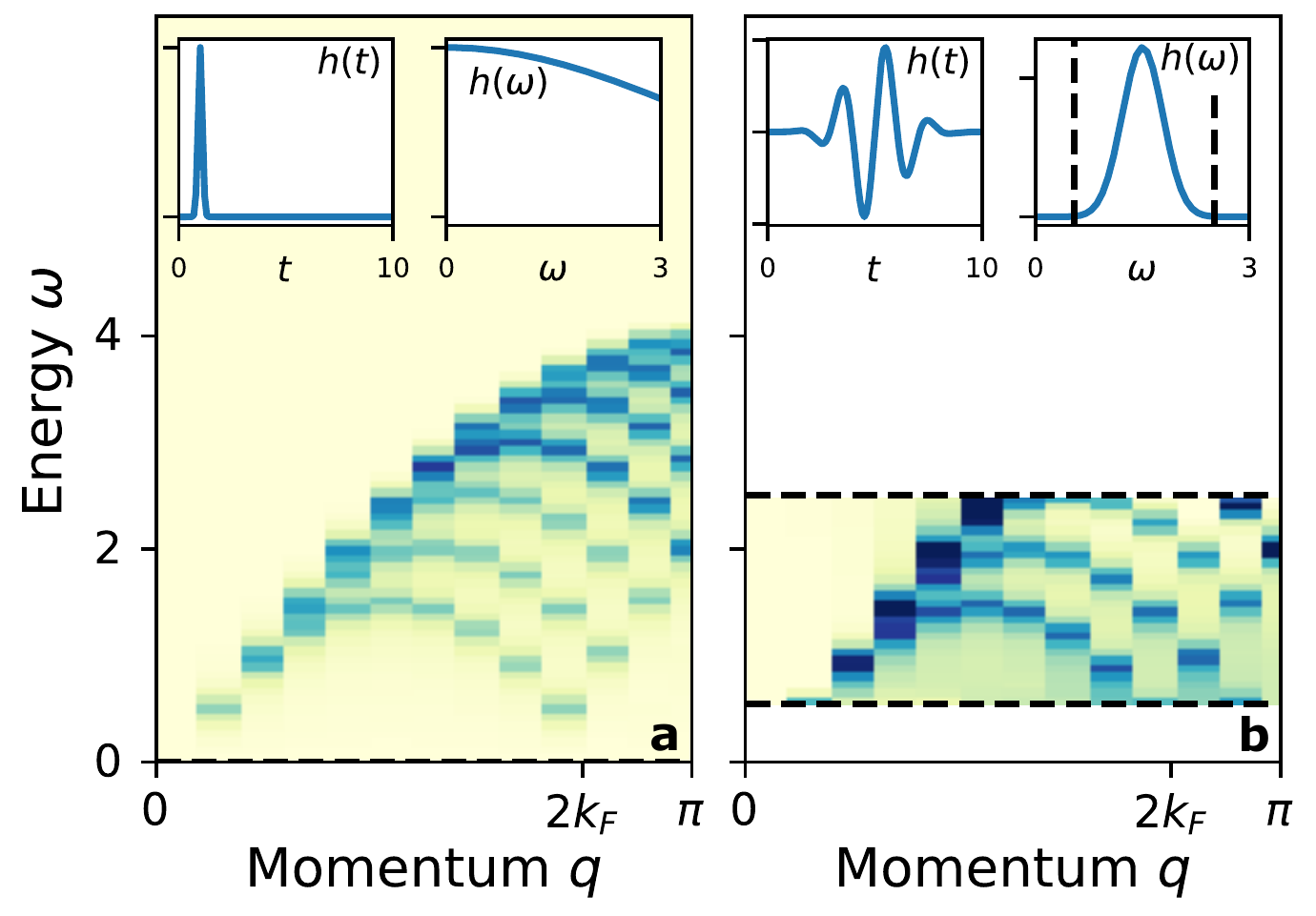}
    \caption{\textbf{Polarizability for the 1D chain.} Both
    panels show $\mathrm{Im}\ \chi(q,\omega)$ in false color. The insets
    show the driving field $h(t)$ and its Fourier transform.
    \textbf{a.} $\chi(q,\omega)$ obtained from the response due to a
    sharp excitation with height 0.1. \textbf{b.} $\chi(q,\omega)$ obtained from the
    response of a frequency selective field.  The dashed lines indicate
    the range where $|h(\omega)|^2 < 10^{-3}$.
    Here, we used $h(t)$ as sinusoid with 
    a Gaussian profile
    of width $\sigma=0.625$, height 0.05, and centered at $\omega=1.5$. 
    }
    \label{fig:charge_corr}
\end{figure}

We next consider the polarizability $\chi(q,\omega)$ of the 1D chain. The 
polarizability is the response of the electronic system to an applied
potential. It plays a critical role in the screening of interactions between
electrons in solids and molecules, and in their electromagnetic properties.
Experimentally, the polarizability can be studied by light absorption or scattering,
or by momentum-resolved electron energy loss spectroscopy (M-EELS).
The polarizability is defined by
\begin{align}
    \chi(r,t) = -i\braket{\psi_0 | \delta n(r,t) \delta n(r=0,t=0) | \psi_0},
\end{align}
i.e. it is a charge-charge correlation function. Here
$\delta n$ is the change in the charge from the equilibrium density. The observable $\op{A}$ is
the charge, and the applied field
$\op{B}$ (which is conjugate to the charge) is a 
potential. 
The excitations are
changes in the density, which are composed of pairs of fermionic operators,
and thus this is a bosonic correlation function.

For this demonstration, $\op{B}$ acts on a single site,
and we classically simulate a partially filled 24-site chain ($\mu=0.9$).
As discussed above, one
of the advantages of the linear response framework is that all 24
correlation functions are obtained with a single calculation.
Fig.~\ref{fig:charge_corr}\textbf{a} shows $\mathrm{Im}\ \chi(q,\omega)$,
which is the double Fourier transform of $\chi(r,t)$ 
obtained from driving a single site with a sharp $h(t)$. $\mathrm{Im}\ \chi(q,\omega)$ has all
the textbook features of the response of a 1D charged system; there is no
response at all at $q=0$ due to charge conservation, there is a narrow dispersive
feature at low $q,\omega$ that broadens with increasing $q$,
and a low-energy turnover with a minimum at $2k_F$.  

Since $h(\omega)$ has support across the entire spectrum of $\chi(q,\omega)$ (shown in the
inset), the entire spectrum can be obtained from this measurement.  This is in contrast to
panel $\textbf{b}$, where we drive with a short-duration sinusoid centered
at $\omega=1.5$. This excitation is \emph{frequency-selective}; that is, it only
excites the system at frequencies where $h(\omega)$ has finite support.
This range of frequencies is indicated by dashed lines in the figure.
With our particular choice of $h(t)$
we
are able to observe some of the middle range of excitations, but are insensitive to the lower frequencies
and the top of the spectrum.
Note that there is no restriction on the Fourier
transform of $\chi(r,t)$ per se; rather, the need to divide by
$h(\omega)$ (see Eq.~\ref{eq:functional_derivative}) limits the applicable window to the
ranges where $h(\omega)$ is finite. 

\section*{Discussion}
\label{sec:conclusion}

The linear-response based formalism is a shift in perspective on quantum simulation; the
measurement process is truly a part of the simulation as an experimental driving field. This is in contrast
to Hadamard-test and other competing approaches, where the simulation is limited to the system,
and the desired observables are extracted either outside of the system qubits and/or from 
a large excitation.  This shift in
perspective and methodology enables a much broader set of observables to be envisioned
and easily calculated, and enables a direct connection to experiment. Moreover, it relies
almost entirely on time evolution, a task for which quantum computers are naturally
suited.

This shift in perspective and the resulting implementation
leads to several clear advantages.
First, mirroring experimental procedure, we can straightforwardly achieve
momentum- and frequency selectivity by focusing the perturbation $h(t)\op{B}$ on a certain momentum or frequency range, without difficulty or additional implementation cost. 
This is enabled on the quantum circuit level by an implementation advantage of the linear response: $\op{B}$ can be chosen to be non-unitary because we apply $e^{-ih(t)\op{B}\Delta t}$, as opposed to the Hadamard test which applies $\op{B}$ on the state. 
The resulting momentum selectivity produces less
noise in the response functions (c.f. Fig.~\ref{fig:simulator})
because the calculation is done via one circuit rather than certain linear combinations of results obtained from structurally different circuits with different noise.  
This advantage is particularly underscored 
for translationally-invariant
systems where momentum is a good quantum number; enforcing the momentum selectivity on $\op{B}$ and running only one circuit
per momentum $k$ is sufficient to measure the response in momentum basis because $\chi^R_{k,k'}$ is diagonal. The resulting signal will thus has a fixed number of frequency peaks for given $k$ value (c.f. Figs.~\ref{fig:ssh_model} and \ref{fig:simulator}), simplifying the signal processing.

Fermionic response functions (anti-correlation functions)
can be obtained
with the same experimentally centered, linear response perspective; this is unlike
other ancilla-free methods\cite{gustafson_large_2021,uhrich} which are limited to
bosonic response functions.
The post-selection method is intuitive, as the
particle number sectors are clearly delineated. On the other hand, the auxiliary
operator method is an unusual perspective; it is sufficient to measure almost the same
operator as for the bosonic correlation function.  The electron Green's function,
for example, is obtained simply by keeping track of the parity as well as the
occupation number measurement.
In either case, this is an important advance since electron Green's functions 
play a key role in physics; as an important measurement
per se\cite{damascelli2003angle}, and as an ingredient in 
embedding theories such as dynamical mean field theory
\cite{georges1996dynamical,Zgid2011,
rungger2019dynamical,keen2020quantum, steckmann2021simulating,jamet2022quantum}. 

While here we have explicitly demonstrated the linear 
response approach in the context of a charge density wave,
it is a general method to obtain response functions, and is not limited to electronic Hamiltonians. It can be applied to spin or bosonic models, or other models from fields where quantum simulation plays a role,
including chemistry and high energy physics. Different choices of $\op{A}$ and $\op{B}$ extend the method to a wide
variety of observables. For example, the conductivity is a current-current correlation function, for which $h(t)$
is an applied electric field. A $zz$-spin susceptibility can
be obtained with $h(t)$ as a $z$-axis magnetic field,
and the operators $\op{A}=\op{B}=S_z$.
Moving forward, the functional derivative formalism can be extended to
higher order derivatives 
that involve multiple driving fields. One notable
application is resonant inelasic X-ray scattering (RIXS), which is a four-point correlation
function\cite{ament2011resonant}, which is very challenging to calculate via diagrammatics.  In addition, and aside from direct experimental probes, pairing vertices in
superconductors and other ordered phenomena also fall into this class of observables.
We reserve these discussions for future work.

This approach is a quantum-inspired paradigm, which can also be applied in conventional computation. Rather than having to measure all of the different matrix elements needed for the Lehmann formula, this approach requires simulating time-evolution and then measuring the expectation value of a single operator. As such, it is likely to be much more efficient than currently used methods.

\section*{Materials and Methods}

The data shown in Fig.~\ref{fig:ssh_model} was calculated on 
{\em ibm\_auckland}.
For each $k$
and $\delta$ we collected 3 data sets with 8,000 shots each, yielding 24,000
shots total per curve. While no measurement error mitigation was used, we incorporated
dynamical decoupling and Pauli twirling as implemented in the {\em qiskit\_research}
package.
The raw data is shown in the supplementary material in Fig.~\ref{fig:GF_fulldata}.
The calibration data is shown
in tables~\ref{table:auckland} and~\ref{table:auckland2}.

\begin{acknowledgments}
We acknowledge helpful discussions with Erik Gustafson and Vito Scarola.
We acknowledge the use of IBM Q via the IBM Q Hub at North Carolina (NC) State for this paper. The views expressed are those of the authors and do not reflect the official policy or position of the IBM Q Hub at NC State, IBM or the IBM Q team. 
We acknowledge the use of the QISKIT software package \cite{Qiskit} for performing the quantum simulations.
\textbf{Funding:} This work was supported by the Department of Energy, Office of Basic Energy Sciences, Division of Materials Sciences and Engineering under grant no. DE-SC0023231. J.K.F. was also supported by the McDevitt bequest at Georgetown.
Author contributions: A.F.K. and J.K.F. conceptualized the project. A.F.K. developed the methodology,
performed the quantum computer experiments, and ran the polarizability calculations.
E.K. contributed to the mathematical development for fermionic response
functions and designed the quantum circuits.
H.A.L. ran the noisy quantum simulator calculations.
All authors discussed the results and contributed to the development of the manuscript. 
\textbf{Competing interests:} The authors declare that they have no competing interests. 
\textbf{Data and materials availability:} All data needed to evaluate the conclusions in the paper are present in the paper and/or the Supplementary Materials. 
\end{acknowledgments}

\bibliography{ref,qst}

\begin{thebibliography}{44}%
\makeatletter
\providecommand \@ifxundefined [1]{%
 \@ifx{#1\undefined}
}%
\providecommand \@ifnum [1]{%
 \ifnum #1\expandafter \@firstoftwo
 \else \expandafter \@secondoftwo
 \fi
}%
\providecommand \@ifx [1]{%
 \ifx #1\expandafter \@firstoftwo
 \else \expandafter \@secondoftwo
 \fi
}%
\providecommand \natexlab [1]{#1}%
\providecommand \enquote  [1]{``#1''}%
\providecommand \bibnamefont  [1]{#1}%
\providecommand \bibfnamefont [1]{#1}%
\providecommand \citenamefont [1]{#1}%
\providecommand \href@noop [0]{\@secondoftwo}%
\providecommand \href [0]{\begingroup \@sanitize@url \@href}%
\providecommand \@href[1]{\@@startlink{#1}\@@href}%
\providecommand \@@href[1]{\endgroup#1\@@endlink}%
\providecommand \@sanitize@url [0]{\catcode `\\12\catcode `\$12\catcode
  `\&12\catcode `\#12\catcode `\^12\catcode `\_12\catcode `\%12\relax}%
\providecommand \@@startlink[1]{}%
\providecommand \@@endlink[0]{}%
\providecommand \url  [0]{\begingroup\@sanitize@url \@url }%
\providecommand \@url [1]{\endgroup\@href {#1}{\urlprefix }}%
\providecommand \urlprefix  [0]{URL }%
\providecommand \Eprint [0]{\href }%
\providecommand \doibase [0]{http://dx.doi.org/}%
\providecommand \selectlanguage [0]{\@gobble}%
\providecommand \bibinfo  [0]{\@secondoftwo}%
\providecommand \bibfield  [0]{\@secondoftwo}%
\providecommand \translation [1]{[#1]}%
\providecommand \BibitemOpen [0]{}%
\providecommand \bibitemStop [0]{}%
\providecommand \bibitemNoStop [0]{.\EOS\space}%
\providecommand \EOS [0]{\spacefactor3000\relax}%
\providecommand \BibitemShut  [1]{\csname bibitem#1\endcsname}%
\let\auto@bib@innerbib\@empty
\bibitem [{\citenamefont {Mahan}(2010)}]{Mahan}%
  \BibitemOpen
  \bibfield  {author} {\bibinfo {author} {\bibfnamefont {G.~D.}\ \bibnamefont
  {Mahan}},\ }\href@noop {} {\emph {\bibinfo {title} {Many Particle Physics}}}\
  (\bibinfo  {publisher} {Springer},\ \bibinfo {address} {New York, NY10013,
  USA},\ \bibinfo {year} {2010})\BibitemShut {NoStop}%
\bibitem [{\citenamefont {Stefanucci}\ and\ \citenamefont {van
  Leeuwen}(2013)}]{stefanucci_nonequilibrium_2013}%
  \BibitemOpen
  \bibfield  {author} {\bibinfo {author} {\bibfnamefont {G.}~\bibnamefont
  {Stefanucci}}\ and\ \bibinfo {author} {\bibfnamefont {R.}~\bibnamefont {van
  Leeuwen}},\ }\href@noop {} {\emph {\bibinfo {title} {Nonequilibrium
  {Many}-{Body} {Theory} of {Quantum} {Systems}: {A} {Modern}
  {Introduction}}}},\ \bibinfo {edition} {1st}\ ed.\ (\bibinfo  {publisher}
  {Cambridge University Press},\ \bibinfo {year} {2013})\BibitemShut {NoStop}%
\bibitem [{\citenamefont {Georges}\ \emph {et~al.}(1996)\citenamefont
  {Georges}, \citenamefont {Kotliar}, \citenamefont {Krauth},\ and\
  \citenamefont {Rozenberg}}]{georges1996dynamical}%
  \BibitemOpen
  \bibfield  {author} {\bibinfo {author} {\bibfnamefont {A.}~\bibnamefont
  {Georges}}, \bibinfo {author} {\bibfnamefont {G.}~\bibnamefont {Kotliar}},
  \bibinfo {author} {\bibfnamefont {W.}~\bibnamefont {Krauth}}, \ and\ \bibinfo
  {author} {\bibfnamefont {M.~J.}\ \bibnamefont {Rozenberg}},\ }\href@noop {}
  {\bibfield  {journal} {\bibinfo  {journal} {Reviews of Modern Physics}\
  }\textbf {\bibinfo {volume} {68}},\ \bibinfo {pages} {13} (\bibinfo {year}
  {1996})}\BibitemShut {NoStop}%
\bibitem [{\citenamefont {Zgid}\ and\ \citenamefont {Chan}(2011)}]{Zgid2011}%
  \BibitemOpen
  \bibfield  {author} {\bibinfo {author} {\bibfnamefont {D.}~\bibnamefont
  {Zgid}}\ and\ \bibinfo {author} {\bibfnamefont {G.~K.-L.}\ \bibnamefont
  {Chan}},\ }\href@noop {} {\bibfield  {journal} {\bibinfo  {journal} {J. Chem.
  Phys.}\ }\textbf {\bibinfo {volume} {134}} (\bibinfo {year}
  {2011})}\BibitemShut {NoStop}%
\bibitem [{\citenamefont {Rungger}\ \emph {et~al.}(2020)\citenamefont
  {Rungger}, \citenamefont {Fitzpatrick}, \citenamefont {Chen}, \citenamefont
  {Alderete}, \citenamefont {Apel}, \citenamefont {Cowtan}, \citenamefont
  {Patterson}, \citenamefont {Ramo}, \citenamefont {Zhu}, \citenamefont
  {Nguyen}, \citenamefont {Grant}, \citenamefont {Chretien}, \citenamefont
  {Wossnig}, \citenamefont {Linke},\ and\ \citenamefont
  {Duncan}}]{rungger2019dynamical}%
  \BibitemOpen
  \bibfield  {author} {\bibinfo {author} {\bibfnamefont {I.}~\bibnamefont
  {Rungger}}, \bibinfo {author} {\bibfnamefont {N.}~\bibnamefont
  {Fitzpatrick}}, \bibinfo {author} {\bibfnamefont {H.}~\bibnamefont {Chen}},
  \bibinfo {author} {\bibfnamefont {C.~H.}\ \bibnamefont {Alderete}}, \bibinfo
  {author} {\bibfnamefont {H.}~\bibnamefont {Apel}}, \bibinfo {author}
  {\bibfnamefont {A.}~\bibnamefont {Cowtan}}, \bibinfo {author} {\bibfnamefont
  {A.}~\bibnamefont {Patterson}}, \bibinfo {author} {\bibfnamefont {D.~M.}\
  \bibnamefont {Ramo}}, \bibinfo {author} {\bibfnamefont {Y.}~\bibnamefont
  {Zhu}}, \bibinfo {author} {\bibfnamefont {N.~H.}\ \bibnamefont {Nguyen}},
  \bibinfo {author} {\bibfnamefont {E.}~\bibnamefont {Grant}}, \bibinfo
  {author} {\bibfnamefont {S.}~\bibnamefont {Chretien}}, \bibinfo {author}
  {\bibfnamefont {L.}~\bibnamefont {Wossnig}}, \bibinfo {author} {\bibfnamefont
  {N.~M.}\ \bibnamefont {Linke}}, \ and\ \bibinfo {author} {\bibfnamefont
  {R.}~\bibnamefont {Duncan}},\ }\href@noop {} {\enquote {\bibinfo {title}
  {Dynamical mean field theory algorithm and experiment on quantum
  computers},}\ } (\bibinfo {year} {2020}),\ \Eprint
  {http://arxiv.org/abs/1910.04735} {arXiv:1910.04735 [quant-ph]} \BibitemShut
  {NoStop}%
\bibitem [{\citenamefont {Keen}\ \emph {et~al.}(2020)\citenamefont {Keen},
  \citenamefont {Maier}, \citenamefont {Johnston},\ and\ \citenamefont
  {Lougovski}}]{keen2020quantum}%
  \BibitemOpen
  \bibfield  {author} {\bibinfo {author} {\bibfnamefont {T.}~\bibnamefont
  {Keen}}, \bibinfo {author} {\bibfnamefont {T.}~\bibnamefont {Maier}},
  \bibinfo {author} {\bibfnamefont {S.}~\bibnamefont {Johnston}}, \ and\
  \bibinfo {author} {\bibfnamefont {P.}~\bibnamefont {Lougovski}},\ }\href
  {\doibase 10.1088/2058-9565/ab7d4c} {\bibfield  {journal} {\bibinfo
  {journal} {Quantum Science and Technology}\ }\textbf {\bibinfo {volume}
  {5}},\ \bibinfo {pages} {035001} (\bibinfo {year} {2020})}\BibitemShut
  {NoStop}%
\bibitem [{\citenamefont {Steckmann}\ \emph {et~al.}(2021)\citenamefont
  {Steckmann}, \citenamefont {Keen}, \citenamefont {Kemper}, \citenamefont
  {Dumitrescu},\ and\ \citenamefont {Wang}}]{steckmann2021simulating}%
  \BibitemOpen
  \bibfield  {author} {\bibinfo {author} {\bibfnamefont {T.}~\bibnamefont
  {Steckmann}}, \bibinfo {author} {\bibfnamefont {T.}~\bibnamefont {Keen}},
  \bibinfo {author} {\bibfnamefont {A.~F.}\ \bibnamefont {Kemper}}, \bibinfo
  {author} {\bibfnamefont {E.~F.}\ \bibnamefont {Dumitrescu}}, \ and\ \bibinfo
  {author} {\bibfnamefont {Y.}~\bibnamefont {Wang}},\ }\href@noop {} {\bibfield
   {journal} {\bibinfo  {journal} {arXiv preprint arXiv:2112.05688}\ }
  (\bibinfo {year} {2021})}\BibitemShut {NoStop}%
\bibitem [{\citenamefont {Jamet}\ \emph {et~al.}(2022)\citenamefont {Jamet},
  \citenamefont {Agarwal},\ and\ \citenamefont {Rungger}}]{jamet2022quantum}%
  \BibitemOpen
  \bibfield  {author} {\bibinfo {author} {\bibfnamefont {F.}~\bibnamefont
  {Jamet}}, \bibinfo {author} {\bibfnamefont {A.}~\bibnamefont {Agarwal}}, \
  and\ \bibinfo {author} {\bibfnamefont {I.}~\bibnamefont {Rungger}},\
  }\href@noop {} {\bibfield  {journal} {\bibinfo  {journal} {arXiv preprint
  arXiv:2205.00094}\ } (\bibinfo {year} {2022})}\BibitemShut {NoStop}%
\bibitem [{\citenamefont {Chiesa}\ \emph {et~al.}(2019)\citenamefont {Chiesa},
  \citenamefont {Tacchino}, \citenamefont {Grossi}, \citenamefont {Santini},
  \citenamefont {Tavernelli}, \citenamefont {Gerace},\ and\ \citenamefont
  {Carretta}}]{chiesa2019quantum}%
  \BibitemOpen
  \bibfield  {author} {\bibinfo {author} {\bibfnamefont {A.}~\bibnamefont
  {Chiesa}}, \bibinfo {author} {\bibfnamefont {F.}~\bibnamefont {Tacchino}},
  \bibinfo {author} {\bibfnamefont {M.}~\bibnamefont {Grossi}}, \bibinfo
  {author} {\bibfnamefont {P.}~\bibnamefont {Santini}}, \bibinfo {author}
  {\bibfnamefont {I.}~\bibnamefont {Tavernelli}}, \bibinfo {author}
  {\bibfnamefont {D.}~\bibnamefont {Gerace}}, \ and\ \bibinfo {author}
  {\bibfnamefont {S.}~\bibnamefont {Carretta}},\ }\href@noop {} {\bibfield
  {journal} {\bibinfo  {journal} {Nature Physics}\ }\textbf {\bibinfo {volume}
  {15}},\ \bibinfo {pages} {455} (\bibinfo {year} {2019})}\BibitemShut
  {NoStop}%
\bibitem [{\citenamefont {Roggero}\ and\ \citenamefont
  {Carlson}(2019)}]{Roggero19}%
  \BibitemOpen
  \bibfield  {author} {\bibinfo {author} {\bibfnamefont {A.}~\bibnamefont
  {Roggero}}\ and\ \bibinfo {author} {\bibfnamefont {J.}~\bibnamefont
  {Carlson}},\ }\href {\doibase 10.1103/PhysRevC.100.034610} {\bibfield
  {journal} {\bibinfo  {journal} {Phys. Rev. C}\ }\textbf {\bibinfo {volume}
  {100}},\ \bibinfo {pages} {034610} (\bibinfo {year} {2019})}\BibitemShut
  {NoStop}%
\bibitem [{\citenamefont {Francis}\ \emph {et~al.}(2020)\citenamefont
  {Francis}, \citenamefont {Freericks},\ and\ \citenamefont
  {Kemper}}]{francis2020quantum}%
  \BibitemOpen
  \bibfield  {author} {\bibinfo {author} {\bibfnamefont {A.}~\bibnamefont
  {Francis}}, \bibinfo {author} {\bibfnamefont {J.~K.}\ \bibnamefont
  {Freericks}}, \ and\ \bibinfo {author} {\bibfnamefont {A.~F.}\ \bibnamefont
  {Kemper}},\ }\href {\doibase 10.1103/PhysRevB.101.014411} {\bibfield
  {journal} {\bibinfo  {journal} {Phys. Rev. B}\ }\textbf {\bibinfo {volume}
  {101}},\ \bibinfo {pages} {014411} (\bibinfo {year} {2020})}\BibitemShut
  {NoStop}%
\bibitem [{\citenamefont {Kosugi}\ and\ \citenamefont
  {Matsushita}(2020{\natexlab{a}})}]{Kosugi20}%
  \BibitemOpen
  \bibfield  {author} {\bibinfo {author} {\bibfnamefont {T.}~\bibnamefont
  {Kosugi}}\ and\ \bibinfo {author} {\bibfnamefont {Y.~I.}\ \bibnamefont
  {Matsushita}},\ }\href {\doibase 10.1103/PhysRevA.101.012330} {\bibfield
  {journal} {\bibinfo  {journal} {Physical Review A}\ }\textbf {\bibinfo
  {volume} {101}},\ \bibinfo {pages} {1} (\bibinfo {year}
  {2020}{\natexlab{a}})}\BibitemShut {NoStop}%
\bibitem [{\citenamefont {Kosugi}\ and\ \citenamefont
  {Matsushita}(2020{\natexlab{b}})}]{Kosugi20linear}%
  \BibitemOpen
  \bibfield  {author} {\bibinfo {author} {\bibfnamefont {T.}~\bibnamefont
  {Kosugi}}\ and\ \bibinfo {author} {\bibfnamefont {Y.-i.}\ \bibnamefont
  {Matsushita}},\ }\href {\doibase 10.1103/PhysRevResearch.2.033043} {\bibfield
   {journal} {\bibinfo  {journal} {Phys. Rev. Research}\ }\textbf {\bibinfo
  {volume} {2}},\ \bibinfo {pages} {033043} (\bibinfo {year}
  {2020}{\natexlab{b}})}\BibitemShut {NoStop}%
\bibitem [{\citenamefont {Endo}\ \emph {et~al.}(2020)\citenamefont {Endo},
  \citenamefont {Kurata},\ and\ \citenamefont
  {Nakagawa}}]{endo2020calculation}%
  \BibitemOpen
  \bibfield  {author} {\bibinfo {author} {\bibfnamefont {S.}~\bibnamefont
  {Endo}}, \bibinfo {author} {\bibfnamefont {I.}~\bibnamefont {Kurata}}, \ and\
  \bibinfo {author} {\bibfnamefont {Y.~O.}\ \bibnamefont {Nakagawa}},\ }\href
  {\doibase 10.1103/PhysRevResearch.2.033281} {\bibfield  {journal} {\bibinfo
  {journal} {Phys. Rev. Research}\ }\textbf {\bibinfo {volume} {2}},\ \bibinfo
  {pages} {033281} (\bibinfo {year} {2020})}\BibitemShut {NoStop}%
\bibitem [{\citenamefont {Libbi}\ \emph {et~al.}(2022)\citenamefont {Libbi},
  \citenamefont {Rizzo}, \citenamefont {Tacchino}, \citenamefont {Marzari},\
  and\ \citenamefont {Tavernelli}}]{libbi2022effective}%
  \BibitemOpen
  \bibfield  {author} {\bibinfo {author} {\bibfnamefont {F.}~\bibnamefont
  {Libbi}}, \bibinfo {author} {\bibfnamefont {J.}~\bibnamefont {Rizzo}},
  \bibinfo {author} {\bibfnamefont {F.}~\bibnamefont {Tacchino}}, \bibinfo
  {author} {\bibfnamefont {N.}~\bibnamefont {Marzari}}, \ and\ \bibinfo
  {author} {\bibfnamefont {I.}~\bibnamefont {Tavernelli}},\ }\href@noop {}
  {\bibfield  {journal} {\bibinfo  {journal} {arXiv preprint arXiv:2203.12372}\
  } (\bibinfo {year} {2022})}\BibitemShut {NoStop}%
\bibitem [{\citenamefont {Chen}\ \emph {et~al.}(2021)\citenamefont {Chen},
  \citenamefont {Nusspickel}, \citenamefont {Tilly}, \citenamefont {Booth}
  \emph {et~al.}}]{chen2021variational}%
  \BibitemOpen
  \bibfield  {author} {\bibinfo {author} {\bibfnamefont {H.}~\bibnamefont
  {Chen}}, \bibinfo {author} {\bibfnamefont {M.}~\bibnamefont {Nusspickel}},
  \bibinfo {author} {\bibfnamefont {J.}~\bibnamefont {Tilly}}, \bibinfo
  {author} {\bibfnamefont {G.~H.}\ \bibnamefont {Booth}},  \emph {et~al.},\
  }\href@noop {} {\bibfield  {journal} {\bibinfo  {journal} {Physical Review
  A}\ }\textbf {\bibinfo {volume} {104}},\ \bibinfo {pages} {032405} (\bibinfo
  {year} {2021})}\BibitemShut {NoStop}%
\bibitem [{\citenamefont {Gyawali}\ and\ \citenamefont
  {Lawler}(2021)}]{gyawali2021insights}%
  \BibitemOpen
  \bibfield  {author} {\bibinfo {author} {\bibfnamefont {G.}~\bibnamefont
  {Gyawali}}\ and\ \bibinfo {author} {\bibfnamefont {M.~J.}\ \bibnamefont
  {Lawler}},\ }\href@noop {} {\bibfield  {journal} {\bibinfo  {journal} {arXiv
  preprint arXiv:2109.12126}\ } (\bibinfo {year} {2021})}\BibitemShut {NoStop}%
\bibitem [{\citenamefont {Lee}\ \emph {et~al.}(2022{\natexlab{a}})\citenamefont
  {Lee}, \citenamefont {Zhang}, \citenamefont {Hsieh}, \citenamefont {Zhang},\
  and\ \citenamefont {Shi}}]{lee2022variational}%
  \BibitemOpen
  \bibfield  {author} {\bibinfo {author} {\bibfnamefont {C.~K.}\ \bibnamefont
  {Lee}}, \bibinfo {author} {\bibfnamefont {S.-X.}\ \bibnamefont {Zhang}},
  \bibinfo {author} {\bibfnamefont {C.-Y.}\ \bibnamefont {Hsieh}}, \bibinfo
  {author} {\bibfnamefont {S.}~\bibnamefont {Zhang}}, \ and\ \bibinfo {author}
  {\bibfnamefont {L.}~\bibnamefont {Shi}},\ }\href
  {https://arxiv.org/abs/2206.05571} {\bibfield  {journal} {\bibinfo  {journal}
  {arXiv preprint arXiv:2206.05571}\ } (\bibinfo {year}
  {2022}{\natexlab{a}})}\BibitemShut {NoStop}%
\bibitem [{\citenamefont {Jensen}\ \emph {et~al.}(2022)\citenamefont {Jensen},
  \citenamefont {Johnson},\ and\ \citenamefont {Kunitsa}}]{jensen2022nearterm}%
  \BibitemOpen
  \bibfield  {author} {\bibinfo {author} {\bibfnamefont {P.~W.~K.}\
  \bibnamefont {Jensen}}, \bibinfo {author} {\bibfnamefont {P.~D.}\
  \bibnamefont {Johnson}}, \ and\ \bibinfo {author} {\bibfnamefont {A.~A.}\
  \bibnamefont {Kunitsa}},\ }\href {https://arxiv.org/abs/2206.09881}
  {\bibfield  {journal} {\bibinfo  {journal} {arXiv preprint arXiv:2206.09881}\
  } (\bibinfo {year} {2022})}\BibitemShut {NoStop}%
\bibitem [{\citenamefont {Huang}\ \emph {et~al.}(2022)\citenamefont {Huang},
  \citenamefont {Cai}, \citenamefont {Li}, \citenamefont {Ge}, \citenamefont
  {Hou}, \citenamefont {Li}, \citenamefont {Liu}, \citenamefont {Shi},
  \citenamefont {Chen}, \citenamefont {Zheng} \emph
  {et~al.}}]{huang2022variational}%
  \BibitemOpen
  \bibfield  {author} {\bibinfo {author} {\bibfnamefont {K.}~\bibnamefont
  {Huang}}, \bibinfo {author} {\bibfnamefont {X.}~\bibnamefont {Cai}}, \bibinfo
  {author} {\bibfnamefont {H.}~\bibnamefont {Li}}, \bibinfo {author}
  {\bibfnamefont {Z.-Y.}\ \bibnamefont {Ge}}, \bibinfo {author} {\bibfnamefont
  {R.}~\bibnamefont {Hou}}, \bibinfo {author} {\bibfnamefont {H.}~\bibnamefont
  {Li}}, \bibinfo {author} {\bibfnamefont {T.}~\bibnamefont {Liu}}, \bibinfo
  {author} {\bibfnamefont {Y.}~\bibnamefont {Shi}}, \bibinfo {author}
  {\bibfnamefont {C.}~\bibnamefont {Chen}}, \bibinfo {author} {\bibfnamefont
  {D.}~\bibnamefont {Zheng}},  \emph {et~al.},\ }\href@noop {} {\bibfield
  {journal} {\bibinfo  {journal} {The Journal of Physical Chemistry Letters}\
  }\textbf {\bibinfo {volume} {13}},\ \bibinfo {pages} {9114} (\bibinfo {year}
  {2022})}\BibitemShut {NoStop}%
\bibitem [{\citenamefont {Ciavarella}(2020)}]{Ciavarella20}%
  \BibitemOpen
  \bibfield  {author} {\bibinfo {author} {\bibfnamefont {A.}~\bibnamefont
  {Ciavarella}},\ }\href {\doibase 10.1103/PhysRevD.102.094505} {\bibfield
  {journal} {\bibinfo  {journal} {Phys. Rev. D}\ }\textbf {\bibinfo {volume}
  {102}},\ \bibinfo {pages} {094505} (\bibinfo {year} {2020})}\BibitemShut
  {NoStop}%
\bibitem [{\citenamefont {Roggero}(2020)}]{Roggero20}%
  \BibitemOpen
  \bibfield  {author} {\bibinfo {author} {\bibfnamefont {A.}~\bibnamefont
  {Roggero}},\ }\href {\doibase 10.1103/PhysRevA.102.022409} {\bibfield
  {journal} {\bibinfo  {journal} {Phys. Rev. A}\ }\textbf {\bibinfo {volume}
  {102}},\ \bibinfo {pages} {022409} (\bibinfo {year} {2020})}\BibitemShut
  {NoStop}%
\bibitem [{\citenamefont {Keen}\ \emph {et~al.}(2021)\citenamefont {Keen},
  \citenamefont {Dumitrescu},\ and\ \citenamefont {Wang}}]{keen2021quantum}%
  \BibitemOpen
  \bibfield  {author} {\bibinfo {author} {\bibfnamefont {T.}~\bibnamefont
  {Keen}}, \bibinfo {author} {\bibfnamefont {E.}~\bibnamefont {Dumitrescu}}, \
  and\ \bibinfo {author} {\bibfnamefont {Y.}~\bibnamefont {Wang}},\ }\href@noop
  {} {\bibfield  {journal} {\bibinfo  {journal} {arXiv preprint
  arXiv:2112.05731}\ } (\bibinfo {year} {2021})}\BibitemShut {NoStop}%
\bibitem [{\citenamefont {Tong}\ \emph {et~al.}(2021)\citenamefont {Tong},
  \citenamefont {An}, \citenamefont {Wiebe},\ and\ \citenamefont
  {Lin}}]{Tong21}%
  \BibitemOpen
  \bibfield  {author} {\bibinfo {author} {\bibfnamefont {Y.}~\bibnamefont
  {Tong}}, \bibinfo {author} {\bibfnamefont {D.}~\bibnamefont {An}}, \bibinfo
  {author} {\bibfnamefont {N.}~\bibnamefont {Wiebe}}, \ and\ \bibinfo {author}
  {\bibfnamefont {L.}~\bibnamefont {Lin}},\ }\href {\doibase
  10.1103/PhysRevA.104.032422} {\bibfield  {journal} {\bibinfo  {journal}
  {Phys. Rev. A}\ }\textbf {\bibinfo {volume} {104}},\ \bibinfo {pages}
  {032422} (\bibinfo {year} {2021})}\BibitemShut {NoStop}%
\bibitem [{\citenamefont {Gily{\'e}n}\ \emph {et~al.}(2019)\citenamefont
  {Gily{\'e}n}, \citenamefont {Su}, \citenamefont {Low},\ and\ \citenamefont
  {Wiebe}}]{gilyen2019quantum}%
  \BibitemOpen
  \bibfield  {author} {\bibinfo {author} {\bibfnamefont {A.}~\bibnamefont
  {Gily{\'e}n}}, \bibinfo {author} {\bibfnamefont {Y.}~\bibnamefont {Su}},
  \bibinfo {author} {\bibfnamefont {G.~H.}\ \bibnamefont {Low}}, \ and\
  \bibinfo {author} {\bibfnamefont {N.}~\bibnamefont {Wiebe}},\ }in\ \href@noop
  {} {\emph {\bibinfo {booktitle} {Proceedings of the 51st Annual ACM SIGACT
  Symposium on Theory of Computing}}}\ (\bibinfo {year} {2019})\ pp.\ \bibinfo
  {pages} {193--204}\BibitemShut {NoStop}%
\bibitem [{\citenamefont {Su}\ \emph {et~al.}(1979)\citenamefont {Su},
  \citenamefont {Schrieffer},\ and\ \citenamefont {Heeger}}]{ssh1}%
  \BibitemOpen
  \bibfield  {author} {\bibinfo {author} {\bibfnamefont {W.~P.}\ \bibnamefont
  {Su}}, \bibinfo {author} {\bibfnamefont {J.~R.}\ \bibnamefont {Schrieffer}},
  \ and\ \bibinfo {author} {\bibfnamefont {A.~J.}\ \bibnamefont {Heeger}},\
  }\href {\doibase 10.1103/PhysRevLett.42.1698} {\bibfield  {journal} {\bibinfo
   {journal} {Phys. Rev. Lett.}\ }\textbf {\bibinfo {volume} {42}},\ \bibinfo
  {pages} {1698} (\bibinfo {year} {1979})}\BibitemShut {NoStop}%
\bibitem [{\citenamefont {Bruus}\ and\ \citenamefont
  {Flensberg}(2004)}]{bruus}%
  \BibitemOpen
  \bibfield  {author} {\bibinfo {author} {\bibfnamefont {H.}~\bibnamefont
  {Bruus}}\ and\ \bibinfo {author} {\bibfnamefont {K.}~\bibnamefont
  {Flensberg}},\ }\href@noop {} {\emph {\bibinfo {title} {Many-body quantum
  theory in condensed matter physics: an introduction}}}\ (\bibinfo
  {publisher} {OUP Oxford},\ \bibinfo {year} {2004})\BibitemShut {NoStop}%
\bibitem [{\citenamefont {Gustafson}\ \emph {et~al.}(2021)\citenamefont
  {Gustafson}, \citenamefont {Holzman}, \citenamefont {Kowalkowski},
  \citenamefont {Lamm}, \citenamefont {Li}, \citenamefont {Perdue},
  \citenamefont {Boixo}, \citenamefont {Isakov}, \citenamefont {Martin},
  \citenamefont {Thomson}, \citenamefont {Heidweiller}, \citenamefont {Beall},
  \citenamefont {Ganahl}, \citenamefont {Vidal},\ and\ \citenamefont
  {Peters}}]{gustafson_large_2021}%
  \BibitemOpen
  \bibfield  {author} {\bibinfo {author} {\bibfnamefont {E.}~\bibnamefont
  {Gustafson}}, \bibinfo {author} {\bibfnamefont {B.}~\bibnamefont {Holzman}},
  \bibinfo {author} {\bibfnamefont {J.}~\bibnamefont {Kowalkowski}}, \bibinfo
  {author} {\bibfnamefont {H.}~\bibnamefont {Lamm}}, \bibinfo {author}
  {\bibfnamefont {A.~C.~Y.}\ \bibnamefont {Li}}, \bibinfo {author}
  {\bibfnamefont {G.}~\bibnamefont {Perdue}}, \bibinfo {author} {\bibfnamefont
  {S.}~\bibnamefont {Boixo}}, \bibinfo {author} {\bibfnamefont
  {S.}~\bibnamefont {Isakov}}, \bibinfo {author} {\bibfnamefont
  {O.}~\bibnamefont {Martin}}, \bibinfo {author} {\bibfnamefont
  {R.}~\bibnamefont {Thomson}}, \bibinfo {author} {\bibfnamefont {C.~V.}\
  \bibnamefont {Heidweiller}}, \bibinfo {author} {\bibfnamefont
  {J.}~\bibnamefont {Beall}}, \bibinfo {author} {\bibfnamefont
  {M.}~\bibnamefont {Ganahl}}, \bibinfo {author} {\bibfnamefont
  {G.}~\bibnamefont {Vidal}}, \ and\ \bibinfo {author} {\bibfnamefont
  {E.}~\bibnamefont {Peters}},\ }\href {\doibase 10.48550/ARXIV.2110.07482} {\
  (\bibinfo {year} {2021}),\ 10.48550/ARXIV.2110.07482}\BibitemShut {NoStop}%
\bibitem [{\citenamefont {Uhrich}\ \emph {et~al.}(2017)\citenamefont {Uhrich},
  \citenamefont {Castrignano}, \citenamefont {Uys},\ and\ \citenamefont
  {Kastner}}]{uhrich}%
  \BibitemOpen
  \bibfield  {author} {\bibinfo {author} {\bibfnamefont {P.}~\bibnamefont
  {Uhrich}}, \bibinfo {author} {\bibfnamefont {S.}~\bibnamefont {Castrignano}},
  \bibinfo {author} {\bibfnamefont {H.}~\bibnamefont {Uys}}, \ and\ \bibinfo
  {author} {\bibfnamefont {M.}~\bibnamefont {Kastner}},\ }\href {\doibase
  10.1103/PhysRevA.96.022127} {\bibfield  {journal} {\bibinfo  {journal} {Phys.
  Rev. A}\ }\textbf {\bibinfo {volume} {96}},\ \bibinfo {pages} {022127}
  (\bibinfo {year} {2017})}\BibitemShut {NoStop}%
\bibitem [{\citenamefont {White}(2009)}]{white2009minimally}%
  \BibitemOpen
  \bibfield  {author} {\bibinfo {author} {\bibfnamefont {S.~R.}\ \bibnamefont
  {White}},\ }\href@noop {} {\bibfield  {journal} {\bibinfo  {journal}
  {Physical review letters}\ }\textbf {\bibinfo {volume} {102}},\ \bibinfo
  {pages} {190601} (\bibinfo {year} {2009})}\BibitemShut {NoStop}%
\bibitem [{\citenamefont {Verdon}\ \emph {et~al.}(2019)\citenamefont {Verdon},
  \citenamefont {Marks}, \citenamefont {Nanda}, \citenamefont {Leichenauer},\
  and\ \citenamefont {Hidary}}]{verdon2019quantum}%
  \BibitemOpen
  \bibfield  {author} {\bibinfo {author} {\bibfnamefont {G.}~\bibnamefont
  {Verdon}}, \bibinfo {author} {\bibfnamefont {J.}~\bibnamefont {Marks}},
  \bibinfo {author} {\bibfnamefont {S.}~\bibnamefont {Nanda}}, \bibinfo
  {author} {\bibfnamefont {S.}~\bibnamefont {Leichenauer}}, \ and\ \bibinfo
  {author} {\bibfnamefont {J.}~\bibnamefont {Hidary}},\ }\href@noop {}
  {\bibfield  {journal} {\bibinfo  {journal} {arXiv preprint arXiv:1910.02071}\
  } (\bibinfo {year} {2019})}\BibitemShut {NoStop}%
\bibitem [{\citenamefont {Cohn}\ \emph {et~al.}(2020)\citenamefont {Cohn},
  \citenamefont {Yang}, \citenamefont {Najafi}, \citenamefont {Jones},\ and\
  \citenamefont {Freericks}}]{cohn2020minimal}%
  \BibitemOpen
  \bibfield  {author} {\bibinfo {author} {\bibfnamefont {J.}~\bibnamefont
  {Cohn}}, \bibinfo {author} {\bibfnamefont {F.}~\bibnamefont {Yang}}, \bibinfo
  {author} {\bibfnamefont {K.}~\bibnamefont {Najafi}}, \bibinfo {author}
  {\bibfnamefont {B.}~\bibnamefont {Jones}}, \ and\ \bibinfo {author}
  {\bibfnamefont {J.~K.}\ \bibnamefont {Freericks}},\ }\href {\doibase
  10.1103/PhysRevA.102.022622} {\bibfield  {journal} {\bibinfo  {journal}
  {Phys. Rev. A}\ }\textbf {\bibinfo {volume} {102}},\ \bibinfo {pages}
  {022622} (\bibinfo {year} {2020})}\BibitemShut {NoStop}%
\bibitem [{\citenamefont {Poulin}\ and\ \citenamefont
  {Wocjan}(2009)}]{poulin2009sampling}%
  \BibitemOpen
  \bibfield  {author} {\bibinfo {author} {\bibfnamefont {D.}~\bibnamefont
  {Poulin}}\ and\ \bibinfo {author} {\bibfnamefont {P.}~\bibnamefont
  {Wocjan}},\ }\href@noop {} {\bibfield  {journal} {\bibinfo  {journal}
  {Physical review letters}\ }\textbf {\bibinfo {volume} {103}},\ \bibinfo
  {pages} {220502} (\bibinfo {year} {2009})}\BibitemShut {NoStop}%
\bibitem [{\citenamefont {Motta}\ \emph {et~al.}(2020)\citenamefont {Motta},
  \citenamefont {Sun}, \citenamefont {Tan}, \citenamefont {O’Rourke},
  \citenamefont {Ye}, \citenamefont {Minnich}, \citenamefont {Brand{\~a}o},\
  and\ \citenamefont {Chan}}]{motta2020determining}%
  \BibitemOpen
  \bibfield  {author} {\bibinfo {author} {\bibfnamefont {M.}~\bibnamefont
  {Motta}}, \bibinfo {author} {\bibfnamefont {C.}~\bibnamefont {Sun}}, \bibinfo
  {author} {\bibfnamefont {A.~T.}\ \bibnamefont {Tan}}, \bibinfo {author}
  {\bibfnamefont {M.~J.}\ \bibnamefont {O’Rourke}}, \bibinfo {author}
  {\bibfnamefont {E.}~\bibnamefont {Ye}}, \bibinfo {author} {\bibfnamefont
  {A.~J.}\ \bibnamefont {Minnich}}, \bibinfo {author} {\bibfnamefont {F.~G.}\
  \bibnamefont {Brand{\~a}o}}, \ and\ \bibinfo {author} {\bibfnamefont
  {G.~K.-L.}\ \bibnamefont {Chan}},\ }\href@noop {} {\bibfield  {journal}
  {\bibinfo  {journal} {Nature Physics}\ }\textbf {\bibinfo {volume} {16}},\
  \bibinfo {pages} {205} (\bibinfo {year} {2020})}\BibitemShut {NoStop}%
\bibitem [{\citenamefont {Metcalf}\ \emph {et~al.}(2020)\citenamefont
  {Metcalf}, \citenamefont {Moussa}, \citenamefont {de~Jong},\ and\
  \citenamefont {Sarovar}}]{metcalf2020engineered}%
  \BibitemOpen
  \bibfield  {author} {\bibinfo {author} {\bibfnamefont {M.}~\bibnamefont
  {Metcalf}}, \bibinfo {author} {\bibfnamefont {J.~E.}\ \bibnamefont {Moussa}},
  \bibinfo {author} {\bibfnamefont {W.~A.}\ \bibnamefont {de~Jong}}, \ and\
  \bibinfo {author} {\bibfnamefont {M.}~\bibnamefont {Sarovar}},\ }\href@noop
  {} {\bibfield  {journal} {\bibinfo  {journal} {Physical Review Research}\
  }\textbf {\bibinfo {volume} {2}},\ \bibinfo {pages} {023214} (\bibinfo {year}
  {2020})}\BibitemShut {NoStop}%
\bibitem [{\citenamefont {Polla}\ \emph {et~al.}(2019)\citenamefont {Polla},
  \citenamefont {Herasymenko},\ and\ \citenamefont
  {O'Brien}}]{polla2019quantum}%
  \BibitemOpen
  \bibfield  {author} {\bibinfo {author} {\bibfnamefont {S.}~\bibnamefont
  {Polla}}, \bibinfo {author} {\bibfnamefont {Y.}~\bibnamefont {Herasymenko}},
  \ and\ \bibinfo {author} {\bibfnamefont {T.~E.}\ \bibnamefont {O'Brien}},\
  }\href@noop {} {\bibfield  {journal} {\bibinfo  {journal} {arXiv preprint
  arXiv:1909.10538}\ } (\bibinfo {year} {2019})}\BibitemShut {NoStop}%
\bibitem [{\citenamefont {Zhang}\ \emph {et~al.}(2020)\citenamefont {Zhang},
  \citenamefont {Zhang}, \citenamefont {Xue}, \citenamefont {Zhu},\ and\
  \citenamefont {Wang}}]{zhang2020continuous}%
  \BibitemOpen
  \bibfield  {author} {\bibinfo {author} {\bibfnamefont {D.-B.}\ \bibnamefont
  {Zhang}}, \bibinfo {author} {\bibfnamefont {G.-Q.}\ \bibnamefont {Zhang}},
  \bibinfo {author} {\bibfnamefont {Z.-Y.}\ \bibnamefont {Xue}}, \bibinfo
  {author} {\bibfnamefont {S.-L.}\ \bibnamefont {Zhu}}, \ and\ \bibinfo
  {author} {\bibfnamefont {Z.}~\bibnamefont {Wang}},\ }\href@noop {} {\bibfield
   {journal} {\bibinfo  {journal} {arXiv preprint arXiv:2006.00471}\ }
  (\bibinfo {year} {2020})}\BibitemShut {NoStop}%
\bibitem [{\citenamefont {Metcalf}\ \emph {et~al.}(2021)\citenamefont
  {Metcalf}, \citenamefont {Stone}, \citenamefont {Klymko}, \citenamefont
  {Kemper}, \citenamefont {Sarovar},\ and\ \citenamefont
  {de~Jong}}]{metcalf2021quantum}%
  \BibitemOpen
  \bibfield  {author} {\bibinfo {author} {\bibfnamefont {M.}~\bibnamefont
  {Metcalf}}, \bibinfo {author} {\bibfnamefont {E.}~\bibnamefont {Stone}},
  \bibinfo {author} {\bibfnamefont {K.}~\bibnamefont {Klymko}}, \bibinfo
  {author} {\bibfnamefont {A.~F.}\ \bibnamefont {Kemper}}, \bibinfo {author}
  {\bibfnamefont {M.}~\bibnamefont {Sarovar}}, \ and\ \bibinfo {author}
  {\bibfnamefont {W.~A.}\ \bibnamefont {de~Jong}},\ }\href@noop {} {\bibfield
  {journal} {\bibinfo  {journal} {arXiv preprint arXiv:2103.03207}\ } (\bibinfo
  {year} {2021})}\BibitemShut {NoStop}%
\bibitem [{\citenamefont {K{\"o}kc{\"u}}\ \emph {et~al.}(2022)\citenamefont
  {K{\"o}kc{\"u}}, \citenamefont {Camps}, \citenamefont {Bassman},
  \citenamefont {Freericks}, \citenamefont {de~Jong}, \citenamefont
  {Van~Beeumen},\ and\ \citenamefont {Kemper}}]{kokcu2022algebraic}%
  \BibitemOpen
  \bibfield  {author} {\bibinfo {author} {\bibfnamefont {E.}~\bibnamefont
  {K{\"o}kc{\"u}}}, \bibinfo {author} {\bibfnamefont {D.}~\bibnamefont
  {Camps}}, \bibinfo {author} {\bibfnamefont {L.}~\bibnamefont {Bassman}},
  \bibinfo {author} {\bibfnamefont {J.~K.}\ \bibnamefont {Freericks}}, \bibinfo
  {author} {\bibfnamefont {W.~A.}\ \bibnamefont {de~Jong}}, \bibinfo {author}
  {\bibfnamefont {R.}~\bibnamefont {Van~Beeumen}}, \ and\ \bibinfo {author}
  {\bibfnamefont {A.~F.}\ \bibnamefont {Kemper}},\ }\href@noop {} {\bibfield
  {journal} {\bibinfo  {journal} {Physical Review A}\ }\textbf {\bibinfo
  {volume} {105}},\ \bibinfo {pages} {032420} (\bibinfo {year}
  {2022})}\BibitemShut {NoStop}%
\bibitem [{\citenamefont {Camps}\ \emph {et~al.}(2022)\citenamefont {Camps},
  \citenamefont {K{\"o}kc{\"u}}, \citenamefont {Bassman}, \citenamefont
  {de~Jong}, \citenamefont {Kemper},\ and\ \citenamefont
  {Beeumen}}]{camps2022algebraic}%
  \BibitemOpen
  \bibfield  {author} {\bibinfo {author} {\bibfnamefont {D.}~\bibnamefont
  {Camps}}, \bibinfo {author} {\bibfnamefont {E.}~\bibnamefont
  {K{\"o}kc{\"u}}}, \bibinfo {author} {\bibfnamefont {L.}~\bibnamefont
  {Bassman}}, \bibinfo {author} {\bibfnamefont {W.~A.}\ \bibnamefont
  {de~Jong}}, \bibinfo {author} {\bibfnamefont {A.~F.}\ \bibnamefont {Kemper}},
  \ and\ \bibinfo {author} {\bibfnamefont {R.~V.}\ \bibnamefont {Beeumen}},\
  }\href@noop {} {\bibfield  {journal} {\bibinfo  {journal} {SIAM Journal on
  Matrix Analysis and Applications}\ }\textbf {\bibinfo {volume} {43}},\
  \bibinfo {pages} {1084} (\bibinfo {year} {2022})}\BibitemShut {NoStop}%
\bibitem [{\citenamefont {Lee}\ \emph {et~al.}(2022{\natexlab{b}})\citenamefont
  {Lee}, \citenamefont {Scott},\ and\ \citenamefont
  {Scarola}}]{lee2022compact}%
  \BibitemOpen
  \bibfield  {author} {\bibinfo {author} {\bibfnamefont {W.-R.}\ \bibnamefont
  {Lee}}, \bibinfo {author} {\bibfnamefont {R.}~\bibnamefont {Scott}}, \ and\
  \bibinfo {author} {\bibfnamefont {V.}~\bibnamefont {Scarola}},\ }\href@noop
  {} {\bibfield  {journal} {\bibinfo  {journal} {arXiv preprint
  arXiv:2212.14039}\ } (\bibinfo {year} {2022}{\natexlab{b}})}\BibitemShut
  {NoStop}%
\bibitem [{\citenamefont {ANIS}\ \emph {et~al.}(2021)\citenamefont {ANIS},
  \citenamefont {Abby-Mitchell}, \citenamefont {Abraham}, \citenamefont
  {AduOffei}, \citenamefont {Agarwal}, \citenamefont {Agliardi}, \citenamefont
  {Aharoni}, \citenamefont {Akhalwaya}, \citenamefont {Aleksandrowicz},
  \citenamefont {Alexander}, \citenamefont {Amy}, \citenamefont {Anagolum},
  \citenamefont {Anthony-Gandon}, \citenamefont {Arbel}, \citenamefont {Asfaw},
  \citenamefont {Athalye}, \citenamefont {Avkhadiev}, \citenamefont {Azaustre},
  \citenamefont {BHOLE}, \citenamefont {Banerjee}, \citenamefont {Banerjee},
  \citenamefont {Bang}, \citenamefont {Bansal}, \citenamefont {Barkoutsos},
  \citenamefont {Barnawal}, \citenamefont {Barron}, \citenamefont {Barron},
  \citenamefont {Bello}, \citenamefont {Ben-Haim}, \citenamefont {Bennett},
  \citenamefont {Bevenius}, \citenamefont {Bhatnagar}, \citenamefont {Bhobe},
  \citenamefont {Bianchini}, \citenamefont {Bishop}, \citenamefont {Blank},
  \citenamefont {Bolos}, \citenamefont {Bopardikar}, \citenamefont {Bosch},
  \citenamefont {Brandhofer}, \citenamefont {Brandon}, \citenamefont {Bravyi},
  \citenamefont {Bronn}, \citenamefont {Bryce-Fuller}, \citenamefont {Bucher},
  \citenamefont {Burov}, \citenamefont {Cabrera}, \citenamefont {Calpin},
  \citenamefont {Capelluto}, \citenamefont {Carballo}, \citenamefont
  {Carrascal}, \citenamefont {Carriker}, \citenamefont {Carvalho},
  \citenamefont {Chen}, \citenamefont {Chen}, \citenamefont {Chen},
  \citenamefont {Chen}, \citenamefont {Chen}, \citenamefont {Chevallier},
  \citenamefont {Chinda}, \citenamefont {Cholarajan}, \citenamefont {Chow},
  \citenamefont {Churchill}, \citenamefont {CisterMoke}, \citenamefont {Claus},
  \citenamefont {Clauss}, \citenamefont {Clothier}, \citenamefont {Cocking},
  \citenamefont {Cocuzzo}, \citenamefont {Connor}, \citenamefont {Correa},
  \citenamefont {Crockett}, \citenamefont {Cross}, \citenamefont {Cross},
  \citenamefont {Cross}, \citenamefont {Cruz-Benito}, \citenamefont {Culver},
  \citenamefont {C{\'o}rcoles-Gonzales}, \citenamefont {D}, \citenamefont
  {Dague}, \citenamefont {Dandachi}, \citenamefont {Dangwal}, \citenamefont
  {Daniel}, \citenamefont {Daniels}, \citenamefont {Dartiailh}, \citenamefont
  {Davila}, \citenamefont {Debouni}, \citenamefont {Dekusar}, \citenamefont
  {Deshmukh}, \citenamefont {Deshpande}, \citenamefont {Ding}, \citenamefont
  {Doi}, \citenamefont {Dow}, \citenamefont {Downing}, \citenamefont
  {Drechsler}, \citenamefont {Dumitrescu}, \citenamefont {Dumon}, \citenamefont
  {Duran}, \citenamefont {EL-Safty}, \citenamefont {Eastman}, \citenamefont
  {Eberle}, \citenamefont {Ebrahimi}, \citenamefont {Eendebak}, \citenamefont
  {Egger}, \citenamefont {ElePT}, \citenamefont {Emilio}, \citenamefont
  {Espiricueta}, \citenamefont {Everitt}, \citenamefont {Facoetti},
  \citenamefont {Farida}, \citenamefont {Fern{\'a}ndez}, \citenamefont
  {Ferracin}, \citenamefont {Ferrari}, \citenamefont {Ferrera}, \citenamefont
  {Fouilland}, \citenamefont {Frisch}, \citenamefont {Fuhrer}, \citenamefont
  {Fuller}, \citenamefont {GEORGE}, \citenamefont {Gacon}, \citenamefont
  {Gago}, \citenamefont {Gambella}, \citenamefont {Gambetta}, \citenamefont
  {Gammanpila}, \citenamefont {Garcia}, \citenamefont {Garg}, \citenamefont
  {Garion}, \citenamefont {Garrison}, \citenamefont {Garrison}, \citenamefont
  {Gates}, \citenamefont {Georgiev}, \citenamefont {Gil}, \citenamefont
  {Gilliam}, \citenamefont {Giridharan}, \citenamefont {Gomez-Mosquera},
  \citenamefont {Gonzalo}, \citenamefont {de~la Puente~Gonz{\'a}lez},
  \citenamefont {Gorzinski}, \citenamefont {Gould}, \citenamefont {Greenberg},
  \citenamefont {Grinko}, \citenamefont {Guan}, \citenamefont {Guijo},
  \citenamefont {Gunnels}, \citenamefont {Gupta}, \citenamefont {Gupta},
  \citenamefont {G{\"u}nther}, \citenamefont {Haglund}, \citenamefont {Haide},
  \citenamefont {Hamamura}, \citenamefont {Hamido}, \citenamefont {Harkins},
  \citenamefont {Hartman}, \citenamefont {Hasan}, \citenamefont {Havlicek},
  \citenamefont {Hellmers}, \citenamefont {Herok}, \citenamefont {Hillmich},
  \citenamefont {Horii}, \citenamefont {Howington}, \citenamefont {Hu},
  \citenamefont {Hu}, \citenamefont {Huang}, \citenamefont {Huisman},
  \citenamefont {Imai}, \citenamefont {Imamichi}, \citenamefont {Ishizaki},
  \citenamefont {Ishwor}, \citenamefont {Iten}, \citenamefont {Itoko},
  \citenamefont {Ivrii}, \citenamefont {Javadi}, \citenamefont {Javadi-Abhari},
  \citenamefont {Javed}, \citenamefont {Jianhua}, \citenamefont {Jivrajani},
  \citenamefont {Johns}, \citenamefont {Johnstun}, \citenamefont
  {Jonathan-Shoemaker}, \citenamefont {JosDenmark}, \citenamefont {JoshDumo},
  \citenamefont {Judge}, \citenamefont {Kachmann}, \citenamefont {Kale},
  \citenamefont {Kanazawa}, \citenamefont {Kane}, \citenamefont {Kang-Bae},
  \citenamefont {Kapila}, \citenamefont {Karazeev}, \citenamefont {Kassebaum},
  \citenamefont {Kehrer}, \citenamefont {Kelso}, \citenamefont {Kelso},
  \citenamefont {Khanderao}, \citenamefont {King}, \citenamefont {Kobayashi},
  \citenamefont {Kovi11Day}, \citenamefont {Kovyrshin}, \citenamefont
  {Krishnakumar}, \citenamefont {Krishnan}, \citenamefont {Krsulich},
  \citenamefont {Kumkar}, \citenamefont {Kus}, \citenamefont {LaRose},
  \citenamefont {Lacal}, \citenamefont {Lambert}, \citenamefont {Landa},
  \citenamefont {Lapeyre}, \citenamefont {Latone}, \citenamefont {Lawrence},
  \citenamefont {Lee}, \citenamefont {Li}, \citenamefont {Lishman},
  \citenamefont {Liu}, \citenamefont {Liu}, \citenamefont {Lolcroc},
  \citenamefont {M}, \citenamefont {Madden}, \citenamefont {Maeng},
  \citenamefont {Maheshkar}, \citenamefont {Majmudar}, \citenamefont
  {Malyshev}, \citenamefont {Mandouh}, \citenamefont {Manela}, \citenamefont
  {Manjula}, \citenamefont {Marecek}, \citenamefont {Marques}, \citenamefont
  {Marwaha}, \citenamefont {Maslov}, \citenamefont {Maszota}, \citenamefont
  {Mathews}, \citenamefont {Matsuo}, \citenamefont {Mazhandu}, \citenamefont
  {McClure}, \citenamefont {McElaney}, \citenamefont {McGarry}, \citenamefont
  {McKay}, \citenamefont {McPherson}, \citenamefont {Meesala}, \citenamefont
  {Meirom}, \citenamefont {Mendell}, \citenamefont {Metcalfe}, \citenamefont
  {Mevissen}, \citenamefont {Meyer}, \citenamefont {Mezzacapo}, \citenamefont
  {Midha}, \citenamefont {Miller}, \citenamefont {Minev}, \citenamefont
  {Mitchell}, \citenamefont {Moll}, \citenamefont {Montanez}, \citenamefont
  {Monteiro}, \citenamefont {Mooring}, \citenamefont {Morales}, \citenamefont
  {Moran}, \citenamefont {Morcuende}, \citenamefont {Mostafa}, \citenamefont
  {Motta}, \citenamefont {Moyard}, \citenamefont {Murali}, \citenamefont
  {Murata}, \citenamefont {M{\"u}ggenburg}, \citenamefont {NEMOZ},
  \citenamefont {Nadlinger}, \citenamefont {Nakanishi}, \citenamefont
  {Nannicini}, \citenamefont {Nation}, \citenamefont {Navarro}, \citenamefont
  {Naveh}, \citenamefont {Neagle}, \citenamefont {Neuweiler}, \citenamefont
  {Ngoueya}, \citenamefont {Nguyen}, \citenamefont {Nicander}, \citenamefont
  {Nick-Singstock}, \citenamefont {Niroula}, \citenamefont {Norlen},
  \citenamefont {NuoWenLei}, \citenamefont {O'Riordan}, \citenamefont
  {Ogunbayo}, \citenamefont {Ollitrault}, \citenamefont {Onodera},
  \citenamefont {Otaolea}, \citenamefont {Oud}, \citenamefont {Padilha},
  \citenamefont {Paik}, \citenamefont {Pal}, \citenamefont {Pang},
  \citenamefont {Panigrahi}, \citenamefont {Pascuzzi}, \citenamefont
  {Perriello}, \citenamefont {Peterson}, \citenamefont {Phan}, \citenamefont
  {Pilch}, \citenamefont {Piro}, \citenamefont {Pistoia}, \citenamefont
  {Piveteau}, \citenamefont {Plewa}, \citenamefont {Pocreau}, \citenamefont
  {Pozas-Kerstjens}, \citenamefont {Pracht}, \citenamefont {Prokop},
  \citenamefont {Prutyanov}, \citenamefont {Puri}, \citenamefont {Puzzuoli},
  \citenamefont {P{\'e}rez}, \citenamefont {Quant02}, \citenamefont {Quintiii},
  \citenamefont {Rahman}, \citenamefont {Raja}, \citenamefont {Rajeev},
  \citenamefont {Rajput}, \citenamefont {Ramagiri}, \citenamefont {Rao},
  \citenamefont {Raymond}, \citenamefont {Reardon-Smith}, \citenamefont
  {Redondo}, \citenamefont {Reuter}, \citenamefont {Rice}, \citenamefont
  {Riedemann}, \citenamefont {Rietesh}, \citenamefont {Risinger}, \citenamefont
  {Rocca}, \citenamefont {Rodr{\'\i}guez}, \citenamefont {RohithKarur},
  \citenamefont {Rosand}, \citenamefont {Rossmannek}, \citenamefont {Ryu},
  \citenamefont {SAPV}, \citenamefont {Sa}, \citenamefont {Saha}, \citenamefont
  {Ash-Saki}, \citenamefont {Sanand}, \citenamefont {Sandberg}, \citenamefont
  {Sandesara}, \citenamefont {Sapra}, \citenamefont {Sargsyan}, \citenamefont
  {Sarkar}, \citenamefont {Sathaye}, \citenamefont {Schmitt}, \citenamefont
  {Schnabel}, \citenamefont {Schoenfeld}, \citenamefont {Scholten},
  \citenamefont {Schoute}, \citenamefont {Schulterbrandt}, \citenamefont
  {Schwarm}, \citenamefont {Seaward}, \citenamefont {Sergi}, \citenamefont
  {Sertage}, \citenamefont {Setia}, \citenamefont {Shah}, \citenamefont
  {Shammah}, \citenamefont {Sharma}, \citenamefont {Shi}, \citenamefont
  {Shoemaker}, \citenamefont {Silva}, \citenamefont {Simonetto}, \citenamefont
  {Singh}, \citenamefont {Singh}, \citenamefont {Singh}, \citenamefont
  {Singkanipa}, \citenamefont {Siraichi}, \citenamefont {Siri}, \citenamefont
  {Sistos}, \citenamefont {Sitdikov}, \citenamefont {Sivarajah}, \citenamefont
  {Slavikmew}, \citenamefont {Sletfjerding}, \citenamefont {Smolin},
  \citenamefont {Soeken}, \citenamefont {Sokolov}, \citenamefont {Sokolov},
  \citenamefont {Soloviev}, \citenamefont {SooluThomas}, \citenamefont
  {Starfish}, \citenamefont {Steenken}, \citenamefont {Stypulkoski},
  \citenamefont {Suau}, \citenamefont {Sun}, \citenamefont {Sung},
  \citenamefont {Suwama}, \citenamefont {S{\l}owik}, \citenamefont {Takahashi},
  \citenamefont {Takawale}, \citenamefont {Tavernelli}, \citenamefont {Taylor},
  \citenamefont {Taylour}, \citenamefont {Thomas}, \citenamefont {Tian},
  \citenamefont {Tillet}, \citenamefont {Tod}, \citenamefont {Tomasik},
  \citenamefont {Tornow}, \citenamefont {de~la Torre}, \citenamefont {Toural},
  \citenamefont {Trabing}, \citenamefont {Treinish}, \citenamefont {Trenev},
  \citenamefont {TrishaPe}, \citenamefont {Truger}, \citenamefont
  {Tsilimigkounakis}, \citenamefont {Tulsi}, \citenamefont {Turner},
  \citenamefont {Vaknin}, \citenamefont {Valcarce}, \citenamefont {Varchon},
  \citenamefont {Vartak}, \citenamefont {Vazquez}, \citenamefont
  {Vijaywargiya}, \citenamefont {Villar}, \citenamefont {Vishnu}, \citenamefont
  {Vogt-Lee}, \citenamefont {Vuillot}, \citenamefont {Weaver}, \citenamefont
  {Weidenfeller}, \citenamefont {Wieczorek}, \citenamefont {Wildstrom},
  \citenamefont {Wilson}, \citenamefont {Winston}, \citenamefont
  {WinterSoldier}, \citenamefont {Woehr}, \citenamefont {Woerner},
  \citenamefont {Woo}, \citenamefont {Wood}, \citenamefont {Wood},
  \citenamefont {Wood}, \citenamefont {Wootton}, \citenamefont {Wright},
  \citenamefont {Xing}, \citenamefont {YU}, \citenamefont {Yang}, \citenamefont
  {Yang}, \citenamefont {Yao}, \citenamefont {Yeralin}, \citenamefont
  {Yonekura}, \citenamefont {Yonge-Mallo}, \citenamefont {Yoshida},
  \citenamefont {Young}, \citenamefont {Yu}, \citenamefont {Yu}, \citenamefont
  {Zachow}, \citenamefont {Zdanski}, \citenamefont {Zhang}, \citenamefont
  {Zidaru}, \citenamefont {Zimmermann}, \citenamefont {Zoufal}, \citenamefont
  {aeddins ibm}, \citenamefont {alexzhang13}, \citenamefont {b63},
  \citenamefont {bartek bartlomiej}, \citenamefont {bcamorrison}, \citenamefont
  {brandhsn}, \citenamefont {charmerDark}, \citenamefont {deeplokhande},
  \citenamefont {dekel.meirom}, \citenamefont {dime10}, \citenamefont
  {dlasecki}, \citenamefont {ehchen}, \citenamefont {fanizzamarco},
  \citenamefont {fs1132429}, \citenamefont {gadial}, \citenamefont
  {galeinston}, \citenamefont {georgezhou20}, \citenamefont {georgios ts},
  \citenamefont {gruu}, \citenamefont {hhorii}, \citenamefont {hykavitha},
  \citenamefont {itoko}, \citenamefont {jeppevinkel}, \citenamefont {jessica
  angel7}, \citenamefont {jezerjojo14}, \citenamefont {jliu45}, \citenamefont
  {jscott2}, \citenamefont {klinvill}, \citenamefont {krutik2966},
  \citenamefont {ma5x}, \citenamefont {michelle4654}, \citenamefont {msuwama},
  \citenamefont {nico lgrs}, \citenamefont {nrhawkins}, \citenamefont
  {ntgiwsvp}, \citenamefont {ordmoj}, \citenamefont {sagar pahwa},
  \citenamefont {pritamsinha2304}, \citenamefont {ryancocuzzo}, \citenamefont
  {saktar unr}, \citenamefont {saswati qiskit}, \citenamefont {septembrr},
  \citenamefont {sethmerkel}, \citenamefont {sg495}, \citenamefont {shaashwat},
  \citenamefont {smturro2}, \citenamefont {sternparky}, \citenamefont
  {strickroman}, \citenamefont {tigerjack}, \citenamefont {tsura crisaldo},
  \citenamefont {upsideon}, \citenamefont {vadebayo49}, \citenamefont {welien},
  \citenamefont {willhbang}, \citenamefont {wmurphy collabstar}, \citenamefont
  {yang.luh},\ and\ \citenamefont {{\v{C}}epulkovskis}}]{Qiskit}%
  \BibitemOpen
  \bibfield  {author} {\bibinfo {author} {\bibfnamefont {M.~S.}\ \bibnamefont
  {ANIS}}, \bibinfo {author} {\bibnamefont {Abby-Mitchell}}, \bibinfo {author}
  {\bibfnamefont {H.}~\bibnamefont {Abraham}}, \bibinfo {author} {\bibnamefont
  {AduOffei}}, \bibinfo {author} {\bibfnamefont {R.}~\bibnamefont {Agarwal}},
  \bibinfo {author} {\bibfnamefont {G.}~\bibnamefont {Agliardi}}, \bibinfo
  {author} {\bibfnamefont {M.}~\bibnamefont {Aharoni}}, \bibinfo {author}
  {\bibfnamefont {I.~Y.}\ \bibnamefont {Akhalwaya}}, \bibinfo {author}
  {\bibfnamefont {G.}~\bibnamefont {Aleksandrowicz}}, \bibinfo {author}
  {\bibfnamefont {T.}~\bibnamefont {Alexander}}, \bibinfo {author}
  {\bibfnamefont {M.}~\bibnamefont {Amy}}, \bibinfo {author} {\bibfnamefont
  {S.}~\bibnamefont {Anagolum}}, \bibinfo {author} {\bibnamefont
  {Anthony-Gandon}}, \bibinfo {author} {\bibfnamefont {E.}~\bibnamefont
  {Arbel}}, \bibinfo {author} {\bibfnamefont {A.}~\bibnamefont {Asfaw}},
  \bibinfo {author} {\bibfnamefont {A.}~\bibnamefont {Athalye}}, \bibinfo
  {author} {\bibfnamefont {A.}~\bibnamefont {Avkhadiev}}, \bibinfo {author}
  {\bibfnamefont {C.}~\bibnamefont {Azaustre}}, \bibinfo {author}
  {\bibfnamefont {P.}~\bibnamefont {BHOLE}}, \bibinfo {author} {\bibfnamefont
  {A.}~\bibnamefont {Banerjee}}, \bibinfo {author} {\bibfnamefont
  {S.}~\bibnamefont {Banerjee}}, \bibinfo {author} {\bibfnamefont
  {W.}~\bibnamefont {Bang}}, \bibinfo {author} {\bibfnamefont {A.}~\bibnamefont
  {Bansal}}, \bibinfo {author} {\bibfnamefont {P.}~\bibnamefont {Barkoutsos}},
  \bibinfo {author} {\bibfnamefont {A.}~\bibnamefont {Barnawal}}, \bibinfo
  {author} {\bibfnamefont {G.}~\bibnamefont {Barron}}, \bibinfo {author}
  {\bibfnamefont {G.~S.}\ \bibnamefont {Barron}}, \bibinfo {author}
  {\bibfnamefont {L.}~\bibnamefont {Bello}}, \bibinfo {author} {\bibfnamefont
  {Y.}~\bibnamefont {Ben-Haim}}, \bibinfo {author} {\bibfnamefont {M.~C.}\
  \bibnamefont {Bennett}}, \bibinfo {author} {\bibfnamefont {D.}~\bibnamefont
  {Bevenius}}, \bibinfo {author} {\bibfnamefont {D.}~\bibnamefont {Bhatnagar}},
  \bibinfo {author} {\bibfnamefont {A.}~\bibnamefont {Bhobe}}, \bibinfo
  {author} {\bibfnamefont {P.}~\bibnamefont {Bianchini}}, \bibinfo {author}
  {\bibfnamefont {L.~S.}\ \bibnamefont {Bishop}}, \bibinfo {author}
  {\bibfnamefont {C.}~\bibnamefont {Blank}}, \bibinfo {author} {\bibfnamefont
  {S.}~\bibnamefont {Bolos}}, \bibinfo {author} {\bibfnamefont
  {S.}~\bibnamefont {Bopardikar}}, \bibinfo {author} {\bibfnamefont
  {S.}~\bibnamefont {Bosch}}, \bibinfo {author} {\bibfnamefont
  {S.}~\bibnamefont {Brandhofer}}, \bibinfo {author} {\bibnamefont {Brandon}},
  \bibinfo {author} {\bibfnamefont {S.}~\bibnamefont {Bravyi}}, \bibinfo
  {author} {\bibfnamefont {N.}~\bibnamefont {Bronn}}, \bibinfo {author}
  {\bibnamefont {Bryce-Fuller}}, \bibinfo {author} {\bibfnamefont
  {D.}~\bibnamefont {Bucher}}, \bibinfo {author} {\bibfnamefont
  {A.}~\bibnamefont {Burov}}, \bibinfo {author} {\bibfnamefont
  {F.}~\bibnamefont {Cabrera}}, \bibinfo {author} {\bibfnamefont
  {P.}~\bibnamefont {Calpin}}, \bibinfo {author} {\bibfnamefont
  {L.}~\bibnamefont {Capelluto}}, \bibinfo {author} {\bibfnamefont
  {J.}~\bibnamefont {Carballo}}, \bibinfo {author} {\bibfnamefont
  {G.}~\bibnamefont {Carrascal}}, \bibinfo {author} {\bibfnamefont
  {A.}~\bibnamefont {Carriker}}, \bibinfo {author} {\bibfnamefont
  {I.}~\bibnamefont {Carvalho}}, \bibinfo {author} {\bibfnamefont
  {A.}~\bibnamefont {Chen}}, \bibinfo {author} {\bibfnamefont {C.-F.}\
  \bibnamefont {Chen}}, \bibinfo {author} {\bibfnamefont {E.}~\bibnamefont
  {Chen}}, \bibinfo {author} {\bibfnamefont {J.~C.}\ \bibnamefont {Chen}},
  \bibinfo {author} {\bibfnamefont {R.}~\bibnamefont {Chen}}, \bibinfo {author}
  {\bibfnamefont {F.}~\bibnamefont {Chevallier}}, \bibinfo {author}
  {\bibfnamefont {K.}~\bibnamefont {Chinda}}, \bibinfo {author} {\bibfnamefont
  {R.}~\bibnamefont {Cholarajan}}, \bibinfo {author} {\bibfnamefont {J.~M.}\
  \bibnamefont {Chow}}, \bibinfo {author} {\bibfnamefont {S.}~\bibnamefont
  {Churchill}}, \bibinfo {author} {\bibnamefont {CisterMoke}}, \bibinfo
  {author} {\bibfnamefont {C.}~\bibnamefont {Claus}}, \bibinfo {author}
  {\bibfnamefont {C.}~\bibnamefont {Clauss}}, \bibinfo {author} {\bibfnamefont
  {C.}~\bibnamefont {Clothier}}, \bibinfo {author} {\bibfnamefont
  {R.}~\bibnamefont {Cocking}}, \bibinfo {author} {\bibfnamefont
  {R.}~\bibnamefont {Cocuzzo}}, \bibinfo {author} {\bibfnamefont
  {J.}~\bibnamefont {Connor}}, \bibinfo {author} {\bibfnamefont
  {F.}~\bibnamefont {Correa}}, \bibinfo {author} {\bibfnamefont
  {Z.}~\bibnamefont {Crockett}}, \bibinfo {author} {\bibfnamefont {A.~J.}\
  \bibnamefont {Cross}}, \bibinfo {author} {\bibfnamefont {A.~W.}\ \bibnamefont
  {Cross}}, \bibinfo {author} {\bibfnamefont {S.}~\bibnamefont {Cross}},
  \bibinfo {author} {\bibfnamefont {J.}~\bibnamefont {Cruz-Benito}}, \bibinfo
  {author} {\bibfnamefont {C.}~\bibnamefont {Culver}}, \bibinfo {author}
  {\bibfnamefont {A.~D.}\ \bibnamefont {C{\'o}rcoles-Gonzales}}, \bibinfo
  {author} {\bibfnamefont {N.}~\bibnamefont {D}}, \bibinfo {author}
  {\bibfnamefont {S.}~\bibnamefont {Dague}}, \bibinfo {author} {\bibfnamefont
  {T.~E.}\ \bibnamefont {Dandachi}}, \bibinfo {author} {\bibfnamefont {A.~N.}\
  \bibnamefont {Dangwal}}, \bibinfo {author} {\bibfnamefont {J.}~\bibnamefont
  {Daniel}}, \bibinfo {author} {\bibfnamefont {M.}~\bibnamefont {Daniels}},
  \bibinfo {author} {\bibfnamefont {M.}~\bibnamefont {Dartiailh}}, \bibinfo
  {author} {\bibfnamefont {A.~R.}\ \bibnamefont {Davila}}, \bibinfo {author}
  {\bibfnamefont {F.}~\bibnamefont {Debouni}}, \bibinfo {author} {\bibfnamefont
  {A.}~\bibnamefont {Dekusar}}, \bibinfo {author} {\bibfnamefont
  {A.}~\bibnamefont {Deshmukh}}, \bibinfo {author} {\bibfnamefont
  {M.}~\bibnamefont {Deshpande}}, \bibinfo {author} {\bibfnamefont
  {D.}~\bibnamefont {Ding}}, \bibinfo {author} {\bibfnamefont {J.}~\bibnamefont
  {Doi}}, \bibinfo {author} {\bibfnamefont {E.~M.}\ \bibnamefont {Dow}},
  \bibinfo {author} {\bibfnamefont {P.}~\bibnamefont {Downing}}, \bibinfo
  {author} {\bibfnamefont {E.}~\bibnamefont {Drechsler}}, \bibinfo {author}
  {\bibfnamefont {E.}~\bibnamefont {Dumitrescu}}, \bibinfo {author}
  {\bibfnamefont {K.}~\bibnamefont {Dumon}}, \bibinfo {author} {\bibfnamefont
  {I.}~\bibnamefont {Duran}}, \bibinfo {author} {\bibfnamefont
  {K.}~\bibnamefont {EL-Safty}}, \bibinfo {author} {\bibfnamefont
  {E.}~\bibnamefont {Eastman}}, \bibinfo {author} {\bibfnamefont
  {G.}~\bibnamefont {Eberle}}, \bibinfo {author} {\bibfnamefont
  {A.}~\bibnamefont {Ebrahimi}}, \bibinfo {author} {\bibfnamefont
  {P.}~\bibnamefont {Eendebak}}, \bibinfo {author} {\bibfnamefont
  {D.}~\bibnamefont {Egger}}, \bibinfo {author} {\bibnamefont {ElePT}},
  \bibinfo {author} {\bibnamefont {Emilio}}, \bibinfo {author} {\bibfnamefont
  {A.}~\bibnamefont {Espiricueta}}, \bibinfo {author} {\bibfnamefont
  {M.}~\bibnamefont {Everitt}}, \bibinfo {author} {\bibfnamefont
  {D.}~\bibnamefont {Facoetti}}, \bibinfo {author} {\bibnamefont {Farida}},
  \bibinfo {author} {\bibfnamefont {P.~M.}\ \bibnamefont {Fern{\'a}ndez}},
  \bibinfo {author} {\bibfnamefont {S.}~\bibnamefont {Ferracin}}, \bibinfo
  {author} {\bibfnamefont {D.}~\bibnamefont {Ferrari}}, \bibinfo {author}
  {\bibfnamefont {A.~H.}\ \bibnamefont {Ferrera}}, \bibinfo {author}
  {\bibfnamefont {R.}~\bibnamefont {Fouilland}}, \bibinfo {author}
  {\bibfnamefont {A.}~\bibnamefont {Frisch}}, \bibinfo {author} {\bibfnamefont
  {A.}~\bibnamefont {Fuhrer}}, \bibinfo {author} {\bibfnamefont
  {B.}~\bibnamefont {Fuller}}, \bibinfo {author} {\bibfnamefont
  {M.}~\bibnamefont {GEORGE}}, \bibinfo {author} {\bibfnamefont
  {J.}~\bibnamefont {Gacon}}, \bibinfo {author} {\bibfnamefont {B.~G.}\
  \bibnamefont {Gago}}, \bibinfo {author} {\bibfnamefont {C.}~\bibnamefont
  {Gambella}}, \bibinfo {author} {\bibfnamefont {J.~M.}\ \bibnamefont
  {Gambetta}}, \bibinfo {author} {\bibfnamefont {A.}~\bibnamefont
  {Gammanpila}}, \bibinfo {author} {\bibfnamefont {L.}~\bibnamefont {Garcia}},
  \bibinfo {author} {\bibfnamefont {T.}~\bibnamefont {Garg}}, \bibinfo {author}
  {\bibfnamefont {S.}~\bibnamefont {Garion}}, \bibinfo {author} {\bibfnamefont
  {J.~R.}\ \bibnamefont {Garrison}}, \bibinfo {author} {\bibfnamefont
  {J.}~\bibnamefont {Garrison}}, \bibinfo {author} {\bibfnamefont
  {T.}~\bibnamefont {Gates}}, \bibinfo {author} {\bibfnamefont
  {H.}~\bibnamefont {Georgiev}}, \bibinfo {author} {\bibfnamefont
  {L.}~\bibnamefont {Gil}}, \bibinfo {author} {\bibfnamefont {A.}~\bibnamefont
  {Gilliam}}, \bibinfo {author} {\bibfnamefont {A.}~\bibnamefont {Giridharan}},
  \bibinfo {author} {\bibfnamefont {J.}~\bibnamefont {Gomez-Mosquera}},
  \bibinfo {author} {\bibnamefont {Gonzalo}}, \bibinfo {author} {\bibfnamefont
  {S.}~\bibnamefont {de~la Puente~Gonz{\'a}lez}}, \bibinfo {author}
  {\bibfnamefont {J.}~\bibnamefont {Gorzinski}}, \bibinfo {author}
  {\bibfnamefont {I.}~\bibnamefont {Gould}}, \bibinfo {author} {\bibfnamefont
  {D.}~\bibnamefont {Greenberg}}, \bibinfo {author} {\bibfnamefont
  {D.}~\bibnamefont {Grinko}}, \bibinfo {author} {\bibfnamefont
  {W.}~\bibnamefont {Guan}}, \bibinfo {author} {\bibfnamefont {D.}~\bibnamefont
  {Guijo}}, \bibinfo {author} {\bibfnamefont {J.~A.}\ \bibnamefont {Gunnels}},
  \bibinfo {author} {\bibfnamefont {H.}~\bibnamefont {Gupta}}, \bibinfo
  {author} {\bibfnamefont {N.}~\bibnamefont {Gupta}}, \bibinfo {author}
  {\bibfnamefont {J.~M.}\ \bibnamefont {G{\"u}nther}}, \bibinfo {author}
  {\bibfnamefont {M.}~\bibnamefont {Haglund}}, \bibinfo {author} {\bibfnamefont
  {I.}~\bibnamefont {Haide}}, \bibinfo {author} {\bibfnamefont
  {I.}~\bibnamefont {Hamamura}}, \bibinfo {author} {\bibfnamefont {O.~C.}\
  \bibnamefont {Hamido}}, \bibinfo {author} {\bibfnamefont {F.}~\bibnamefont
  {Harkins}}, \bibinfo {author} {\bibfnamefont {K.}~\bibnamefont {Hartman}},
  \bibinfo {author} {\bibfnamefont {A.}~\bibnamefont {Hasan}}, \bibinfo
  {author} {\bibfnamefont {V.}~\bibnamefont {Havlicek}}, \bibinfo {author}
  {\bibfnamefont {J.}~\bibnamefont {Hellmers}}, \bibinfo {author}
  {\bibfnamefont {{\L}.}~\bibnamefont {Herok}}, \bibinfo {author}
  {\bibfnamefont {S.}~\bibnamefont {Hillmich}}, \bibinfo {author}
  {\bibfnamefont {H.}~\bibnamefont {Horii}}, \bibinfo {author} {\bibfnamefont
  {C.}~\bibnamefont {Howington}}, \bibinfo {author} {\bibfnamefont
  {S.}~\bibnamefont {Hu}}, \bibinfo {author} {\bibfnamefont {W.}~\bibnamefont
  {Hu}}, \bibinfo {author} {\bibfnamefont {J.}~\bibnamefont {Huang}}, \bibinfo
  {author} {\bibfnamefont {R.}~\bibnamefont {Huisman}}, \bibinfo {author}
  {\bibfnamefont {H.}~\bibnamefont {Imai}}, \bibinfo {author} {\bibfnamefont
  {T.}~\bibnamefont {Imamichi}}, \bibinfo {author} {\bibfnamefont
  {K.}~\bibnamefont {Ishizaki}}, \bibinfo {author} {\bibnamefont {Ishwor}},
  \bibinfo {author} {\bibfnamefont {R.}~\bibnamefont {Iten}}, \bibinfo {author}
  {\bibfnamefont {T.}~\bibnamefont {Itoko}}, \bibinfo {author} {\bibfnamefont
  {A.}~\bibnamefont {Ivrii}}, \bibinfo {author} {\bibfnamefont
  {A.}~\bibnamefont {Javadi}}, \bibinfo {author} {\bibfnamefont
  {A.}~\bibnamefont {Javadi-Abhari}}, \bibinfo {author} {\bibfnamefont
  {W.}~\bibnamefont {Javed}}, \bibinfo {author} {\bibfnamefont
  {Q.}~\bibnamefont {Jianhua}}, \bibinfo {author} {\bibfnamefont
  {M.}~\bibnamefont {Jivrajani}}, \bibinfo {author} {\bibfnamefont
  {K.}~\bibnamefont {Johns}}, \bibinfo {author} {\bibfnamefont
  {S.}~\bibnamefont {Johnstun}}, \bibinfo {author} {\bibnamefont
  {Jonathan-Shoemaker}}, \bibinfo {author} {\bibnamefont {JosDenmark}},
  \bibinfo {author} {\bibnamefont {JoshDumo}}, \bibinfo {author} {\bibfnamefont
  {J.}~\bibnamefont {Judge}}, \bibinfo {author} {\bibfnamefont
  {T.}~\bibnamefont {Kachmann}}, \bibinfo {author} {\bibfnamefont
  {A.}~\bibnamefont {Kale}}, \bibinfo {author} {\bibfnamefont {N.}~\bibnamefont
  {Kanazawa}}, \bibinfo {author} {\bibfnamefont {J.}~\bibnamefont {Kane}},
  \bibinfo {author} {\bibnamefont {Kang-Bae}}, \bibinfo {author} {\bibfnamefont
  {A.}~\bibnamefont {Kapila}}, \bibinfo {author} {\bibfnamefont
  {A.}~\bibnamefont {Karazeev}}, \bibinfo {author} {\bibfnamefont
  {P.}~\bibnamefont {Kassebaum}}, \bibinfo {author} {\bibfnamefont
  {T.}~\bibnamefont {Kehrer}}, \bibinfo {author} {\bibfnamefont
  {J.}~\bibnamefont {Kelso}}, \bibinfo {author} {\bibfnamefont
  {S.}~\bibnamefont {Kelso}}, \bibinfo {author} {\bibfnamefont
  {V.}~\bibnamefont {Khanderao}}, \bibinfo {author} {\bibfnamefont
  {S.}~\bibnamefont {King}}, \bibinfo {author} {\bibfnamefont {Y.}~\bibnamefont
  {Kobayashi}}, \bibinfo {author} {\bibnamefont {Kovi11Day}}, \bibinfo {author}
  {\bibfnamefont {A.}~\bibnamefont {Kovyrshin}}, \bibinfo {author}
  {\bibfnamefont {R.}~\bibnamefont {Krishnakumar}}, \bibinfo {author}
  {\bibfnamefont {V.}~\bibnamefont {Krishnan}}, \bibinfo {author}
  {\bibfnamefont {K.}~\bibnamefont {Krsulich}}, \bibinfo {author}
  {\bibfnamefont {P.}~\bibnamefont {Kumkar}}, \bibinfo {author} {\bibfnamefont
  {G.}~\bibnamefont {Kus}}, \bibinfo {author} {\bibfnamefont {R.}~\bibnamefont
  {LaRose}}, \bibinfo {author} {\bibfnamefont {E.}~\bibnamefont {Lacal}},
  \bibinfo {author} {\bibfnamefont {R.}~\bibnamefont {Lambert}}, \bibinfo
  {author} {\bibfnamefont {H.}~\bibnamefont {Landa}}, \bibinfo {author}
  {\bibfnamefont {J.}~\bibnamefont {Lapeyre}}, \bibinfo {author} {\bibfnamefont
  {J.}~\bibnamefont {Latone}}, \bibinfo {author} {\bibfnamefont
  {S.}~\bibnamefont {Lawrence}}, \bibinfo {author} {\bibfnamefont
  {C.}~\bibnamefont {Lee}}, \bibinfo {author} {\bibfnamefont {G.}~\bibnamefont
  {Li}}, \bibinfo {author} {\bibfnamefont {J.}~\bibnamefont {Lishman}},
  \bibinfo {author} {\bibfnamefont {D.}~\bibnamefont {Liu}}, \bibinfo {author}
  {\bibfnamefont {P.}~\bibnamefont {Liu}}, \bibinfo {author} {\bibnamefont
  {Lolcroc}}, \bibinfo {author} {\bibfnamefont {A.~K.}\ \bibnamefont {M}},
  \bibinfo {author} {\bibfnamefont {L.}~\bibnamefont {Madden}}, \bibinfo
  {author} {\bibfnamefont {Y.}~\bibnamefont {Maeng}}, \bibinfo {author}
  {\bibfnamefont {S.}~\bibnamefont {Maheshkar}}, \bibinfo {author}
  {\bibfnamefont {K.}~\bibnamefont {Majmudar}}, \bibinfo {author}
  {\bibfnamefont {A.}~\bibnamefont {Malyshev}}, \bibinfo {author}
  {\bibfnamefont {M.~E.}\ \bibnamefont {Mandouh}}, \bibinfo {author}
  {\bibfnamefont {J.}~\bibnamefont {Manela}}, \bibinfo {author} {\bibnamefont
  {Manjula}}, \bibinfo {author} {\bibfnamefont {J.}~\bibnamefont {Marecek}},
  \bibinfo {author} {\bibfnamefont {M.}~\bibnamefont {Marques}}, \bibinfo
  {author} {\bibfnamefont {K.}~\bibnamefont {Marwaha}}, \bibinfo {author}
  {\bibfnamefont {D.}~\bibnamefont {Maslov}}, \bibinfo {author} {\bibfnamefont
  {P.}~\bibnamefont {Maszota}}, \bibinfo {author} {\bibfnamefont
  {D.}~\bibnamefont {Mathews}}, \bibinfo {author} {\bibfnamefont
  {A.}~\bibnamefont {Matsuo}}, \bibinfo {author} {\bibfnamefont
  {F.}~\bibnamefont {Mazhandu}}, \bibinfo {author} {\bibfnamefont
  {D.}~\bibnamefont {McClure}}, \bibinfo {author} {\bibfnamefont
  {M.}~\bibnamefont {McElaney}}, \bibinfo {author} {\bibfnamefont
  {C.}~\bibnamefont {McGarry}}, \bibinfo {author} {\bibfnamefont
  {D.}~\bibnamefont {McKay}}, \bibinfo {author} {\bibfnamefont
  {D.}~\bibnamefont {McPherson}}, \bibinfo {author} {\bibfnamefont
  {S.}~\bibnamefont {Meesala}}, \bibinfo {author} {\bibfnamefont
  {D.}~\bibnamefont {Meirom}}, \bibinfo {author} {\bibfnamefont
  {C.}~\bibnamefont {Mendell}}, \bibinfo {author} {\bibfnamefont
  {T.}~\bibnamefont {Metcalfe}}, \bibinfo {author} {\bibfnamefont
  {M.}~\bibnamefont {Mevissen}}, \bibinfo {author} {\bibfnamefont
  {A.}~\bibnamefont {Meyer}}, \bibinfo {author} {\bibfnamefont
  {A.}~\bibnamefont {Mezzacapo}}, \bibinfo {author} {\bibfnamefont
  {R.}~\bibnamefont {Midha}}, \bibinfo {author} {\bibfnamefont
  {D.}~\bibnamefont {Miller}}, \bibinfo {author} {\bibfnamefont
  {Z.}~\bibnamefont {Minev}}, \bibinfo {author} {\bibfnamefont
  {A.}~\bibnamefont {Mitchell}}, \bibinfo {author} {\bibfnamefont
  {N.}~\bibnamefont {Moll}}, \bibinfo {author} {\bibfnamefont {A.}~\bibnamefont
  {Montanez}}, \bibinfo {author} {\bibfnamefont {G.}~\bibnamefont {Monteiro}},
  \bibinfo {author} {\bibfnamefont {M.~D.}\ \bibnamefont {Mooring}}, \bibinfo
  {author} {\bibfnamefont {R.}~\bibnamefont {Morales}}, \bibinfo {author}
  {\bibfnamefont {N.}~\bibnamefont {Moran}}, \bibinfo {author} {\bibfnamefont
  {D.}~\bibnamefont {Morcuende}}, \bibinfo {author} {\bibfnamefont
  {S.}~\bibnamefont {Mostafa}}, \bibinfo {author} {\bibfnamefont
  {M.}~\bibnamefont {Motta}}, \bibinfo {author} {\bibfnamefont
  {R.}~\bibnamefont {Moyard}}, \bibinfo {author} {\bibfnamefont
  {P.}~\bibnamefont {Murali}}, \bibinfo {author} {\bibfnamefont
  {D.}~\bibnamefont {Murata}}, \bibinfo {author} {\bibfnamefont
  {J.}~\bibnamefont {M{\"u}ggenburg}}, \bibinfo {author} {\bibfnamefont
  {T.}~\bibnamefont {NEMOZ}}, \bibinfo {author} {\bibfnamefont
  {D.}~\bibnamefont {Nadlinger}}, \bibinfo {author} {\bibfnamefont
  {K.}~\bibnamefont {Nakanishi}}, \bibinfo {author} {\bibfnamefont
  {G.}~\bibnamefont {Nannicini}}, \bibinfo {author} {\bibfnamefont
  {P.}~\bibnamefont {Nation}}, \bibinfo {author} {\bibfnamefont
  {E.}~\bibnamefont {Navarro}}, \bibinfo {author} {\bibfnamefont
  {Y.}~\bibnamefont {Naveh}}, \bibinfo {author} {\bibfnamefont {S.~W.}\
  \bibnamefont {Neagle}}, \bibinfo {author} {\bibfnamefont {P.}~\bibnamefont
  {Neuweiler}}, \bibinfo {author} {\bibfnamefont {A.}~\bibnamefont {Ngoueya}},
  \bibinfo {author} {\bibfnamefont {T.}~\bibnamefont {Nguyen}}, \bibinfo
  {author} {\bibfnamefont {J.}~\bibnamefont {Nicander}}, \bibinfo {author}
  {\bibnamefont {Nick-Singstock}}, \bibinfo {author} {\bibfnamefont
  {P.}~\bibnamefont {Niroula}}, \bibinfo {author} {\bibfnamefont
  {H.}~\bibnamefont {Norlen}}, \bibinfo {author} {\bibnamefont {NuoWenLei}},
  \bibinfo {author} {\bibfnamefont {L.~J.}\ \bibnamefont {O'Riordan}}, \bibinfo
  {author} {\bibfnamefont {O.}~\bibnamefont {Ogunbayo}}, \bibinfo {author}
  {\bibfnamefont {P.}~\bibnamefont {Ollitrault}}, \bibinfo {author}
  {\bibfnamefont {T.}~\bibnamefont {Onodera}}, \bibinfo {author} {\bibfnamefont
  {R.}~\bibnamefont {Otaolea}}, \bibinfo {author} {\bibfnamefont
  {S.}~\bibnamefont {Oud}}, \bibinfo {author} {\bibfnamefont {D.}~\bibnamefont
  {Padilha}}, \bibinfo {author} {\bibfnamefont {H.}~\bibnamefont {Paik}},
  \bibinfo {author} {\bibfnamefont {S.}~\bibnamefont {Pal}}, \bibinfo {author}
  {\bibfnamefont {Y.}~\bibnamefont {Pang}}, \bibinfo {author} {\bibfnamefont
  {A.}~\bibnamefont {Panigrahi}}, \bibinfo {author} {\bibfnamefont {V.~R.}\
  \bibnamefont {Pascuzzi}}, \bibinfo {author} {\bibfnamefont {S.}~\bibnamefont
  {Perriello}}, \bibinfo {author} {\bibfnamefont {E.}~\bibnamefont {Peterson}},
  \bibinfo {author} {\bibfnamefont {A.}~\bibnamefont {Phan}}, \bibinfo {author}
  {\bibfnamefont {K.}~\bibnamefont {Pilch}}, \bibinfo {author} {\bibfnamefont
  {F.}~\bibnamefont {Piro}}, \bibinfo {author} {\bibfnamefont {M.}~\bibnamefont
  {Pistoia}}, \bibinfo {author} {\bibfnamefont {C.}~\bibnamefont {Piveteau}},
  \bibinfo {author} {\bibfnamefont {J.}~\bibnamefont {Plewa}}, \bibinfo
  {author} {\bibfnamefont {P.}~\bibnamefont {Pocreau}}, \bibinfo {author}
  {\bibfnamefont {A.}~\bibnamefont {Pozas-Kerstjens}}, \bibinfo {author}
  {\bibfnamefont {R.}~\bibnamefont {Pracht}}, \bibinfo {author} {\bibfnamefont
  {M.}~\bibnamefont {Prokop}}, \bibinfo {author} {\bibfnamefont
  {V.}~\bibnamefont {Prutyanov}}, \bibinfo {author} {\bibfnamefont
  {S.}~\bibnamefont {Puri}}, \bibinfo {author} {\bibfnamefont {D.}~\bibnamefont
  {Puzzuoli}}, \bibinfo {author} {\bibfnamefont {J.}~\bibnamefont {P{\'e}rez}},
  \bibinfo {author} {\bibnamefont {Quant02}}, \bibinfo {author} {\bibnamefont
  {Quintiii}}, \bibinfo {author} {\bibfnamefont {R.~I.}\ \bibnamefont
  {Rahman}}, \bibinfo {author} {\bibfnamefont {A.}~\bibnamefont {Raja}},
  \bibinfo {author} {\bibfnamefont {R.}~\bibnamefont {Rajeev}}, \bibinfo
  {author} {\bibfnamefont {I.}~\bibnamefont {Rajput}}, \bibinfo {author}
  {\bibfnamefont {N.}~\bibnamefont {Ramagiri}}, \bibinfo {author}
  {\bibfnamefont {A.}~\bibnamefont {Rao}}, \bibinfo {author} {\bibfnamefont
  {R.}~\bibnamefont {Raymond}}, \bibinfo {author} {\bibfnamefont
  {O.}~\bibnamefont {Reardon-Smith}}, \bibinfo {author} {\bibfnamefont
  {R.~M.-C.}\ \bibnamefont {Redondo}}, \bibinfo {author} {\bibfnamefont
  {M.}~\bibnamefont {Reuter}}, \bibinfo {author} {\bibfnamefont
  {J.}~\bibnamefont {Rice}}, \bibinfo {author} {\bibfnamefont {M.}~\bibnamefont
  {Riedemann}}, \bibinfo {author} {\bibnamefont {Rietesh}}, \bibinfo {author}
  {\bibfnamefont {D.}~\bibnamefont {Risinger}}, \bibinfo {author}
  {\bibfnamefont {M.~L.}\ \bibnamefont {Rocca}}, \bibinfo {author}
  {\bibfnamefont {D.~M.}\ \bibnamefont {Rodr{\'\i}guez}}, \bibinfo {author}
  {\bibnamefont {RohithKarur}}, \bibinfo {author} {\bibfnamefont
  {B.}~\bibnamefont {Rosand}}, \bibinfo {author} {\bibfnamefont
  {M.}~\bibnamefont {Rossmannek}}, \bibinfo {author} {\bibfnamefont
  {M.}~\bibnamefont {Ryu}}, \bibinfo {author} {\bibfnamefont {T.}~\bibnamefont
  {SAPV}}, \bibinfo {author} {\bibfnamefont {N.~R.~C.}\ \bibnamefont {Sa}},
  \bibinfo {author} {\bibfnamefont {A.}~\bibnamefont {Saha}}, \bibinfo {author}
  {\bibfnamefont {A.}~\bibnamefont {Ash-Saki}}, \bibinfo {author}
  {\bibfnamefont {S.}~\bibnamefont {Sanand}}, \bibinfo {author} {\bibfnamefont
  {M.}~\bibnamefont {Sandberg}}, \bibinfo {author} {\bibfnamefont
  {H.}~\bibnamefont {Sandesara}}, \bibinfo {author} {\bibfnamefont
  {R.}~\bibnamefont {Sapra}}, \bibinfo {author} {\bibfnamefont
  {H.}~\bibnamefont {Sargsyan}}, \bibinfo {author} {\bibfnamefont
  {A.}~\bibnamefont {Sarkar}}, \bibinfo {author} {\bibfnamefont
  {N.}~\bibnamefont {Sathaye}}, \bibinfo {author} {\bibfnamefont
  {B.}~\bibnamefont {Schmitt}}, \bibinfo {author} {\bibfnamefont
  {C.}~\bibnamefont {Schnabel}}, \bibinfo {author} {\bibfnamefont
  {Z.}~\bibnamefont {Schoenfeld}}, \bibinfo {author} {\bibfnamefont {T.~L.}\
  \bibnamefont {Scholten}}, \bibinfo {author} {\bibfnamefont {E.}~\bibnamefont
  {Schoute}}, \bibinfo {author} {\bibfnamefont {M.}~\bibnamefont
  {Schulterbrandt}}, \bibinfo {author} {\bibfnamefont {J.}~\bibnamefont
  {Schwarm}}, \bibinfo {author} {\bibfnamefont {J.}~\bibnamefont {Seaward}},
  \bibinfo {author} {\bibnamefont {Sergi}}, \bibinfo {author} {\bibfnamefont
  {I.~F.}\ \bibnamefont {Sertage}}, \bibinfo {author} {\bibfnamefont
  {K.}~\bibnamefont {Setia}}, \bibinfo {author} {\bibfnamefont
  {F.}~\bibnamefont {Shah}}, \bibinfo {author} {\bibfnamefont {N.}~\bibnamefont
  {Shammah}}, \bibinfo {author} {\bibfnamefont {R.}~\bibnamefont {Sharma}},
  \bibinfo {author} {\bibfnamefont {Y.}~\bibnamefont {Shi}}, \bibinfo {author}
  {\bibfnamefont {J.}~\bibnamefont {Shoemaker}}, \bibinfo {author}
  {\bibfnamefont {A.}~\bibnamefont {Silva}}, \bibinfo {author} {\bibfnamefont
  {A.}~\bibnamefont {Simonetto}}, \bibinfo {author} {\bibfnamefont
  {D.}~\bibnamefont {Singh}}, \bibinfo {author} {\bibfnamefont
  {D.}~\bibnamefont {Singh}}, \bibinfo {author} {\bibfnamefont
  {P.}~\bibnamefont {Singh}}, \bibinfo {author} {\bibfnamefont
  {P.}~\bibnamefont {Singkanipa}}, \bibinfo {author} {\bibfnamefont
  {Y.}~\bibnamefont {Siraichi}}, \bibinfo {author} {\bibnamefont {Siri}},
  \bibinfo {author} {\bibfnamefont {J.}~\bibnamefont {Sistos}}, \bibinfo
  {author} {\bibfnamefont {I.}~\bibnamefont {Sitdikov}}, \bibinfo {author}
  {\bibfnamefont {S.}~\bibnamefont {Sivarajah}}, \bibinfo {author}
  {\bibnamefont {Slavikmew}}, \bibinfo {author} {\bibfnamefont {M.~B.}\
  \bibnamefont {Sletfjerding}}, \bibinfo {author} {\bibfnamefont {J.~A.}\
  \bibnamefont {Smolin}}, \bibinfo {author} {\bibfnamefont {M.}~\bibnamefont
  {Soeken}}, \bibinfo {author} {\bibfnamefont {I.~O.}\ \bibnamefont {Sokolov}},
  \bibinfo {author} {\bibfnamefont {I.}~\bibnamefont {Sokolov}}, \bibinfo
  {author} {\bibfnamefont {V.~P.}\ \bibnamefont {Soloviev}}, \bibinfo {author}
  {\bibnamefont {SooluThomas}}, \bibinfo {author} {\bibnamefont {Starfish}},
  \bibinfo {author} {\bibfnamefont {D.}~\bibnamefont {Steenken}}, \bibinfo
  {author} {\bibfnamefont {M.}~\bibnamefont {Stypulkoski}}, \bibinfo {author}
  {\bibfnamefont {A.}~\bibnamefont {Suau}}, \bibinfo {author} {\bibfnamefont
  {S.}~\bibnamefont {Sun}}, \bibinfo {author} {\bibfnamefont {K.~J.}\
  \bibnamefont {Sung}}, \bibinfo {author} {\bibfnamefont {M.}~\bibnamefont
  {Suwama}}, \bibinfo {author} {\bibfnamefont {O.}~\bibnamefont {S{\l}owik}},
  \bibinfo {author} {\bibfnamefont {H.}~\bibnamefont {Takahashi}}, \bibinfo
  {author} {\bibfnamefont {T.}~\bibnamefont {Takawale}}, \bibinfo {author}
  {\bibfnamefont {I.}~\bibnamefont {Tavernelli}}, \bibinfo {author}
  {\bibfnamefont {C.}~\bibnamefont {Taylor}}, \bibinfo {author} {\bibfnamefont
  {P.}~\bibnamefont {Taylour}}, \bibinfo {author} {\bibfnamefont
  {S.}~\bibnamefont {Thomas}}, \bibinfo {author} {\bibfnamefont
  {K.}~\bibnamefont {Tian}}, \bibinfo {author} {\bibfnamefont {M.}~\bibnamefont
  {Tillet}}, \bibinfo {author} {\bibfnamefont {M.}~\bibnamefont {Tod}},
  \bibinfo {author} {\bibfnamefont {M.}~\bibnamefont {Tomasik}}, \bibinfo
  {author} {\bibfnamefont {C.}~\bibnamefont {Tornow}}, \bibinfo {author}
  {\bibfnamefont {E.}~\bibnamefont {de~la Torre}}, \bibinfo {author}
  {\bibfnamefont {J.~L.~S.}\ \bibnamefont {Toural}}, \bibinfo {author}
  {\bibfnamefont {K.}~\bibnamefont {Trabing}}, \bibinfo {author} {\bibfnamefont
  {M.}~\bibnamefont {Treinish}}, \bibinfo {author} {\bibfnamefont
  {D.}~\bibnamefont {Trenev}}, \bibinfo {author} {\bibnamefont {TrishaPe}},
  \bibinfo {author} {\bibfnamefont {F.}~\bibnamefont {Truger}}, \bibinfo
  {author} {\bibfnamefont {G.}~\bibnamefont {Tsilimigkounakis}}, \bibinfo
  {author} {\bibfnamefont {D.}~\bibnamefont {Tulsi}}, \bibinfo {author}
  {\bibfnamefont {W.}~\bibnamefont {Turner}}, \bibinfo {author} {\bibfnamefont
  {Y.}~\bibnamefont {Vaknin}}, \bibinfo {author} {\bibfnamefont {C.~R.}\
  \bibnamefont {Valcarce}}, \bibinfo {author} {\bibfnamefont {F.}~\bibnamefont
  {Varchon}}, \bibinfo {author} {\bibfnamefont {A.}~\bibnamefont {Vartak}},
  \bibinfo {author} {\bibfnamefont {A.~C.}\ \bibnamefont {Vazquez}}, \bibinfo
  {author} {\bibfnamefont {P.}~\bibnamefont {Vijaywargiya}}, \bibinfo {author}
  {\bibfnamefont {V.}~\bibnamefont {Villar}}, \bibinfo {author} {\bibfnamefont
  {B.}~\bibnamefont {Vishnu}}, \bibinfo {author} {\bibfnamefont
  {D.}~\bibnamefont {Vogt-Lee}}, \bibinfo {author} {\bibfnamefont
  {C.}~\bibnamefont {Vuillot}}, \bibinfo {author} {\bibfnamefont
  {J.}~\bibnamefont {Weaver}}, \bibinfo {author} {\bibfnamefont
  {J.}~\bibnamefont {Weidenfeller}}, \bibinfo {author} {\bibfnamefont
  {R.}~\bibnamefont {Wieczorek}}, \bibinfo {author} {\bibfnamefont {J.~A.}\
  \bibnamefont {Wildstrom}}, \bibinfo {author} {\bibfnamefont {J.}~\bibnamefont
  {Wilson}}, \bibinfo {author} {\bibfnamefont {E.}~\bibnamefont {Winston}},
  \bibinfo {author} {\bibnamefont {WinterSoldier}}, \bibinfo {author}
  {\bibfnamefont {J.~J.}\ \bibnamefont {Woehr}}, \bibinfo {author}
  {\bibfnamefont {S.}~\bibnamefont {Woerner}}, \bibinfo {author} {\bibfnamefont
  {R.}~\bibnamefont {Woo}}, \bibinfo {author} {\bibfnamefont {C.~J.}\
  \bibnamefont {Wood}}, \bibinfo {author} {\bibfnamefont {R.}~\bibnamefont
  {Wood}}, \bibinfo {author} {\bibfnamefont {S.}~\bibnamefont {Wood}}, \bibinfo
  {author} {\bibfnamefont {J.}~\bibnamefont {Wootton}}, \bibinfo {author}
  {\bibfnamefont {M.}~\bibnamefont {Wright}}, \bibinfo {author} {\bibfnamefont
  {L.}~\bibnamefont {Xing}}, \bibinfo {author} {\bibfnamefont {J.}~\bibnamefont
  {YU}}, \bibinfo {author} {\bibfnamefont {B.}~\bibnamefont {Yang}}, \bibinfo
  {author} {\bibfnamefont {U.}~\bibnamefont {Yang}}, \bibinfo {author}
  {\bibfnamefont {J.}~\bibnamefont {Yao}}, \bibinfo {author} {\bibfnamefont
  {D.}~\bibnamefont {Yeralin}}, \bibinfo {author} {\bibfnamefont
  {R.}~\bibnamefont {Yonekura}}, \bibinfo {author} {\bibfnamefont
  {D.}~\bibnamefont {Yonge-Mallo}}, \bibinfo {author} {\bibfnamefont
  {R.}~\bibnamefont {Yoshida}}, \bibinfo {author} {\bibfnamefont
  {R.}~\bibnamefont {Young}}, \bibinfo {author} {\bibfnamefont
  {J.}~\bibnamefont {Yu}}, \bibinfo {author} {\bibfnamefont {L.}~\bibnamefont
  {Yu}}, \bibinfo {author} {\bibfnamefont {C.}~\bibnamefont {Zachow}}, \bibinfo
  {author} {\bibfnamefont {L.}~\bibnamefont {Zdanski}}, \bibinfo {author}
  {\bibfnamefont {H.}~\bibnamefont {Zhang}}, \bibinfo {author} {\bibfnamefont
  {I.}~\bibnamefont {Zidaru}}, \bibinfo {author} {\bibfnamefont
  {B.}~\bibnamefont {Zimmermann}}, \bibinfo {author} {\bibfnamefont
  {C.}~\bibnamefont {Zoufal}}, \bibinfo {author} {\bibnamefont {aeddins ibm}},
  \bibinfo {author} {\bibnamefont {alexzhang13}}, \bibinfo {author}
  {\bibnamefont {b63}}, \bibinfo {author} {\bibnamefont {bartek bartlomiej}},
  \bibinfo {author} {\bibnamefont {bcamorrison}}, \bibinfo {author}
  {\bibnamefont {brandhsn}}, \bibinfo {author} {\bibnamefont {charmerDark}},
  \bibinfo {author} {\bibnamefont {deeplokhande}}, \bibinfo {author}
  {\bibnamefont {dekel.meirom}}, \bibinfo {author} {\bibnamefont {dime10}},
  \bibinfo {author} {\bibnamefont {dlasecki}}, \bibinfo {author} {\bibnamefont
  {ehchen}}, \bibinfo {author} {\bibnamefont {fanizzamarco}}, \bibinfo {author}
  {\bibnamefont {fs1132429}}, \bibinfo {author} {\bibnamefont {gadial}},
  \bibinfo {author} {\bibnamefont {galeinston}}, \bibinfo {author}
  {\bibnamefont {georgezhou20}}, \bibinfo {author} {\bibnamefont {georgios
  ts}}, \bibinfo {author} {\bibnamefont {gruu}}, \bibinfo {author}
  {\bibnamefont {hhorii}}, \bibinfo {author} {\bibnamefont {hykavitha}},
  \bibinfo {author} {\bibnamefont {itoko}}, \bibinfo {author} {\bibnamefont
  {jeppevinkel}}, \bibinfo {author} {\bibnamefont {jessica angel7}}, \bibinfo
  {author} {\bibnamefont {jezerjojo14}}, \bibinfo {author} {\bibnamefont
  {jliu45}}, \bibinfo {author} {\bibnamefont {jscott2}}, \bibinfo {author}
  {\bibnamefont {klinvill}}, \bibinfo {author} {\bibnamefont {krutik2966}},
  \bibinfo {author} {\bibnamefont {ma5x}}, \bibinfo {author} {\bibnamefont
  {michelle4654}}, \bibinfo {author} {\bibnamefont {msuwama}}, \bibinfo
  {author} {\bibnamefont {nico lgrs}}, \bibinfo {author} {\bibnamefont
  {nrhawkins}}, \bibinfo {author} {\bibnamefont {ntgiwsvp}}, \bibinfo {author}
  {\bibnamefont {ordmoj}}, \bibinfo {author} {\bibnamefont {sagar pahwa}},
  \bibinfo {author} {\bibnamefont {pritamsinha2304}}, \bibinfo {author}
  {\bibnamefont {ryancocuzzo}}, \bibinfo {author} {\bibnamefont {saktar unr}},
  \bibinfo {author} {\bibnamefont {saswati qiskit}}, \bibinfo {author}
  {\bibnamefont {septembrr}}, \bibinfo {author} {\bibnamefont {sethmerkel}},
  \bibinfo {author} {\bibnamefont {sg495}}, \bibinfo {author} {\bibnamefont
  {shaashwat}}, \bibinfo {author} {\bibnamefont {smturro2}}, \bibinfo {author}
  {\bibnamefont {sternparky}}, \bibinfo {author} {\bibnamefont {strickroman}},
  \bibinfo {author} {\bibnamefont {tigerjack}}, \bibinfo {author} {\bibnamefont
  {tsura crisaldo}}, \bibinfo {author} {\bibnamefont {upsideon}}, \bibinfo
  {author} {\bibnamefont {vadebayo49}}, \bibinfo {author} {\bibnamefont
  {welien}}, \bibinfo {author} {\bibnamefont {willhbang}}, \bibinfo {author}
  {\bibnamefont {wmurphy collabstar}}, \bibinfo {author} {\bibnamefont
  {yang.luh}}, \ and\ \bibinfo {author} {\bibfnamefont {M.}~\bibnamefont
  {{\v{C}}epulkovskis}},\ }\href {\doibase 10.5281/zenodo.2573505} {\enquote
  {\bibinfo {title} {Qiskit: An open-source framework for quantum computing},}\
  } (\bibinfo {year} {2021})\BibitemShut {NoStop}%
\bibitem [{\citenamefont {Damascelli}\ \emph {et~al.}(2003)\citenamefont
  {Damascelli}, \citenamefont {Hussain},\ and\ \citenamefont
  {Shen}}]{damascelli2003angle}%
  \BibitemOpen
  \bibfield  {author} {\bibinfo {author} {\bibfnamefont {A.}~\bibnamefont
  {Damascelli}}, \bibinfo {author} {\bibfnamefont {Z.}~\bibnamefont {Hussain}},
  \ and\ \bibinfo {author} {\bibfnamefont {Z.-X.}\ \bibnamefont {Shen}},\
  }\href@noop {} {\bibfield  {journal} {\bibinfo  {journal} {Reviews of modern
  physics}\ }\textbf {\bibinfo {volume} {75}},\ \bibinfo {pages} {473}
  (\bibinfo {year} {2003})}\BibitemShut {NoStop}%
\bibitem [{\citenamefont {Ament}\ \emph {et~al.}(2011)\citenamefont {Ament},
  \citenamefont {Van~Veenendaal}, \citenamefont {Devereaux}, \citenamefont
  {Hill},\ and\ \citenamefont {Van Den~Brink}}]{ament2011resonant}%
  \BibitemOpen
  \bibfield  {author} {\bibinfo {author} {\bibfnamefont {L.~J.}\ \bibnamefont
  {Ament}}, \bibinfo {author} {\bibfnamefont {M.}~\bibnamefont
  {Van~Veenendaal}}, \bibinfo {author} {\bibfnamefont {T.~P.}\ \bibnamefont
  {Devereaux}}, \bibinfo {author} {\bibfnamefont {J.~P.}\ \bibnamefont {Hill}},
  \ and\ \bibinfo {author} {\bibfnamefont {J.}~\bibnamefont {Van Den~Brink}},\
  }\href@noop {} {\bibfield  {journal} {\bibinfo  {journal} {Reviews of Modern
  Physics}\ }\textbf {\bibinfo {volume} {83}},\ \bibinfo {pages} {705}
  (\bibinfo {year} {2011})}\BibitemShut {NoStop}%
\end{thebibliography}%

\clearpage
\onecolumngrid
\appendix

\renewcommand\thefigure{S\arabic{figure}}  
\renewcommand\thetable{S\arabic{table}}  
\setcounter{figure}{0}

\section{Raw data and analysis for the electronic Green's function}\label{app:raw_data}

In this section, we provide the full data for $\mathcal{L}_k(t)$ and $|\mathcal{L}_k(\omega)|^2$ obtained via momentum selective linear response applied on SSH model. While the data in the bottom row is shown in Fig.~\ref{fig:ssh_model} as false color plots, here we provide line plots of the same data for clarity. For each $k$
and $\delta$ we collected 3 data sets with 8,000 shots each, yielding 24,000
shots total per curve. As discussed in the main text, $\mu=5$, $V_{nn}=1$ and the amplitude of the signal $\eta \Delta t = 0.04$. While obtaining the data we incorporated
dynamical decoupling and Pauli twirling as implemented in the {\em qiskit\_research} package, and did not apply any measurement error mitigation method.

\begin{figure*}[htpb]
    \centering
    \includegraphics[width=0.95\textwidth]{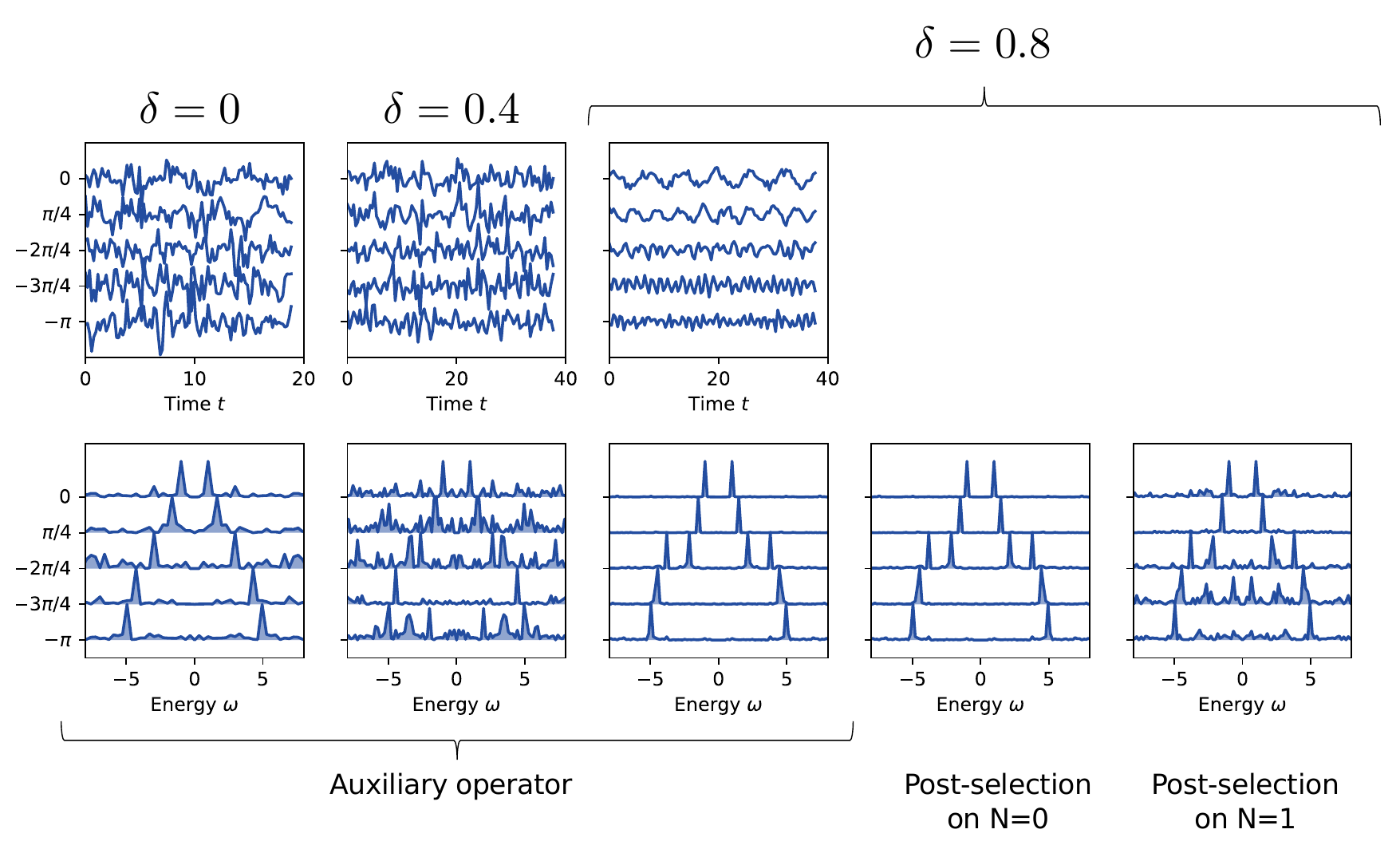}
    \caption{Data from {\em ibm\_auckland} for the three values of $\delta$ presented in the
    main text, as well as the corresponding $|\mathcal{L}_k(\omega)|^2$. As in the main text,
    $\mu=5$ and $V_{nn}=1$. Note that two
    copies of the Green's function appear at positive and negative energies (see
    text for discussion). }
    \label{fig:GF_fulldata}
\end{figure*}

\section{Raw data and analysis for the comparison of momentum-selective linear response, real space linear response, and Hadamard test}\label{app:comparison}

In order to make a comparison between the linear
response method in real and momentum space as well as the Hadamard test method, we performed noisy simulations for each.
We constructed a noise model by adding adjustable quantum errors to single and multi qubits gates. The model mainly depends on adding depolarizing quantum channels that mainly decohere qubits; the decoherence is either a result of phase flip or a bit flip or both. We added a fixed single-qubit depolarizing error with a 0.1\% rate and a 2-qubit depolarizing error once with a 10\% rate and once with a 20\% rate. 
In performing the calculations, we have forced
the noisy simulator to respect the linear
connectivity found on IBM quantum
computers. 

\begin{figure}[htpb]
    \centering
    \includegraphics[width=0.95\textwidth]{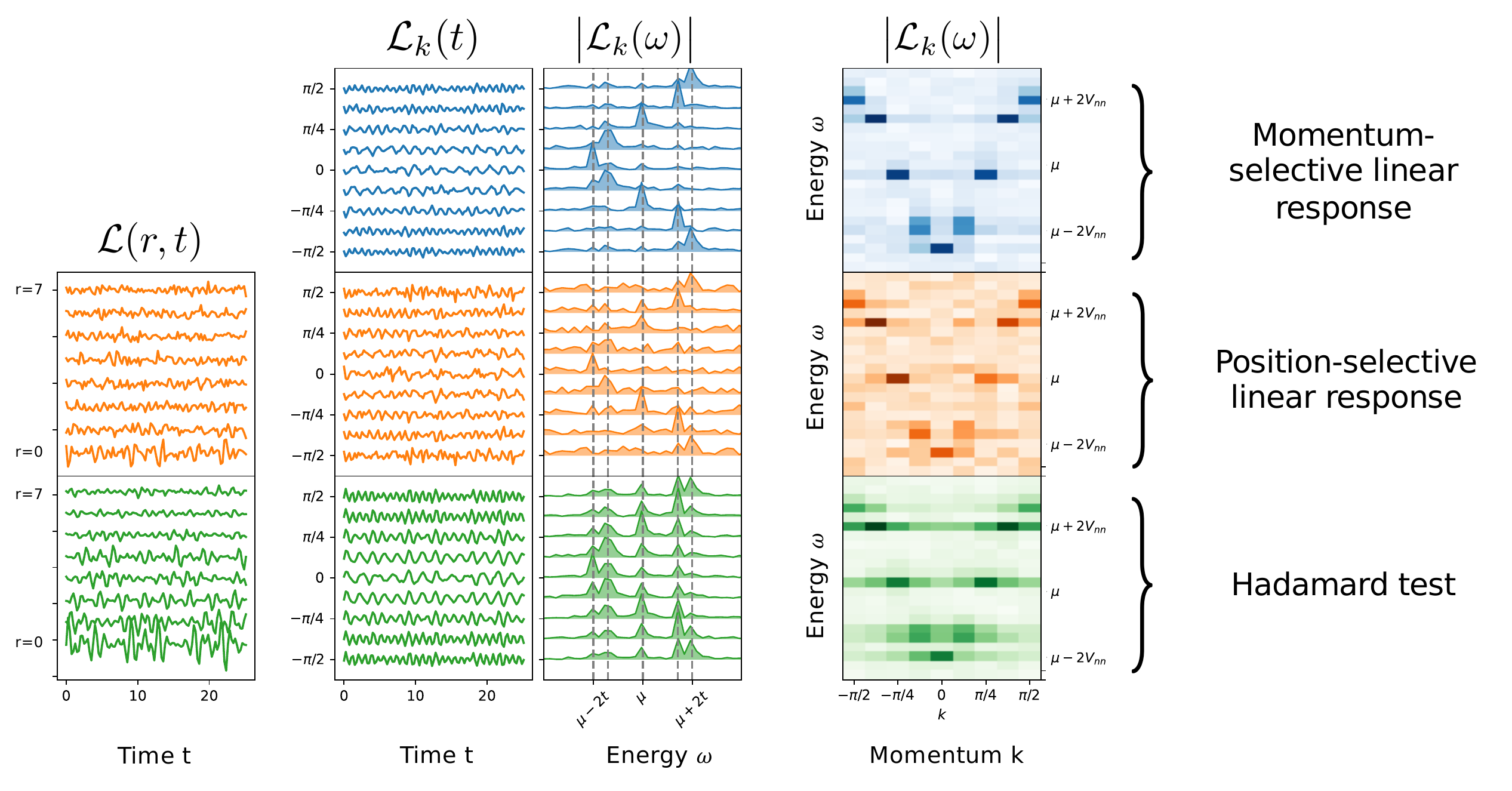}
    \caption{\textbf{Left:} Noisy simulator data of
    $\mathcal{L}(r,t)$. Note that the momentum-selective
    method avoids this step.
    \textbf{Center:} momentum-space Green's function
    as a function of time $t$ or frequency $\omega$.  \textbf{Right:} false-color plot
    of $|\mathcal{L}_k(\omega)|$.
    }
    \label{fig:simulator_detail}
\end{figure}

The results of the simulations, which are {$\mathcal{L}_k(t)$} for the momentum-selective linear response and {$\mathcal{L}(r,t)$} for the others, are shown in Fig.~\ref{fig:simulator_detail}.
The latter two are Fourier transformed to
{$\mathcal{L}_k(t)$} as well, and all three are further
transformed to $\mathcal{L}_k(\omega)$. As discussed
in the main text, and as is clear from the
both the line and false-color plots
of $|\mathcal{L}_k(\omega)|$, the momentum-selective 
linear response method outperforms the other
two in terms of signal to noise ratio.

\section{Derivation for obtaining $G^>$ and $G^<$ via post-selection}
\label{app:postselection}

\subsection{Post selection for a particle conserving Hamiltoniain for an $N$-particle initial state}

We will demonstrate that the lesser (occupied) and greater
(unoccupied) Green's functions can be directly obtained from the measurements
by post-selecting on the particle number. 
In order to do so, we will recast the circuit calculation in fermionic language.
Starting from an $N$-particle state $\ket{\Psi}$, we apply the momentum
creation operator $\mathcal{K}=\exp(-i 2\eta \sum_m \alpha_m \tilde{X}_m)$ where $\tilde{X}_m = Z_1...Z_{m-1}X_m$ which is equal to $(1/2)\left(c_m + c_m^\dagger\right)$ after a Jordan Wigner transformation, to find (to first order in $\eta$),
\begin{align}
    \mathcal{K} \ket{\Psi} \approx 
    \ket{\Psi} - i\eta \alpha_m c_m \ket{\Psi} - i\eta \alpha_m c_m^\dagger \ket{\Psi}.
    \label{eq:psi1}
\end{align}
Moreover, for notational clarity
we have suppressed internal sums over $m$ by using Einstein summation convention . We next apply
the time evolution operator $\mathcal{U}$, and since
we will be measuring the 1st qubit in the $X$ basis, we rotate it by about $y$,
\begin{align}
    R_{1y}(\pi/4) = \frac{1}{\sqrt{2}}\left( 1 + c_1^\dagger - c_1 \right).
\end{align}
Applying this to Eq.~\eqref{eq:psi1}, we find
\begin{align}
    \ket{\Phi^y} := R_{1y}(\pi/4)\mathcal{U} \mathcal{K} \ket{\Psi} = 
    \frac{1}{\sqrt{2}} & \Bigg(
    \mathcal{U}
     - i\eta \alpha_m \mathcal{U}c_m  - i\eta \alpha_m \mathcal{U} c_m^\dagger 
    \Bigg)\ket{\Psi} 
    \nonumber \\
    + \frac{1}{\sqrt{2}} & \Bigg(
    c^\dagger_1 \mathcal{U} - i\eta \alpha_m c^\dagger_1 \mathcal{U} c_m - i\eta \alpha_m c^\dagger_1 \mathcal{U} c_m^\dagger
    \Bigg)  \ket{\Psi} 
    \nonumber \\
    - \frac{1}{\sqrt{2}} & \Bigg(
    c_1 \mathcal{U} - i\eta \alpha_m c_1 \mathcal{U} c_m  - i\eta \alpha_m c_1 \mathcal{U} c_m^\dagger 
    \Bigg)\ket{\Psi}.
\end{align}
At this point, we can read off the particle number for each term.  
Since $\ket{\Psi}$ has $N$ particles and the Hamiltonian is particle conserving, counting the number of annihilation and creation operators we see that the resulting state is a superposition of $N-2, N-1, N, N+1$ and $N+2$ particle states. These states are
\begin{align}
    \ket{\Phi_{N-2}^y} &= i\frac{\eta}{\sqrt{2}} \alpha_m c_1 \mathcal{U} c_m \ket{\Psi}, \nonumber \\
    \ket{\Phi_{N-1}^y} &= \frac{1}{\sqrt{2}} \Big( - c_1 \mathcal{U} -i \eta \alpha_m \mathcal{U}c_m
     \Big) \ket{\Psi}, \nonumber \\
    \ket{\Phi_{N}^y} &=\frac{1}{\sqrt{2}}\Big( \mathcal{U}  + i \eta \alpha_m c_1 \mathcal{U}c_m^\dagger - i \eta \alpha_m c_1 ^\dagger \mathcal{U}c_m \Big) \ket{\Psi}, \nonumber \\
    \ket{\Phi_{N+1}^y} &=\frac{1}{\sqrt{2}}\Big(  c^\dagger_1 \mathcal{U} -i \eta \alpha_m \mathcal{U}c^\dagger_m \Big) \ket{\Psi}, \nonumber\\
    \ket{\Phi_{N+2}^y} &=-i\frac{\eta}{\sqrt{2}} \alpha_m c_1^\dagger \mathcal{U} c_m^\dagger \ket{\Psi}.
\end{align}
We will be measuring expectation values with these states. It will be mainly their norms and expectation value of $c^\dagger_1 c_1$ which can be obtained via $Z_1$ measurement. Observing that $\ket{\Phi_{N \pm 2}} \sim \eta$, they will not contribute up to linear order in $\eta$. 
Independent quantities to linear order in $\eta$ are
\begin{align}
\begin{split}
    \braket{\Phi_{N-1}^y|\Phi_{N-1}^y} &= \frac{1}{2} \braket{n_1(t)} + \frac{i\eta \alpha_m}{2} \left( \braket{c_1^\dagger(t) c_m} - \braket{c_m^\dagger c_1(t)} \right) = \frac{1}{2} \braket{n_1(t)} - \eta \alpha_m \mathrm{Re} \: G^<_{1m}(t)   \\
    \braket{\Phi_{N+1}^y|\Phi_{N+1}^y} &= \frac{1}{2} - \frac{1}{2} \braket{n_1(t)} + \frac{i\eta \alpha_m}{2} \left( \braket{c_m c_1^\dagger(t) } - \braket{c_1(t) c_m^\dagger} \right) = \frac{1}{2} - \frac{1}{2} \braket{n_1(t)} + \eta \alpha_m \mathrm{Re} \: G^>_{1m}(t) \\
    \braket{\Phi_{N}^y|c_1^\dagger c_1|\Phi_{N}^y} &= \frac{1}{2}\braket{n_1(t)} + \frac{i\eta \alpha_m}{2} \left( \braket{c_m^\dagger c_1(t)} - \braket{c_1^\dagger(t) c_m} \right) = \frac{1}{2} \braket{n_1(t)} + \eta \alpha_m \mathrm{Re} \: G^<_{1m}(t)
\end{split}
\end{align}
This leads to first two equations of Eq.\eqref{eq:postselection}.
Instead, if we apply a rotation around $x$ we get:
\begin{align}
    R_{1x}(\pi/4) = \frac{1}{\sqrt{2}}\left( 1 + ic_1^\dagger +i c_1 \right),
\end{align}
then
\begin{align}
    \ket{\Phi^x} := R_{1x}(\pi/4)\mathcal{U} \mathcal{K} \ket{\Psi} = 
    \frac{1}{\sqrt{2}} & \Bigg(
    \mathcal{U}
     - i\eta \alpha_m \mathcal{U}c_m  - i\eta \alpha_m \mathcal{U} c_m^\dagger 
    \Bigg)\ket{\Psi} 
    \nonumber \\
    + \frac{i}{\sqrt{2}} & \Bigg(
    c^\dagger_1 \mathcal{U} - i\eta \alpha_m c^\dagger_1 \mathcal{U} c_m - i\eta \alpha_m c^\dagger_1 \mathcal{U} c_m^\dagger
    \Bigg)  \ket{\Psi} 
    \nonumber \\
    + \frac{i}{\sqrt{2}} & \Bigg(
    c_1 \mathcal{U} - i\eta \alpha_m c_1 \mathcal{U} c_m  - i\eta \alpha_m c_1 \mathcal{U} c_m^\dagger 
    \Bigg)\ket{\Psi}.
\end{align}
Then, the components with different particle number are
\begin{subequations}
\begin{align}
    \ket{\Phi_{N-2}^x} &= \frac{\eta}{\sqrt{2}} \alpha_m c_1 \mathcal{U} c_m \ket{\Psi},  \\
    \ket{\Phi_{N-1}^x} &= \frac{1}{\sqrt{2}} \Big( i c_1 \mathcal{U} -i \eta \alpha_m \mathcal{U}c_m
     \Big) \ket{\Psi},  \\
    \ket{\Phi_{N}^x} &=\frac{1}{\sqrt{2}}\Big( \mathcal{U}  +  \eta \alpha_m c_1 \mathcal{U}c_m^\dagger + \eta \alpha_m c_1 ^\dagger \mathcal{U}c_m \Big) \ket{\Psi},  \\
    \ket{\Phi_{N+1}^x} &=\frac{1}{\sqrt{2}}\Big(  ic^\dagger_1 \mathcal{U} -i \eta \alpha_m \mathcal{U}c^\dagger_m \Big) \ket{\Psi}, \\
    \ket{\Phi_{N+2}^x} &=\frac{\eta}{\sqrt{2}} \alpha_m c_1^\dagger \mathcal{U} c_m^\dagger \ket{\Psi}.
\end{align}
\end{subequations}
To linear order in $\eta$, independent quantities that can be derived from norms and the expectation value of $Z_1$ are
\begin{align}
\begin{split}
    \braket{\Phi_{N-1}^x|\Phi_{N-1}^x} &= \frac{1}{2} \braket{n_1(t)} - \frac{\eta \alpha_m}{2} \left( \braket{c_1^\dagger(t) c_m} + \braket{c_m^\dagger c_1(t)} \right) 
    \nonumber \\
    &= \frac{1}{2} \braket{n_1(t)} - \eta \alpha_m \mathrm{Im}\: G^<_{1m}(t)  \\
    \braket{\Phi_{N+1}^x|\Phi_{N+1}^x} &= \frac{1}{2} - \frac{1}{2} \braket{n_1(t)} - \frac{\eta \alpha_m}{2} \left( \braket{c_m c_1^\dagger(t) } + \braket{c_1(t) c_m^\dagger} \right) 
    \nonumber \\
    &= \frac{1}{2} - \frac{1}{2} \braket{n_1(t)} + {\eta \alpha_m} \mathrm{Im}\:G^>_{1m}(t) \\
    \braket{\Phi_{N}^x|c_1^\dagger c_1|\Phi_{N}^x} &= \frac{1}{2}\braket{n_1(t)} + \frac{\eta \alpha_m}{2} \left( \braket{c_m^\dagger c_1(t)} + \braket{c_1^\dagger(t) c_m} \right)
    \nonumber \\
    &= \frac{1}{2} \braket{n_1(t)} + \eta \alpha_m \mathrm{Im}\: G^<_{1m}(t) 
\end{split}
\end{align}
These can be linearly combined to obtain
the final two equations of Eq.\eqref{eq:postselection} in
the main text, and shows that retarded, lesser and greater fermionic Green's functions can be calculated via post selection.

\subsection{Post selection for SSH model for 0-particle initial state}\label{sec:ssh_0_particle}

In this case our calculation simplifies drastically, since we cannot annihilate a particle from a 0-particle state, and thus the only contribution will come from 0, 1 and 2 particle states:
\begin{align}
    \ket{\Phi_{0}^y} &=\frac{1}{\sqrt{2}}\Big( \mathcal{U}  + i \eta \alpha_m c_1 \mathcal{U}c_m^\dagger - i \eta \alpha_m c_1 ^\dagger \mathcal{U}c_m \Big) \ket{0} \nonumber \\
    &=\frac{1}{\sqrt{2}}\Big( \mathcal{U}  + i \eta \alpha_m c_1 \mathcal{U}c_m^\dagger \Big) \ket{0}, \nonumber \\
    \ket{\Phi_{1}^y} &=\frac{1}{\sqrt{2}}\Big(  c^\dagger_1 \mathcal{U} -i \eta \alpha_m \mathcal{U}c^\dagger_m \Big) \ket{0}, \nonumber\\
    \ket{\Phi_{2}^y} &=-i\frac{\eta}{\sqrt{2}} \alpha_m c_1^\dagger \mathcal{U} c_m^\dagger \ket{0}.
\end{align}
The norm of the 2-particle contribution is $O(\eta^2)$ and is neglected. The norms of 0- and 1-particle contributions are
\begin{align}
    \braket{\Phi_{0}^x|\Phi_{0}^y} &= \frac{1}{2} - \eta \alpha_m \mathrm{Re\:}G^>_{1m}(t), \nonumber \\
    \braket{\Phi_{1}^x|\Phi_{1}^y} &= \frac{1}{2} + \eta \alpha_m \mathrm{Re\:}G^>_{1m}(t). \nonumber
\end{align}
Because the lesser Green's function of the 0-particle state $\ket{0}$ is zero, we can replace the greater Green's functions with the retarded ones:
\begin{align}
    \braket{\Phi_{0}^x|\Phi_{0}^y} &= \frac{1}{2} - \eta \alpha_m \mathrm{Re\:}G^R_{1m}(t), \nonumber \\
    \braket{\Phi_{1}^x|\Phi_{1}^y} &= \frac{1}{2} + \eta \alpha_m \mathrm{Re\:}G^R_{1m}(t), \nonumber
\end{align}
and therefore these partial norms contain information about the single-particle energy spectrum.

\section{Quantum circuit for the SSH Model
Green's function}\label{app:circuit}

The circuit in Fig.~\ref{fig:ssh_model}\textbf{b} mainly consists of three parts: the applied field $\op{B}h(t)$, the time evolution and the measurement of $\op{A}$. Here we will discuss how to use the parity operator as an auxiliary 
to measure fermionic Green's functions.

\subsection{Measurement of $\mathrm{Re\:}G_k(\omega)$}

In this work, we measure the following quantity for SSH model on no particle state $\ket{0}$
\begin{align}
    \mathcal{L}_k(t) = -i \braket{0 | \Big[ X_0(t), \sum_r \cos(kr) X_r \Big] |0}.
\end{align}
We will slowly change this into an expression given in terms of Green's functions, and show that $\mathcal{L}_k(t)$ contains information about the one particle spectral weight.

First observe that $\tilde{X}_r \ket{0} = X_r \ket{0}$. In addition, for $P = Z_0 Z_1 ... Z_n$ we have $P \ket{0} = \ket{0}$, therefore 
\begin{align}\label{seq:reverse_auxiliary}
\begin{split}
    \mathcal{L}_k(t) &= -i \braket{0 | \Big[ {X}_0(t), \sum_r \cos(kr) \tilde{X}_r \Big] P |0} \\
    &= i\braket{0 | \Big\{ {X}_0(t) P, \sum_r \cos(kr) \tilde{X}_r \Big\} |0} \\
    &= i\braket{0 | \Big\{ {X}_0(t) P(t), \sum_r \cos(kr) \tilde{X}_r \Big\} |0}.
\end{split}
\end{align}
On the last line, we have used $P(t) = P$. {Eq. \ref{seq:reverse_auxiliary} is essentially the auxiliary operator method given in the manuscript applied in reverse way to transform commutator into anti-commutator.} 
Now $X_0 P = -i Y_0 Z_1 ... Z_{n-1}$, and $Z_1 ... Z_{n-1} \ket{0} = \ket{0}$. In addition, $\tilde{Y}_0 = Y_0$, then  
\begin{align}
    \mathcal{L}_k(t) &= \braket{0 | \Big\{ \tilde{Y}_0(t), \sum_r \cos(kr) \tilde{X}_r \Big\} |0}.
\end{align}
Applying the Jordan-Wigner transformation to get the Fermionic operators back, let us plug in $\tilde{X}_r = c_r + c^\dagger_r$ and $\tilde{Y}_0 = i (c_0^\dagger - c_0)$:
\begin{align}
\begin{split}
    \sum_r \cos(kr) \tilde{X}_r &= \frac{1}{2} \sum_r (e^{ikr} + e^{-ikr}) (c_r + c^\dagger_r) 
   = \frac{\sqrt{n}}{2} ( c_k + c_{-k} + c^\dagger_k + c^\dagger_{-k})\\
    \tilde{Y}_0 &= i(c^\dagger_0 - c_0) = \frac{i}{\sqrt{n}} \sum_q (c^\dagger_q - c_q).
\end{split}
\end{align}
With these, we obtain
\begin{align}
\begin{split}
    \mathcal{L}_k(t) &= -\frac{i}{2}\braket{0 | \Big\{ \sum_{q} (c_q(t) - c^\dagger_q(t)), ( c_k + c_{-k} + c^\dagger_k + c^\dagger_{-k})  \Big\} |0}. \\
    &= -\frac{i}{2} \sum_{q} \Big( \braket{0 | \{ c_q(t), c^\dagger_k + c^\dagger_{-k} \} |0} - \braket{0 | \{ c^\dagger_q(t),  c_k + c_{-k}\} |0} \Big). \\
\end{split}
\end{align}
The sum can be handled directly because momentum is conserved due to translational invariance of the SSH model, then the creation/annihilation operators anticommute when momentum values are not matched. Since we assume $t>0$, we can plug in $\theta(t) = 1$ in the definition of $G^R_k(t)$, and then obtain
\begin{align}
\begin{split}
    \mathcal{L}_k(t) &= \frac{1}{2} \big( G^R_k(t) + G^R_{-k}(t) + G^R_k(t)^* + G^R_{-k}(t)^*\big) \\
    &= \: \mathrm{Re} \big( G^R_k(t) + G^R_{-k}(t)\big).
\end{split}
\end{align}
The SSH model is symmetric under spatial reflection, thus $G^R_k = G^R_{-k}$ and we get
\begin{align}
    \mathcal{L}_k(t) = 2 \:\mathrm{Re}\: G^R_k(t)
\end{align}

Let us look at this in the frequency basis:
\begin{align}
\begin{split}
    \mathcal{L}_k(\omega) &= \int \mathrm{d}t\: \mathcal{L}_k(t) e^{i \omega t} \\
    &= 2 \int \mathrm{d}t \: \mathrm{Re}\: G^R_k(t) e^{i \omega t} \\
    &=  \int \mathrm{d}t \: \big(G^R_k(t) + G^R_k(t)^* \big) e^{i \omega t} \\
    &= \: G^R_k(\omega) + \: G^R_k(-\omega)^*
\end{split}
\end{align}
Then we get
\begin{align}
\begin{split}
    \mathrm{Im}\:\mathcal{L}_k(\omega) &= \mathrm{Im}\:G^R_k(\omega)-\mathrm{Im}\:G^R_k(-\omega), \\
    \mathrm{Re}\:\mathcal{L}_k(\omega) &= \mathrm{Re}\:G^R_k(\omega)+\mathrm{Re}\:G^R_k(-\omega),
\end{split}
\end{align}
which means that choosing $\op{A}$ and $\op{B}$ as given in the beginning of this subsection, we can get single particle spectral weight. 

Both momentum selective and position selective methods can be used to measure $\mathcal{L}_k(t)$. The momentum selective method is to excite the state $\ket{0}$ with $\op{B} = \sum_r \cos(kr) X_r$ and measuring $X_0$ after time evolution, which can measure $\mathcal{L}_k(t)$ with one circuit. The position selective method is to measure $X_0$ after exciting the state with $\op{B} = X_r$ and time evolving for all $r=1,2,...,n$  values.

\begin{figure}[htpb]
    \centering
    \includegraphics[ width=.65\textwidth]{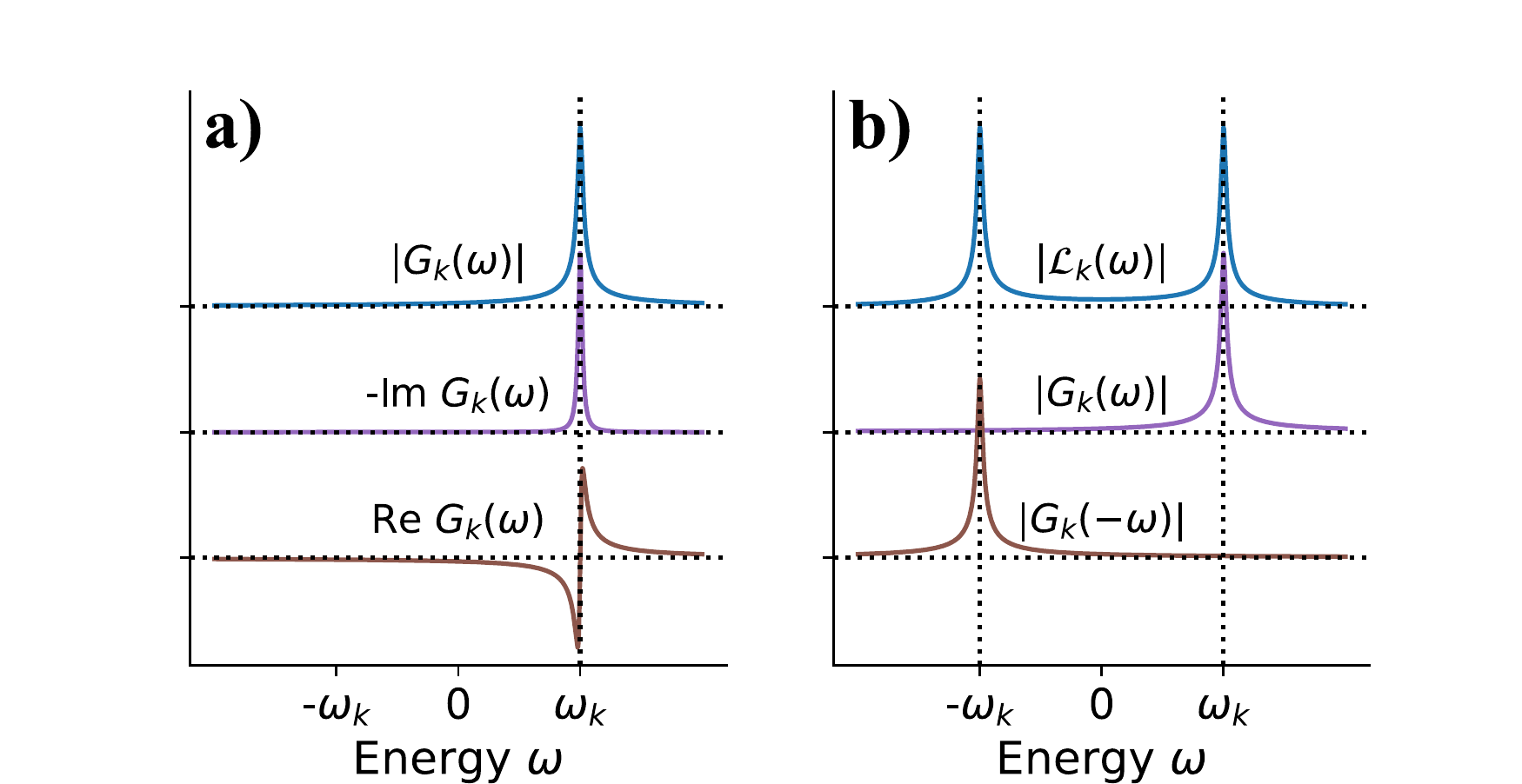}
    \caption{Panel \textbf{a} illustrates that $|G_k^R(\omega)|$ has the same spectral behaviour as $\mathrm{Im}\:G_k^R(\omega)$, and therefore carries information about the single particle spectral weight. Panel \textbf{b} illustrates that this spectral information can be extracted from $|\mathcal{L}_k(\omega)|$ since it has two distinctive peaks, one coming from $|G_k^R(\omega)|$, the other from $|G_k^R(-\omega)|$.
    }
    \label{fig:GvsL}
\end{figure}

For the data shown in Figs.~\ref{fig:ssh_model} and \ref{fig:simulator}, we {only} measure $\mathcal{L}_k(t)$ in linear response methods with momentum and position selectivity, {to run fewer circuits}. The plots show
$|\mathcal{L}_k(\omega)|^2$, which is related to the
retarded Green's function as
\begin{align}\label{seq:interference}
\begin{split}
    |\mathcal{L}_k(\omega)|^2 =& \Big( G_k^R(\omega) + G_k^R(-\omega)^* \Big)\Big( G_k^R(\omega)^* + G_k^R(-\omega) \Big), \\ 
    =& |G_k^R(\omega)|^2 + |G_k^R(-\omega)|^2 + 2 \mathrm{\,Re} \Big( G_k^R(\omega) G_k^R(-\omega)^* \Big).
\end{split}
\end{align} 
Because $G^R_k(\omega)$ is strongly peaked near the single-particle
excitation energy,
the interference term is negligible compared to the absolute squares of the terms $G^R_k(\omega)$ and $G^R_k(-\omega)$. Therefore $|\mathcal{L}_k(\omega)|^2$ contains $|G_k^R(\omega)|^2$ and its mirror image in the plots. Due to our chemical potential choice $\mu = 5$, in the positive frequencies, we only see one of these images, which gives us the information about the single particle spectrum.
{Fig.~\ref{fig:GvsL} illustrates this point by showing that $|\mathcal{L}_k(\omega)|$ tracks the quasi-particle peaks in $|G^R_k(\omega)|$ and $\mathrm{Im}\:G^R_k(\omega)$. In the figure, we compare the real part, imaginary part and absolute value of $G^R_k(\omega)$
\begin{align}
    G^R_k(\omega) = \frac{1}{\omega - \omega_k + i\epsilon},
\end{align}
where $\epsilon = 0.1$, and we picked $\omega_k = 5$.
As it can be seen on panel \textbf{a}, $|G^R(\omega)|$ is peaked at $\omega_k$ just as
$\mathrm{Im}\:G^R_k(\omega)$, with a slightly broader peak.. Panel \textbf{b} shows that $|\mathcal{L}_k(\omega)|$ is small except at $\omega=\pm\omega_k$ energies, which are the peaks of $|G^R(\omega)|$ and $|G^R(-\omega)|$. This provides an illustration of the fact that the interference term in Eq.~\ref{seq:interference} is indeed negligible.}

\subsection{Time Evolution Circuit}\label{app:time_evolution_circuit}

The SSH model is a free fermionic model and thus its time evolution can be compressed into a fixed depth circuit with $O(n^2)$ CNOTs and $O(n)$ depth, where $n$ is the system size, via the algebraic compression method given in  \cite{kokcu2022algebraic,camps2022algebraic}. The method is limited to free fermionic systems
in 1D --- here we use a generalization to
1D periodic systems, which will be detailed in
a forthcoming publication.

\begin{figure}[htpb]
    \centering
    \includegraphics[width = 0.3\columnwidth]{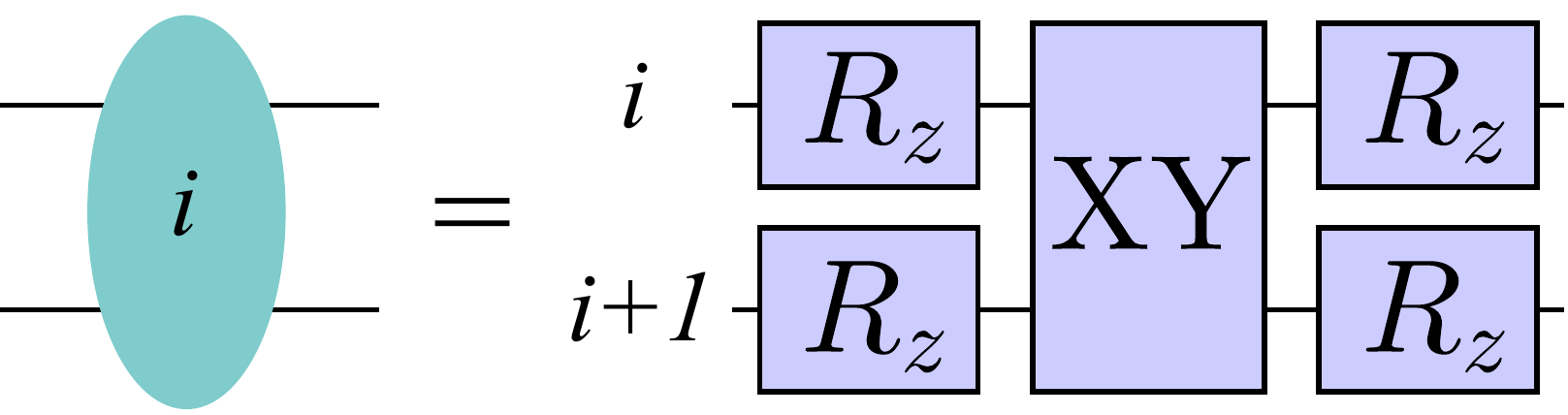}
    \caption{Block given in \eqref{seq:tfxyblock} represented as a 2 qubit gate. $XY$ indicates a rotation about $XX$ followed by $YY$\cite{kokcu2022algebraic,camps2022algebraic}.  $R_Z$, and the $XX$ and $YY$ rotations have independent rotation angles.}
    \label{fig:block_definition}
\end{figure}

\begin{figure}[htpb]
    \centering
    \includegraphics[width = 0.9\columnwidth]{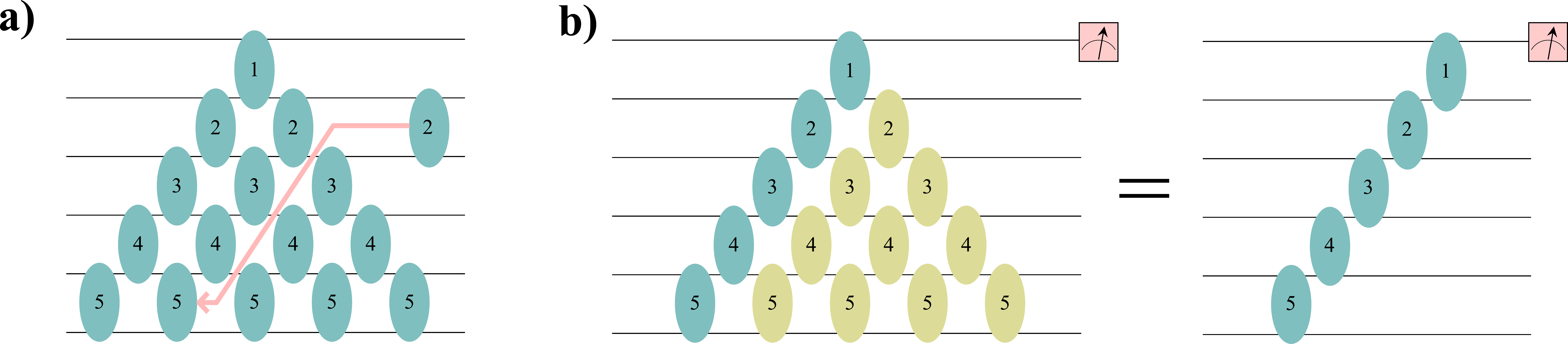}
    \caption{Panel \textbf{a} demonstrates the triangle structure and how it can absorb a block. Block with index 2 goes down with a series of turnover operations, and ends up merging the block at the end of the arrow. Panel \textbf{b} demonstrates the additional simplification due to measurement. Blocks with gold color has no effect on the measurement on the $0$th qubit, and therefore can be discarded, reducing the CNOT count from $O(n^2)$ to $O(n)$.}
    \label{fig:compressionn}
\end{figure}

For completeness, we will summarize the method for the open 1-D chain. The method relies on a structure called a ``block", and is given as the following for free fermionic models (after performing the  Jordan-Wigner transformation):
\begin{align}\label{seq:tfxyblock}
    B_{i}(\vec{\theta}) \equiv& e^{-i \theta_1\: Z_i}e^{-i \theta_2\: Z_{i+1}}e^{-i \theta_3\: X_i X_{i+1}}e^{-i \theta_4\: Y_i Y_{i+1}}\:e^{-i \theta_5\: Z_i}e^{-i \theta_6\: Z_{i+1}}. 
\end{align}
We represent it as the diagram shown in Fig.~\ref{fig:block_definition}.
In Ref.~\onlinecite{kokcu2022algebraic} it is proven that  $B_i(\vec{\theta})$ satisfies the following properties:
\begin{enumerate}
  \item \textbf{Fusion:} for any set of parameters $\vec{\alpha}$ and $\vec{\beta}$, there exists an $\vec{a}$ such that
        \begin{equation}
          B_i(\vec{\alpha}) \, B_i(\vec{\beta}) = B_i(\vec{a}),
        \end{equation}
  \item \textbf{Commutation:} for any set of parameters $\vec{\alpha}$ and $\vec{\beta}$, we have
        \begin{equation}
          B_i(\vec{\alpha}) \, B_j(\vec{\beta}) = B_j(\vec{\beta}) \, B_i(\vec{\alpha}), \qquad |i-j|>1,
        \end{equation}
  \item \textbf{Turnover:} for any set of parameters $\vec{\alpha}$, $\vec{\beta}$ and $\vec{\gamma}$ there exist $\vec{a}$, $\vec{b}$ and $\vec{c}$ such that 
        \begin{equation}
          B_i(\vec{\alpha}) \, B_{i+1}(\vec{\beta}) \, B_i(\vec{\gamma}) = B_{i+1}(\vec{a}) \, B_i(\vec{b}) \, B_{i+1}(\vec{c}).
        \end{equation}
\end{enumerate}
These properties can be exploited to 
build the triangle structure shown in Fig.~\ref{fig:compressionn} \textbf{a}, which can absorb any additional block by simple parameter changes. Calculation of the parameters can be done directly via linear algebra operations without any variational calculation, the details of which are given in \cite{camps2022algebraic}.

For the momentum selective case we only need to measure the $0$th site, and thus the information on the other qubits are not relevant. As shown in Fig.~\ref{fig:compressionn} \textbf{b}, because the measurement is on qubit 0, blocks that do not affect qubit zero can be pushed after the measurement, and therefore can be ignored. 
Although post-selection requires measurement of all qubits, this simplification can still be done simply because the only information used
from the other qubits is the particle number,
and the TFXY blocks do
not change the particle number.

The triangle structure CNOT count is $n(n-1)/2$, which is 28 for the $n=8$ calculations
presented in the main text. After this measurement simplification, the CNOT count decreases to $2(n-1)$, or 14 for our calculations.

\section{Hardware Calibration Details}
\label{sec:hardware}

The results from the quantum computer shown in 
Fig.~\ref{fig:ssh_model} were run on
{\em ibmq\_auckland} via the IBM Quantum Experience.
Calibration information for the two dates we collected data are shown in Tables \ref{table:auckland} and \ref{table:auckland2} and were obtained from the Qiskit API~\cite{Qiskit}.

\begin{table}[htpb]
\centering
\begin{tabular}{|c| c| c |c| c |c| } 
 \hline
 Qubits & T1 ($\mu s$) & T2 ($\mu s$) & readout &  CNOT & CNOT   \\
  & &  & error (\%) & connection & error (\%) \\
  \hline \hline
 13 & 140 & 27.6 & 0.56 & 13-12 &  0.536\\ 
 \hline
 12 & 256 & 232 & 1.84 & 12-10 & 0.835 \\ 
 \hline
 10 & 225 & 49.2 & 0.91 & 10-7 & 0.544 \\ 
 \hline
 7 & 130 & 218 & 0.92 & 7-4 & 0.933 \\ 
 \hline
 4 & 176 & 164 & 1.85 & 4-1 & 1.08 \\ 
 \hline
 1 & 52.7 & 135 & 0.95 & 1-2 & 0.594 \\ 
   \hline
 2 & 173 & 177 & 1.61 &  2-3 & 0.516 \\ 
 \hline
 3 & 104 & 67.6 & 1.55 &  &  \\ 
 \hline
\end{tabular}
\caption{Calibration data for \emph{ibmq\_auckland} on September 19th, 2022.}
\label{table:auckland}
\end{table}
\begin{table}[htpb]
\centering
\begin{tabular}{|c| c| c |c| c |c| } 
 \hline
 Qubits & T1 ($\mu s$) & T2 ($\mu s$) & readout &  CNOT & CNOT   \\
  & &  & error (\%) & connection & error (\%) \\
  \hline \hline
 0 & 372 & 420 & 0.80 & 0-1 &  0.504\\ 
 \hline
 1 & 398 & 285 & 1.02& 1-2 & 0.763 \\ 
 \hline
 2 & 528 & 413 & 0.86 & 2-3 & 0.382 \\ 
 \hline
 3 & 365 & 131 & 1.29 & 3-5 & 0.340 \\ 
 \hline
 5 & 253 & 371 & 21.7 & 5-8 & 0.428 \\ 
 \hline
 8 & 353 & 109 & 1.67 & 8-11 & 0.371 \\ 
   \hline
 11 & 146 & 309 & 1.10 &  11-14 & 0.441 \\ 
 \hline
 14 & 468 & 129 & 22.9 &  &  \\ 
 \hline
\end{tabular}
\caption{Calibration data for \emph{ibmq\_auckland} on September 30th, 2022.}
\label{table:auckland2}
\end{table}

\end{document}